# A guide to choosing and implementing reference models for social network analysis

Elizabeth A. Hobson[1,*], Matthew J. Silk[2,*], Nina H. Fefferman[3], Daniel B. Larremore[4], Puck Rombach[5], Saray Shai[6], and Noa Pinter-Wollman[7]

[1]Department of Biological Sciences, University of Cincinnati, Ohio USA

[2]Centre for Ecology and Conservation, University of Exeter Penryn Campus, Penryn, Cornwall, UK

[3]Departments of Ecology and Evolutionary Biology & Mathematics, University of Tennessee, Knoxville, TN, USA

[4]Department of Computer Science & BioFrontiers Institute, University of Colorado Boulder, Boulder, CO, USA

[5]Department of Mathematics & Statistics, University of Vermont, VT, USA

[6]Department of Mathematics and Computer Science, Wesleyan University, Middletown, CT, USA

[7]Department of Ecology and Evolutionary Biology, University of California, Los Angeles, CA, USA

[*]Equal contribution

## Abstract

Analyzing social networks is challenging. Key features of relational data require the use of non-standard statistical methods such as developing system-specific null, or reference, models that randomize one or more components of the observed data. Here we review a variety of randomization procedures that generate reference models for social network analysis. Reference models provide an expectation for hypothesis-testing when analyzing network data. We outline the key stages in producing an effective reference model and detail four approaches for generating reference distributions: permutation, resampling, sampling from a distribution, and generative models. We highlight when each type of approach would be appropriate and note potential pitfalls for researchers to avoid. Throughout, we illustrate our points with examples from a simulated social system. Our aim is to provide social network researchers with a deeper understanding of analytical approaches to enhance their confidence when tailoring reference models to specific research questions.

# Contents

## 1. Introduction

Individuals interact with each other in many ways but determining why they interact and uncovering the function of social patterns, i.e., the social network, is challenging. Network theory has provided useful tools to quantify patterns of social interactions (Borgatti et al., 2009; Croft et al., 2008; Wasserman and Faust, 1994). The analysis of social networks is complicated by the fact that applying statistical inference using standard methods is often not appropriate because of the inherent dependence of individuals within a network (for example, the actions of one individual are linked to the actions of another) (Croft et al., 2011). Network methods have emerged as a powerful set of tools with which to analyze social systems (Butts, 2008; Cranmer et al., 2017; Farine and Whitehead, 2015; Fisher et al., 2017; Pinter-Wollman et al., 2014; Silk et al., 2017a,b; Sosa et al., 2020; Wey et al., 2008). Using network tools can often be difficult because important assumptions can be cryptic and do not apply universally across all suites of research questions or data types. Network analyses have been used to address many diverse questions in social and behavioral sciences (Bruch and Newman, 2019; Clauset et al., 2015; Crabtree et al., 2017; Croft et al., 2016; Power, 2017; Ripperger et al., 2019; Sih et al., 2018; Webber and Vander Wal, 2019). This diversity of questions and approaches, especially in an interdisciplinary field like network science, has led many researchers to develop tools customized to a particular use case. Careful consideration of the math underlying each approach can help understand the similarities and differences between alternative methods, can ensure that researchers are correctly testing their hypothesis, and can help researchers avoid violating the assumptions of particular methods.

In this paper, we describe methods for drawing statistical inferences about patterns of sociality, focusing on the underlying math and using simulated examples to illustrate each approach (see Supplementary Material). We begin by explaining the concept of a reference (null) model, outlining when these reference models are required, discussing the key considerations facing researchers when using them, and outlining some of the potential pitfalls that may arise. We then introduce different approaches to creating reference models. We highlight the benefits of each approach and provide typical research questions for which different reference models are appropriate. We detail particular pitfalls of using the different approaches illustrating potential questions and pitfalls using examples and simulations. We base our simulations on the social system of a mythical animal, the burbil (Section 2.2). Our paper is targeted at those who have some experience in social network analysis and are looking for ways to make statistical inferences about the social systems they are studying. Whitehead (2008), Croft et al. (2008), Farine and Whitehead (2015), Krause et al. (2015), and Newman (2018) provide excellent introductions to the study of social networks as a jumping off point for what follows here.

## 2. Reference models for statistical inference in network data

The essence of any statistical inference is to determine whether empirical observations are meaningful or whether they are the outcomes of chance alone. When data do not meet the assumptions of traditional statistical methods, as is often the case with network data (Cranmer et al., 2017; Croft et al., 2011), researchers can compare their data against chance distributions, i.e., distributions of values that are generated in a random process.

Traditionally, the likelihood of an observation occurring by chance has been referred to as a *null model* (Croft et al., 2011; Good, 2013). However, the term “null” suggests that no patterns of interest are present. While it is correct to assume for statistical inference purposes that nothing is happening in relation to the phenomenon that is being observed, many other processes can still be acting on, or otherwise limiting, the system. For this reason, we advocate for using the term *reference models* (Gauvin et al., 2020) rather than null models or chance distributions. The use of the term “reference” highlights the notion that we are not comparing observations to a completely random scenario that contains no predictable patterns but rather to a system in which certain features of interest are preserved and others are randomized.

To perform statistical inference using reference models, the two most important questions a researcher needs to ask are: ‘How can empirical data be sampled in an unbiased way?’ and ‘What is the likelihood that a given pattern is present by chance?’ These questions are linked because *chance* could be different depending on the sampling approach. For example, if one samples only females, the chance distribution should not include males because the mechanisms that underlie the observed processes could differ between males and females. Identifying the appropriate chance distribution that observations should be compared to is critical for avoiding straw-man hypotheses. Much previous research and development of tools has focused on sampling data in an unbiased way for network analysis or attempting to account for biases in data collection to conduct statistical inference (e.g., Croft, James, & Krause, 2008; Croft, Madden, Franks, & James, 2011; Farine, 2017; Farine & Carter, 2020; Farine & Strandburg-Peshkin, 2015; Franks, Ruxton, & James, 2010; James, Croft, & Krause, 2009). However, more methodological development is needed to expand our statistical inference possibilities and better tune computational methods to answer specific questions, especially when the generating process of the observed social pattern may be complicated or multifaceted.

Importantly, there are inherent differences between observed biological networks and the mathematical constructs that underlie reference distributions. Observed biological networks are finite and therefore may not embody mathematical properties that are guaranteed to hold asymptotically, e.g., after infinite sampling. Therefore, it is important not to attribute meaning to differences between observed networks and reference models that emerge from the difference between the finite nature of the observed network and the general mathematical construct that describes the reference model. Instead, inference of meaning should come from consideration of agreement with, or deviation from, appropriately chosen patterns that reflect the real-world processes that generate and/or constrain them.

## 2.1. The construction, use, and evaluation of reference models

The effective use of a reference model hinges on four key steps, which focus on answering a biological question by comparing empirical observations to randomized or synthetic constructs. In sequence, we suggest that researchers (i) clearly articulate the biological question, (ii) choose an appropriate test statistic,

(iii) generate a reference distribution, and (iv) evaluate whether the biological question was addressed and whether the model behaved as intended. By identifying these discrete steps, we can scrutinize the analysis process to avoid methodological pitfalls (see Section 4).

**Step 1: Articulate the research question and specify the feature of the reference model to be randomized.** A reference model answers a question by connecting an observation to a distribution of hypothetical observations in which some aspect of the data has been shuffled, resampled, or otherwise stochastically altered. Creating a reference distribution by randomizing some aspect of the observed data is an alternative to an experimental manipulation, where an experimental treatment would create a distribution of observations of the system. Choosing which observed feature(s) to randomize in a reference model is as important as designing a carefully controlled experiment: both require combining the research question, domain knowledge, and accessible data to determine what should be held constant and what should be manipulated. Thus, the outcome of Step 1 is a list of network or data properties that are to be (i) randomized or (ii) maintained. Networks are interesting precisely because they capture complex interdependencies between nodes, which means that choosing what to manipulate and what to hold constant is not always trivial.

Although all reference models randomize some aspect of the data while fixing other aspects, both randomization and fixation can be done at different levels of abstraction: (Level 1) **permutation**, in which observations are *swapped* without replacement, (Level 2) **resampling**, in which observations are *sampled* from the observed data with replacement, (Level 3) **distribution-sampling**, in which observations are *drawn* from a fixed distribution, and (Level 4) **generative-processes**, in which synthetic data or networks are *constructed* from stochastic rules (Fig. 1). These levels of model abstraction can be applied to the observed data at different stages of analysis.

**Step 2: Choose a test statistic.** The test statistic is the quantity that will be calculated from both the empirical data and from the reference model. Many summary measures can be used as test statistics (see summaries in Sosa et al. (2020); Wey et al. (2008)). The test statistic should quantify the network feature, or the relationship between features, that is tied directly to the biological question (see Table 1 for an example).

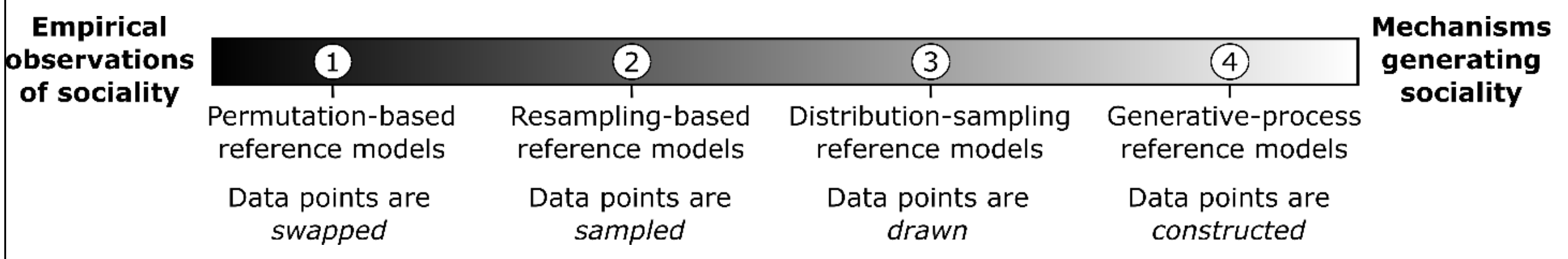

**Figure 1: Methods for creating reference models increase in level of abstraction.** Methods progress from reference models that rely strongly on the empirical observations of sociality (left, Level 1) to methods that make assumptions about the generative processes that underlie the observed sociality and do not use the observed social associations when producing a reference model (right, Level 4).

**Step 3: Generate a reference distribution.** Samples from a reference model constitute a reference dataset and applying the test statistic to each randomized sample in the reference dataset creates a reference distribution. In this way, the samples from the reference model can be compared to the empirical data through the lens of the test statistic. If the test statistic from the observed data is indistinguishable from the distribution of test statistics in the reference distribution, a researcher would not be able to reject their hypothesis. If the test statistic from the observed data falls outside or at the extremities of the reference distribution, then a researcher would conclude that the feature, or relationship, of interest were unlikely to occur by chance and would reject their hypothesis.

**Step 4: Evaluate the process and adjust the reference model approach as needed.** Researchers must carefully evaluate whether the reference model they have built is in alignment with their research question of interest. Researchers also need to determine whether the reference model behaves as intended or whether a different process is needed to test the question of interest. As we show in this paper, there are many ways in which reference models may have hidden biases that can result in misleading outcomes. In evaluating reference models, it is beneficial to separate Step 1 (choosing which features to randomize and which to preserve, and at what level of abstraction) from Step 2 (choosing a test statistic). Such separation will allow researchers to diagnose pitfalls associated with a test statistic versus pitfalls related to the data randomization procedure. Reference models may need several iterations of the construction/evaluation steps to settle on a model that is well-aligned with the research question and biological question and which behaves as intended.

## 2.2 A tangible example of reference model use

To illustrate and provide examples for the different randomization procedures and to summarize some of the key pitfalls that each approach is susceptible to, we created an imaginary social network dataset of the mythical burbils (aka *Burbelis silkensis*). We will refer to this imaginary society throughout the manuscript and provide supporting code for all the examples in the Supplementary Material. Briefly, burbils live in open habitats and exhibit two unique nose-color morphs (red and orange). Individual burbils can be uniquely identified and their sex (male or female) and age (adults, subadults and juveniles) are known. Burbils form fission-fusion societies characterized by large groups that roost together at night but fission into smaller subgroups when foraging during the day. The number of subgroups each day is drawn from a Poisson distribution ($\lambda$= 5) and we suspect that subgroup membership may be assorted by nose colour (see example below). Foraging subgroups from different roosting groups occasionally meet and intermingle, creating opportunities for between-group associations These between-group associations are more likely if the two burbil groups belong to the same “clan” (similar to the vocal clans of killer whales Yurk et al., 2002). Burbil groups differ in size and groups of different sizes might have different social network structures. Within their social groups, burbils are involved in both dominance interactions and affiliative interactions with groupmates and we suspect that these are influenced by age and sex. These interactions can only occur between individuals in the same sub-groups with the number of interactions recorded in each sub-group varying based on the number of individuals recorded. Further information on burbil societies, social network generation, and example analyses are provided in the Supplementary Material.

### 2.2.1 Illustration of several pitfalls in reference model construction and use

To illustrate the need for carefully considering various pitfalls when constructing reference models, we provide an example that compares two reference models, one resulting in more specific outcomes than the other. Specifically, we highlight in this example that carefully articulating the research question (Step 1) has important cascading effects onto the entire analysis. One of the more detrimental cascading effects is a mismatch between the research question and the resulting conclusions. In our example, two teams of researchers (Team 1 and Team 2) set out to study burbil association networks. Both teams have association data from a single group of burbils. Based on these association data, they build a weighted, undirected network (Fig. 2a). The researchers have information on the attributes of the burbils, such as age, sex, and nose color. Team 1 immediately notices that individual burbils differ from one another in their nose color and ask a specific research question related to that trait. Team 2 overlooks the natural history of the burbils and asks a much more general question about burbil social structure. We analyze the process that both teams went through in Table 1, highlighting the pitfalls they each encounter related to the way they defined their research question (R code for both analyses provided in Supplementary Material, Section 3.1.1).

| **Table 1. Example of two research teams and their approach to studying burbil sociality.** |
|---|
| **Step 1a. Articulate research question.** |
| Team 1: Do burbils socially associate by nose-color?<br>Team 2: Do burbils associate at random? |
| **Step 1b. Develop a reference model.** |

Team 1: To determine if burbils associate based on nose color, the researchers decide to preserve the observed network structure (Fig. 2a), i.e. who associates with whom, but randomize it with respect to nose colour. Note that this choice maintains all aspects of burbil social structure - except for nose color - which is the variable the researchers are interested in examining.

Team 2: To determine if burbils associate at random, the researchers generate random networks with the same number of nodes and edges and then for each random network, they draw edge weights from a normal distribution with the same mean and standard deviation as the observed adjacency matrix.

**Step 2. Choose a test statistic.**

Team 1: The researchers use a weighted assortativity coefficient to measure the tendency of burbils to associate with those of the same nose color.

Team 2: The researchers choose a measure of variance of the weighted degree (strength) distribution - coefficient of variance (CV) - as the test statistic to compare the observed and reference networks.

**Step 3. Generate a reference distribution.**

Both teams generate a reference distribution by running 9999 iterations of their randomization procedure to which they compare the observed test statistic. Using 9999 iterations means their full reference dataset (including the observed value) is n=10,000. They use their different algorithms to generate their reference distributions. Both research teams plot the distribution of the 9999 reference test statistics as a histogram and the observed value as a line for visualization (see Fig. 2b for team 1's histogram). They then use a two-tailed comparison to examine if the observed test statistics falls inside or outside the 95% confidence interval (CI) of the reference distribution (i.e., between the 2.5% and 97.5% quantiles or outside this range).

**Step 3a. Network randomization and generating reference test statistic.**

Team 1: After each shuffle of nose color, the weighted assortativity coefficient is calculated to obtain 9999 reference values to compare with the observed value.

Team 2: After the creation of each new interaction network, the CV of the weighted degree distribution is calculated for each simulation to obtain 9999 reference values of simulated weighted degree CV to compare with the observed value.

**Step 3b: Compare reference and observed test statistics.**

**Step 3c: Draw inference from comparison between observed and reference values.**

Team 1: The observed assortativity coefficient falls higher than the 95% confidence interval of the reference distribution indicating that burbils do indeed assort by nose colour - tending to associate more with burbils with the same colour noses (Fig. 2b).

Team 2: The observed weighted degree CV falls inside the 95% interval of the reference distribution, indicating that the network is not different from random with regard to this particular network measure.

**Step 4: Evaluate the process and adjust the reference model approach as needed.**

Team 1 asked a specific question, used a permutation procedure that shuffled only the one aspect of burbil society that was of interest, and they chose a test statistic that was well- matched to their question.

Team 2 asked a vague question (what does it mean for a network to be non-random? What is the biological meaning of ‘random’ and how is it measured?). They found it difficult to define a satisfactory reference model and they chose a test statistic that was not as directly linked to their question. Team 2 is therefore uncertain about the biological conclusions they can draw. Most importantly, they failed to determine how the way in which they generated their reference distribution matches their research question. This failure stems from the lack of specificity of their biological question.

Further, they missed the fact that they included zero values for self-loops in their calculation of the mean and standard deviation of the edge weights when generating their reference networks. These edge weights had a biased representation and inflated their importance compared to the observed edge weights.

**Figure 2. An example of study approaches: Do burbils socially assort by nose-color?** (a) Association network of burbils, with nodes color-coded by nose color and (b) distribution of values based on the permutation procedure of the informed team; observed value of the test statistic shown as a red solid line and the 2.5% and 97.5% quantiles of the reference distribution as blue dashed lines

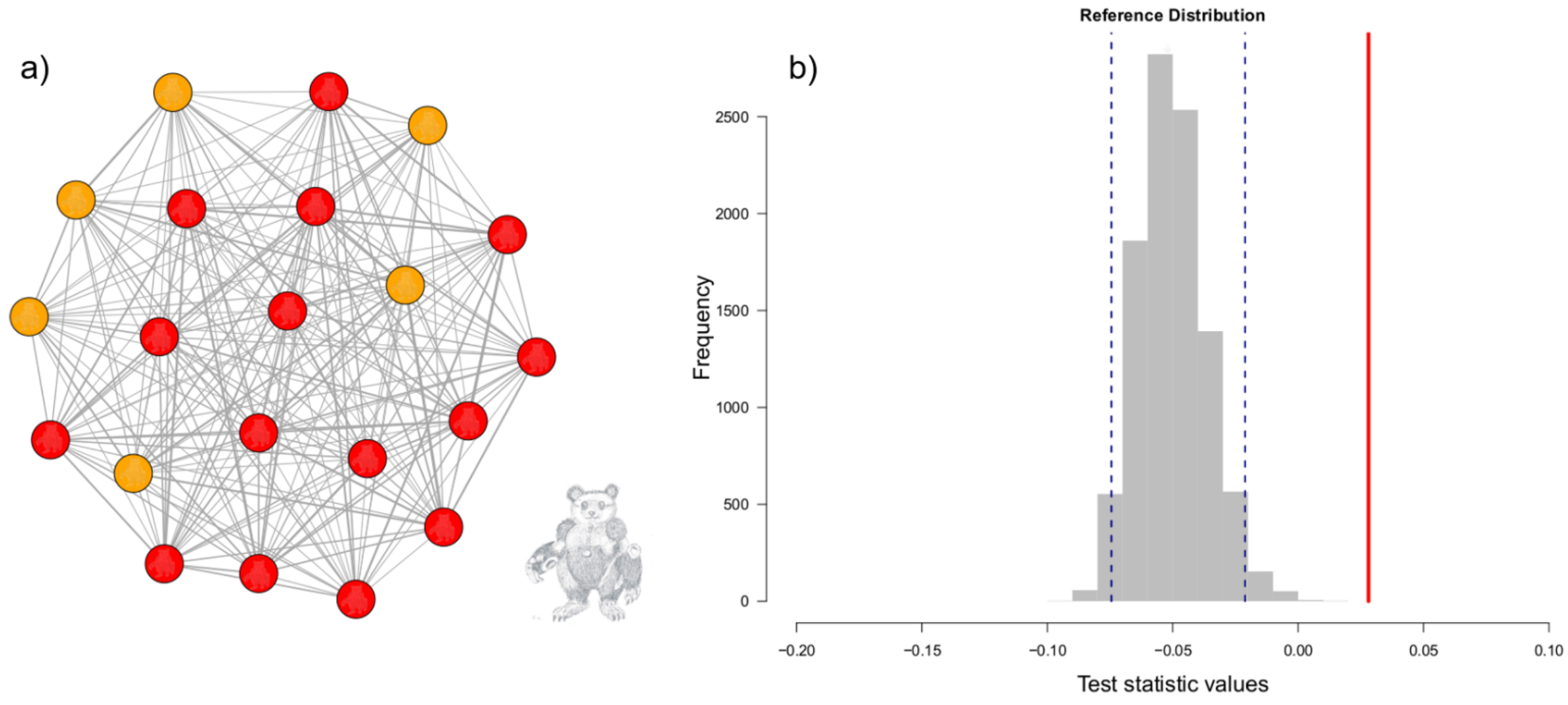

## 3. Do you need a reference model? The importance of distinguishing between exploration- and hypothesis-driven investigation

A reference model functions, in a computational sense, as a control against the observed outcomes in a system. The ‘null hypothesis’ would be that no meaningful differences exist between the calculated reference and the measured results. Our goal in constructing an appropriate reference model is therefore to know confidently when to reject that null hypothesis.

Although we focus on selecting appropriate reference models against which to contrast hypothesized processes or outcomes (i.e., an appropriate control for observational experiment), the idea of a test against a reference model itself relies implicitly on the existence of a known and concrete alternate hypothesis describing either the process from which the observations emerged or describing features of the observed data/structures themselves. One potential (and common) point of complication in the analysis of social networks is that hypothesis generation (i.e., data exploration) and hypothesis testing may be easily conflated. In exploration-driven investigations it is impossible to design an appropriate reference model because it is impossible to decouple a hypothesis from the observations themselves.

There is often a temptation to randomize each pattern of interest in a network with the hope that finding the correct reference model for contrast can allow meaningful interpretation from observations that may not be rich enough, or well understood enough yet, to support it. This is not at all to suggest that exploratory data analysis is inappropriate. It is critical to purposefully differentiate between the *exploration phase* of research (when pattern discovery may itself be the goal and does not require statistical departure from a constructed reference model) and the *hypothesis testing phase* of research (when appropriate reference models are necessary).

### 3.1. Exploration versus hypothesis testing - a case study

Consider the case in which a researcher suspects that individual risk of infection from a contagious disease circulating in a population may be correlated with some measure of the centrality of individuals in the network. There are three potential cases that are all included in this general description:

*Case 1) The mode of transmission of the pathogen is known (e.g., sexually transmitted).* In this case, the network of contacts among individuals that may provide the means for disease transmission is well-defined (in the same example, an edge is drawn between two individuals who have engaged in sexual contact with each other). Given that network, we may hypothesize that a particular centrality measure may correlate with infection risk (for example, eigenvector or betweenness centrality). Calculating the individual centralities of each node in the network and their respective correlation with disease burden observed is a valid endeavor and requires the construction of an appropriate reference model to be able to infer meaning and make appropriate interpretations of the outcome.

*Case 2) The researchers are interested in finding the correlation of one particular centrality measure with disease risk, but the mode of transmission for the infection is not known.* For example, it might transmit by sexual transmission or be transmitted via inhalation of droplets from the respiratory system, so close contact with anyone coughing/sneezing/exhaling while infected is sufficient for potentially successful transmission. In this case the researchers may construct two potential networks: one from observed sexual contact and the other from some spatial proximity index that would reflect exposure to exhaled droplets from others. Here again, calculating the betweenness of the individuals in each of these two different networks and their respective correlations with observed disease burden is also valid and requires an appropriate reference model.

*Case 3) The researchers are unsure of the mode of transmission of the pathogen, nor do they know which centrality measure might correlate with infection risk.* In this case, selecting the combination of network structure and centrality measure that yields the highest correlation with observed disease burden may not be a well-formed question, but an exploratory approach would be appropriate. Therefore, a reference model is not necessary, and no matter how carefully constructed, it would not be able to provide a valid context for interpretation. Because the centrality calculation does not exist in the absence of the structure of the selected network, the "pair" of measure and network that produces the greatest fit to the observed transmission pattern is the logical equivalent of over-fitting a regression. Unfortunately, unlike a simple regression,

because the centralities of individuals depend on the network structure (i.e., factors that are extrinsic to the node itself), validation by sensitivity of the correlation under iterative removal and recalculation (or other common techniques) is not possible.

# 4. Common pitfalls when using reference models

When using reference models to analyze network data, researchers should keep in mind the pitfalls that can arise at each of the above steps. We provide a broad overview here and link these general pitfalls to specific examples that are related to the different approaches to generating reference models, which we detail below.

## 4.1. Pitfalls in matching a reference model to the research question

The most important step when designing a randomization procedure is ensuring that the research question is directly addressed. Researchers may set out to examine a particular question but then randomize the network in a way that misses the question and results in misleading conclusions (see Table 1 for an example). Designing an appropriate randomization procedure can be challenging because changing one property of a network can often change others and imposing too many constraints may lead to computational issues or prevent researchers from answering the desired question. Therefore, having a clear understanding of the types of constraints that can be imposed is important.

Reference models for social networks can be constructed to preserve both or either non-social or social aspects of the animals' biology. *Non-social constraints* are properties of the biological system that are extrinsic to the social processes that underlie the social network but might influence whether or not an interaction occurs. Such constraints might shape how the reference distribution is generated, for example by providing restrictions on possible permutations or resampling. Restricting permutations (or resampling) to specific time windows, for example, could prevent creating interactions between individuals that had not been born yet and ones that have already died, or immigrated away from the study site. Similarly, including spatial constraints in reference models recognizes that some individuals can never meet, for example terrestrial organisms that are separated by a river they cannot cross. Failing to prevent the generation of samples in the reference model that are not naturally feasible may lead to false positive results. Often, imposing these constraints will require knowledge of the study system. *Social constraints* are emergent properties of the network that might be important to maintain when testing particular hypotheses (e.g., the degree distribution, the number of network components or clusters, etc.). These properties are easier to maintain using some approaches to generating reference models than others. For example, *datastream permutation* methods overlook the importance of maintaining specific properties of the social network (more detailed discussion in Section 5.2) (Weiss et al., 2020), which can be important when developing reference models to answer some questions. Not accounting for social constraints can result in reference datasets (networks) that fall within the non-social constraints imposed but which have substantial differences between some key properties of the emergent network structure in the reference dataset and the observed social network. A failure to include social constraints can result in errors in inference (Weiss et al., 2020).

Both social and non-social constraints can cause unintended changes, resulting in a reference model that no longer addresses the original research question or addresses a similar but not identical research question (see Section 5.2). For example, randomization of movement data can alter the social networks constructed from those movements, which may in turn, introduce undesired changes in reference networks that could not be foreseen from the movement data permutations. Thus, randomizing away correlations at one scale (e.g., movement) may introduce correlations at another scale (e.g., social).

Further, while it is important to consider both network social and non-social constraints on the reference datasets, a reference model can include too many constraints (see Section 5.2). In some cases, these constraints may prevent the production of a reference model (too few possible configurations) or make the process too computationally intensive. Applying constraints may also lead to a narrow reference distribution. While this does not have to be a pitfall (and might just be the nature of the biological question), a pitfall arises if these restrictions stop a researcher from randomizing the aspects of the data that are the focus of the research question. Sometimes creating a wide enough reference model is not possible using less abstract approaches (permutations and resampling, see Fig. 1), for example, very small networks have a small, finite, number of possible edge permutations. In such a situation, it might be beneficial to change the randomization approach. We discuss in Section 7 a randomization procedure that can allow researchers to produce wider distributions than those that are obtained by permutation.

## 4.2. Pitfalls in test statistic choice

We cannot emphasize enough the importance of choosing a biologically appropriate test statistic. The data, network structure, or properties of the test statistic may constrain decisions about test statistic use. Understanding the biological meaning of the test statistic that is being compared between observed and reference data will determine whether or not the biological question can be answered. Researchers might be familiar with particular network measures (e.g., degree, strength, betweenness, density, modularity) and use only those to answer all their questions about network structure. However, not all measures are appropriate for answering every research question, and each measure has a different biological meaning that can depend also on the network structure (Brent, 2015; Farine and Whitehead, 2015; Silk et al., 2017a; Sosa et al., 2020; Wey et al., 2008). Therefore, it is important to understand the *biological meaning* of the test statistic. Understanding the biological meaning of the test statistic will prevent testing too many measures (Webber, Schneider, & Vander Wal, 2020). The more test statistics one measures, the more hypotheses are being tested and so the greater the need to account for multiple testing (to prevent false positive errors). Additionally, it is important to consider correlation between test statistics. For example, a researcher might be interested in uncovering the centrality of individuals in a network and would like to use degree, strength, and betweenness. However, it is possible that these three measures are highly correlated with one another (e.g., Borgatti 2005; Farine and Whitehead, 2015; Silk et al., 2017a), and some test statistics may be correlated in unexpected ways (e.g., centrality measures can be correlated with community structure). An additional pitfall is that for some research questions, the randomization procedure can affect the test statistic in unexpected ways, especially if comparing networks of different sizes. There might be ways to adjust a test statistic, but such adjustments can lead to subtle change in the research question being asked and therefore to new inferences (see Supplementary Material 3.2.1).

A related pitfall when comparing networks of different sizes is that the most appropriate normalization approach can depend on the behavioural rules that generate the network. Determining effects of network size on the choice of test statistics may require conducting simulations and/or examination of the literature. We show in the supplementary material (Section 3.2.1) how the generative process that underlies the network can impact the ability to compare networks of different sizes. For example, if we ask how network size influences the average connectivity of individuals, we could compare the mean degree of burbils in huddling networks of two different sized groups. In both cases the same rules underlie network structure. We consider two situations, a random graph or a small-world process, in which individuals are typically connected to nearby nodes with only occasional long-distance connections. When comparing the mean degree of two networks of different sizes a sensible normalization is to divide raw degree values by the number of individuals in the group minus one (i.e., the number of individuals it is possible to be connected to). However, the outcome of doing this depends on whether the network is generated as a random graph or a small-world process. In the former, the normalized mean degree is much more similar between the two groups than the mean of the raw degree values. However, when we do the same for a small-world network the mean of the raw degrees is similar, while the mean of normalized degree values is very different. This example highlights the challenges in testing the similarity of different-sized networks without knowledge of the behaviour that generated them. A similar caveat applies when using resampling based reference models to compare networks of different sizes.

### 4.3. Pitfalls in generating the reference distribution

The process of generating the reference distribution holds a number of potential pitfalls for the unwary.

First, the reference model does not always sample the full parameter space. There might be values that will never appear in the reference distribution because of the structure of the data or the algorithm of the randomization. Under-representation of values in the reference distribution might be important to maintain but could also be an unwanted side-product that could be resolved by using a different randomization procedure, as we explain in Section 4.1. We provide an example of how sampling from different distributions yield different ranges of values in Figure 4. Second, the parameter space needs to be sampled in an unbiased manner. When generating a reference distribution, certain values might be over- or under-represented if the procedure used to generate the model does not explore the entire parameter space or explores it naively. Ideally, the randomization procedure will produce a reference distribution in which values are uniformly distributed or follow a distribution that is appropriate for the network structure. It is important to understand the constraints of the randomization procedure that is being used to determine if such biased distribution may emerge. We provide a detailed example in Section 5.2.

Third, generating a reference distribution can be computationally intensive, to the point that it is not feasible to generate a large enough reference distribution. We offer a range of approaches, some of which (like sampling from distributions, Section 7) are less computationally intensive than others (such as permutations in Section 5). If computational constraints influence the choice of methods, it is important to carefully evaluate what concessions are being made regarding the ability of the randomization procedure to answer the biological question. For example, when using permutations conducting too few swaps can lead to problems with statistical inference (see Section 5).

### 4.4. Pitfalls in failing to comprehensively evaluate the process and adjust the reference model approach as needed

Many of the general pitfalls identified here can be detected by carefully evaluating the approach used to make sure that the reference model is tuned to the research question and is behaving as expected. This step can help identify further potential pitfalls. One important point to consider is whether the reference model is being used to answer a statistically motivated question (i.e., to test a hypothesis) rather than to explore the data in search of significant deviations from the model (as discussed in Section 3). A second potential pitfall is that agreement between observed data and reference model outcomes does not necessarily imply similar causality. If the observed data is similar to the randomized data, this does not necessarily mean that the algorithm underlying the randomization is the same as the biological process that underlies the observed network; with a close match, the algorithm is a plausible generating mechanism for the observed patterns but must be tested further. For example, many observed social networks are characterized by a heavy-tailed degree distribution, such that the network has few individuals with much higher degree than the rest of the individuals, i.e., they can be considered as hubs. Often, researchers model the heavy-tailed degree distribution of such networks as a power law, in which the frequency of nodes with a certain degree $k$ is proportional to $k^{-\alpha}$. Although the algorithm of degree-based preferential attachment (i.e., the Barabási-Albert model; Barabási and Albert 1999) yields a network with a power law degree distribution, so do other algorithms (e.g., the "copy model"; Kleinberg et al. 1999). It is therefore clear that inferring the process by which a network results in a power law degree distribution cannot uniquely rely on agreement with the emergent structure itself. We provide further examples of this pitfall in Section 8.

Finally, not all network analysis requires the use of reference models (see also Section 3). While the use of reference models is often necessary when analyzing features of individuals that are linked to others in a network because of the dependency between individuals, there are questions and methods that do not require the use of reference models. For example, one might use network measures to characterize many groups in a society. Researchers might want to ask if a network measure, for example density, increases with the size of the group. In this case a simple correlation between group size and density would address the research question. If, however, the researchers are interested in the process that underlies the relationship between group size and network density they might use generative models (Section 8) or sample from distributions (Section 7) to produce groups of different sizes using different engagement rules. Note, however, that the second approach addresses the question: 'what are the underlying causes of the observed relationship between group size and density?' rather than answering the original research question: 'is there a relationship between group size and density'?

## 5. Permutation-based reference models

Permutation-based reference models take observed data and shuffle it to produce reference datasets (Good, 2013). The resulting reference models preserve certain attributes of the observed dataset, such as distributions of key network measures or features of the raw data, such as group size. Because data are shuffled and observations are swapped, new values are not necessarily introduced in the reference models (although new values of some measures can be calculated). The most simple permutation-based methods randomize a single feature of the observed data while preserving all other observed features. Statistically, this approach breaks correlations that are shaped by the permuted feature. Permutations can be applied either to the network structure itself (e.g., nodes and edges, or features of them) or to the raw data that underlies the network structure (e.g., movement data, group membership, etc.).

### 5.1. Feature permutation

Permutations can be used on both node features and edge features. In both cases, these permutations involve swapping *attributes* among either the nodes or the edges. Attributes can be any feature of the nodes or the edges. Common node attributes are individual identity (often referred to as the node's label), sex, body size, age, color, or other features. Attributes of edges can be the types of edges connecting two nodes, for example, different types of relationships or interactions, such as aggression and affiliation, or the *direction* of the edge for asymmetric relationships or for directed interactions.

Node feature permutation-based reference models swap attributes among nodes in the network. Node feature swaps preserve the structure of the observed networks but break potential correlations between the structure of the network and node attributes. Comparing observed networks to node attribute permutation reference models allows researchers to test if the attributes of interest are associated with observed patterns of interactions or associations (for an example, see Table 1).

Node feature swaps have been frequently used as reference models in social network analysis (Hamilton et al., 2019; Johnson et al., 2017; Snijders et al., 2018; Wilson-Aggarwal et al., 2019). They are most often used to test associations between measures of social network position and phenotypic traits of individuals (e.g., Ellis et al., 2017; Hamilton et al., 2019; Johnson et al., 2017; Keiser et al., 2016; Wilson-Aggarwal et al., 2019). We provide an example in the Supplementary Material (Section 3.1.2) in which we test the relationship between sex and out-strength in burbil dominance networks. Inference from node swap permutations can be complex if there are underlying processes (e.g., differences in sampling) that may generate patterns of interest. For example, we provide an example (Supplementary Material Section 3.1.2) where we swap a node attribute, nose color, to test if burbils socially assort by nose color when they interact in an affiliative manner. These node swap permutations show that burbil affiliative networks are indeed assorted by nose color. However, interactions can only occur when individuals are associating within the same group, therefore, without taking into account patterns of sub-group formation in the population in the permutation, we are unable to answer whether affiliative interactions are assorted for nose color within subgroups.

Edge feature permutation-based reference models swap attributes of the edges, leaving the node identities, node metadata, and the connections among them intact. Edge feature swaps can involve shuffling the following: (i) *labels of edges* - swapping one type of interaction for another, like aggression to affiliation, (ii) *edge directions* - swapping which individual directs a behavior to which recipient in an interaction, swapping an edge from A to B to go from B to A (De Bacco et al., 2018; Miller et al., 2017), or (iii) *edge weights* - swapping the values that represent the strength, frequency, or duration of interactions among individuals, such as swapping a strong relationship between A and B with a weak relationship between C and D). Note that permuting edge weights can only involve swaps between pairs with non-zero weighted edges otherwise it would become *edge rewiring* as detailed in Section 5.2. We provide an example of edge direction swaps in the Supplementary Material (Section 3.1.3) where we test the hypothesis that adult burbils have higher out-strength in networks of dominance interactions than younger individuals (subadults and juveniles). We swap edge directions at random in an iterative process where we generate a Markov Chain (see Section 5.2). Permuting edge weights can be useful for answering questions about the strength of social ties. We provide an example in which we test the hypothesis that burbils of different sexes have different out-strengths in the network of affiliative interactions (Supplementary Material Section 3.1.3). The affiliative network is highly connected (has a high density of edges and few or no zero-weighted edges) making it suitable to use edge weight permutations in this way. In an iterative process we select pairs of dyads and swap the number of affiliative interactions between them to randomise which edges are associated with which weights, breaking down the correlation between edge weights and node attribute, in this case sex.

Edge feature swaps could be used on raw temporal data in edge list form if each interaction between two individuals is labeled with the time at which the interaction occurred. A possible edge label swap would be to randomize the time at which each interaction occurred (changing the time label but keeping the identities of the pairs that interacted). If edges have further information about the type of interactions (e.g., the type of behavior, such as grooming or fighting) one could also randomize the type of interaction that occurred at each particular time, thus, changing the type of interactions but keeping the individuals involved and the timing or order of the interactions the same as in the observed. In both these examples, the edge label swaps would not lead to reference models that are different from the observed dataset if all time points or all types of social interactions are aggregated. However, network measures that are sensitive to temporal dynamics or to the type of interactions (such as multilayer measures, Finn et al., 2019; Kivelä et al., 2014) can be affected by these feature swaps.

### 5.2. Edge rewiring with permutation

Edge rewiring involves swapping the edges that represent interactions or associations in raw datastreams or swapping edges that connect nodes in a network in an adjacency matrix. For example, edge rewiring may swap the edges *ab* and *cd* to replace them with edges *ad* and *cb*. Edge rewiring results in what is known in network science as the *configuration model* (Bollobás, 1980).

The configuration model is a graph which is sampled uniformly from all graphs of a given degree sequence (with some key technicalities). The degree sequence is the list of all observed degrees in a network, which can be summarized as a degree distribution. Configuration models require appropriate care when making decisions about the specifications of the underlying model (Fosdick et al., 2018). Like edge feature swaps, edge rewiring breaks correlations between the node metadata and the structure of the network to test whether the observed edge arrangement leads to a network structure that is different from a structure that would be achieved by chance, while preserving group size and the metadata of nodes. Edge rewiring can be conducted at different stages, from modifying the raw data (in what are often known as *pre-network permutations* or *datastream permutations*; Farine, 2017) to modifying the group's network structure directly by manipulating the adjacency matrix. In general, rewiring models form what mathematicians call a *Markov Chain*, such that drawing samples by rewiring is equivalent to sampling from a distribution of networks by Markov chain Monte Carlo (MCMC) (Fosdick et al., 2018).

Edge rewiring on raw data is often used in animal social network analysis (*datastream permutations*, where edges often represent each single interaction or association rather than a summarized version of and edge's strength). Importantly, when datastream permutations are used on this raw form of the data, the configuration model that is generated is related to the current format of the data rather than the projected social network that is subsequently analysed. Biologists often use a rewiring approach for association data in group-by-individual matrices, also known as *gambit of the group* data formats, (e.g. (Bejder et al., 1998; Brandl et al., 2019; Croft et al., 2006, 2005; Poirier and Festa-Bianchet, 2018; Zeus et al., 2018)). In this data format, each individual is recorded as present in a particular group and "group" is often defined as an aggregation of animals that are present at the same time and the same place (Franks et al., 2010; Whitehead and Dufault, 1999). This data format is a bipartite network with edges that connect individuals to the groups in which they were observed, i.e., it is a bipartite version of the configuration model that respects the bipartition. When such datastream permutations are applied to group-by-individual matrices the edge rewiring step takes place on this bipartite network rather than on the projected social network that is created subsequently. Similarly, when edge rewiring is used for raw data on behavioural interactions (e.g., Miller et al., 2017; Webber et al., 2016), it is the multigraph that contains all interactions (i.e., a network with multiple rather than weighted edges between nodes) that is rewired, while the network analysed is subsequently treated as a weighted network (with single edges between nodes).

Elaborate rewiring procedures can be used to impose both social and non-social constraints. For example, researchers may constrain rewiring to only swap individuals between groups that occur in the same location or on the same day (non-social constraints). Researchers may further want to impose network constraints, such as forcing the re-wired reference models to preserve the degree distribution of the observed network. The R package igraph (Csardi et al., 2006) can rewire social networks while maintaining a fixed degree sequence, while Chodrow (2019) shows how to preserve both event size (the number of individuals in each grouping event) and the degree of each individual if using datastream permutations to analyse data on animal groups (or equivalent bipartite networks in other fields) and Farine and Carter (2020) propose a double permutation test to help avoid elevated type I errors. Another example of an elaborate rewiring procedure is disconnecting either just one or both end(s) of an edge and re-connecting it to a new individual (or individuals) (e.g., Formica et al., 2016; Hobson and DeDeo, 2015; Hobson et al., 2021). For example, an edge connecting A to B can be disconnected from B and re-wired to now connect A to C. This kind of rewiring results in some changes to both the dyadic relationships between individuals and the network structure, but preserves other features of the networks, such as eigenvector centrality, and can be used to generate reference datasets that are consistent with a desired network constraint (Hobson and DeDeo, 2015; Hobson et al., 2021). This edge rewiring procedure is different from the configuration model, as it does not generally preserve the degree sequence. If the network is directed, this type of rewiring can be used to preserve the sequence of out-degrees, but not in-degrees (or vice versa). As the complexity of the rewiring procedures and the constraints imposed on them increase, these rewiring procedures become more similar to generative models, which we detail in Section 8.

We provide examples of datastream permutations for both association (Section 3.1.4.1) and interaction (Section 3.1.4.2) data in the Supplementary Material. For associations, we generate two reference distributions to test the hypotheses that burbil associations are non-random with or without accounting for assortativity by nose color. Our permutations conduct edge rewiring in the group-by-individual matrix and in both reference models we constrain swaps to occur within the same burbil group and to be between two sub-groups observed on the same day. The first reference model is naive as we already know the burbils are assorted by nose color (Table 1). However, when we additionally constrain swaps so that edges can only be rewired between burbils with the same nose color, we see that association patterns are random between burbils with the same nose colour within a subgroup. This example demonstrates the potential power of using multiple reference models in concert. For interactions, we ask what explains burbil affiliative interactions. Using edge rewiring in the raw interaction data we find that there is no evidence for assortativity by nose color when controlling for subgroup membership. We show this by rewiring interactions within each subgroup so that the nose color of each dyad is randomized. Affiliative interactions are assorted by nose colour only because each subgroup tends to be dominated by one nose color or the other (rather than being an unbiased sample of individuals in the group).

## 5.3. Key pitfalls for permutation-based reference models

For permutation-based methods, a first major potential pitfall to watch for is failing to impose the correct constraints on swaps. In feature swaps (conducted on the adjacency matrix itself), it may not always be possible to constrain swaps as desired. For example, swaps can be constrained to only occur between individuals recorded at the same location (e.g., Shizuka et al., 2014), in the same group (e.g., Ellis et al., 2017), or alive at the same time (e.g., Shizuka and Johnson, 2020). However, it can be challenging to incorporate some constraints. If we test for assortment by nose colour in the burbil network of affiliative interactions, then there is no natural way to restrict swaps on the adjacency matrix to account for burbils only interacting with others in the subgroups they occur in (Supplementary Material Section 3.1.2). However, using an edge rewiring approach we can constrain permutations within each subgroup (Supplementary Material Section 3.1.4.2). For reference models generated by edge rewiring, it is critical to consider both the non-social and social constraints because decisions about which constraints to build into the rewiring procedure affect the resulting configuration model. In many common animal social network rewiring methods, researchers control for unwanted structure in non-social constraints (e.g., sampling biases, differences in gregariousness, etc.). It is less common for researchers to consider social constraints, such as forcing the rewired networks to conform to a particular degree distribution. However, without social constraints, the reference model will approach a random network as the number of rewiring steps increases and can result in misleading, false positive inference (Weiss et al., 2020). Chodrow (2019) shows how one can preserve both the size of interactions (number of animals in each interaction) and the degree of each individual to produce a permutation of the datastream that preserves the degree distribution.

A second pitfall of using permutation-based reference models are computational limitations and potential for biased sampling. Permutation-based approaches are often computationally intensive (e.g., as seen when running the code Supplementary Material Section 3.1). Computational constraints can be exacerbated when increasing the number of constraints on the permutation (social or non- social) because many of the attempted swaps will be rejected. In some cases, over-specifying constraints on the randomization can result in a configuration model with insufficient acceptable states, making it impossible to generate a reference model, especially when examining small networks. Furthermore, it is important to sample from the configuration model in an unbiased manner. This pitfall is especially likely when sampling from a distribution of networks by MCMC, as is often done in edge rewiring approaches. When a swap is rejected (i.e., a suggested swap is not possible within the set of constraints imposed) it is important to resample the current reference network as the next iteration of the Markov Chain (Krause et al., 2009). If such resampling is not done, then the configuration model will be sampled in a biased way (Fig. 3), which could lead to errors in inference. Such rejection of swaps will arise more frequently when there are more constraints imposed on the permutations, and then other potential pitfalls arise: the Markov Chain will a) take longer to become stationary and b) be slower to mix, which could lead to further errors in inference. Addressing this pitfall requires a burn-in period during which the permuted networks are not used in the reference distribution and a thinning interval that equates to permuted networks only being saved as reference datasets after so many iterations in the Markov Chain (e.g., every 10th iteration). We provide examples of these in the Supplementary Material (Sections 3.1.3 and 3.1.4).

A third potential pitfall of permutation-based approaches is that they are often prone to having unanticipated effects on network structure, especially when permutations are conducted on the raw datastream. Consequently, failing to properly evaluate the computational approach is a particularly important pitfall for permutation-based approaches. For example, a completely uniform random rewiring might make the data too ‘unrealistic’ or mean the distribution of the response variable is changed considerably (Weiss et al., 2020). Edge-rewiring of the group-by-individual matrices (as explained above) typically alters degree and edge weight distributions, which can lead to false positive errors because the reference model does not address the question originally asked. Incorporating constraints imposed on the rewiring possibilities (e.g., Chodrow 2019) could help resolve this problem.

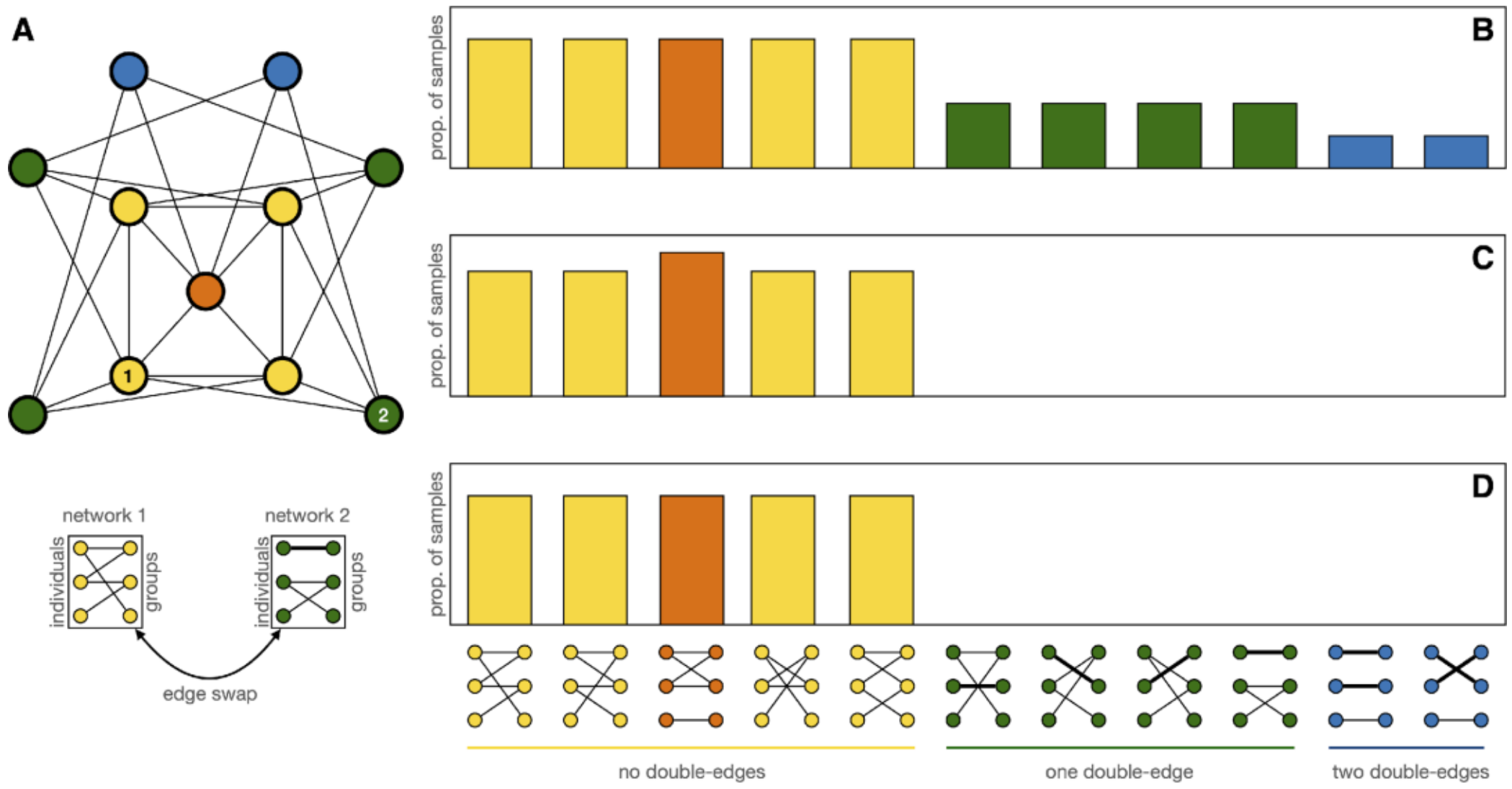

**Figure 3:** An illustration of how incorrect use of MCMC methods can lead to biased sampling from the configuration model when using datastream permutations. When permuting a bipartite group-by- individual network there are 11 possible configurations - depicted under (D). Of these possibilities, 5 (colored in yellow and orange) are acceptable because they do not contain double edges, (as seen in the green and blue possibilities as a thick edge). Double edges indicate that the same individual occurred in the same grouping event twice - which is impossible. In (A) we show the “graph of graphs”, or the Markov chain. (B) is the distribution of samples obtained when permutations are conducted and every state, including those that are impossible (green and blue) are accepted. (C) is the distribution of samples obtained when rejecting swaps that result in double edges and then rewiring a randomized network. Note that a sampling bias arises here - the orange state is oversampled - because it has more routes to other acceptable states as seen in (A). (D) is the distribution of the samples obtained when swaps that make double edges are resampled (i.e., the correct unbiased sampling approach). Note that in (D) the sampling of the five acceptable states is uniform - as it should be.

## 6. Resampling-based reference models

Resampling of network data is a bootstrapping procedure that generates reference models which can be further from the observed data (Fig. 1) than the permutation-based methods we have discussed thus far. While generating reference models using permutations permits each observation to appear only once in the reference model (i.e., sampling without replacement), creating reference models using resampling (i.e., sampling with replacement) results in observations appearing more than once, or not at all, in each simulation iteration. This difference between the two approaches can change which features of the data are maintained and which ones are randomized. For example, if a researcher decides that an important feature of the social structure is the degree distribution, rather than the exact dyadic interactions between individuals, one can produce reference models by resampling from the observed degree sequence (i.e., the list of all observed degrees). Resampling from the degree sequence will produce reference networks with a similar degree distribution to the observed network, but the observed and reference networks might differ in the degree sequence and potentially also in the number of nodes and/or edges. One potential use of resampling-based reference models is the ability to draw reference networks of different sizes and compare them (see Supplementary Material Section 3.2.1 for more details and caveats to using this approach). Resampling can be an effective tool when used with the raw data, however the only network-level properties that can be sampled with replacement are the degree sequence and edge weights (e.g., Supplementary Material Section 3.2.1). Thus, a resampling approach is more specific and limited than other approaches we present.

## 6.1. Resampling raw data

An important utility of the resampling approach in behavioral studies is to resample the raw data that is the foundation of the network, rather than the network itself. For example, researchers of animal social networks often use the spatial positions of animals to infer interactions from co-localization of individuals (two individuals being in the same place at the same time, e.g., Mersch et al., 2013; Pinter-Wollman et al., 2011; Robitaille et al., 2018; Schlägel et al., 2019). A raw data resampling procedure could sample with replacement individuals' locations from the observed locations, thus preserving the physical constraints on these locations. This approach restricts the sampling to biologically feasible locations so, for example, a terrestrial animal could not be resampled in the middle of a lake. We provide an example in the Supplementary Material (Section 3.2.2) of resampling the foraging location of burbil subgroups separately for each of the 16 groups in our main study population. The resulting reference models maintain the observed subgroup memberships and locations are only sampled from within each group's home range.

The way in which the data are resampled could have a large influence on the reference model. For example,

restricting the resampling of locations of particular individuals to only their own set of locations (e.g., Spiegel et al. 2016) will maintain home range sizes and average travel distances, and therefore, it might maintain the number and identity of individuals that each individual interacts with. Such a resampling procedure is more likely to result in reference models that are closer to the observed network structure, especially if non-social rather than social considerations are important in generating this structure. Conversely, if individuals seek out conspecifics to interact with preferentially, then not having network constraints in the resampling procedure means that the resampling will break the temporal overlap between interacting individuals. Consequently, well-designed resampling of locations can be useful to teasing apart non-social and social explanations for network structure (Spiegel et al., 2016). Alternatively, one could allow resampling an individual's position from all observed positions of all individuals in the population. Such a resampling approach would require that it is biologically feasible for animals to move from one position to any other location in which animals were observed. Resampling that breaks the link between the identity of an individual and its movement patterns can produce reference models that differ considerably from the observed networks, for example, in the number of interactions among individuals. These reference models could be used to test the relative importance of non-social factors that may drive interactions.

### 6.2. Pitfalls for resampling-based reference models

The first important pitfall to watch for when resampling network data is that certain re-sampled degree sequences cannot produce a network because they include too many edges or too many nodes. For example, if the sum of all degrees in a network ends up being an odd number after resampling the degree sequence of an unweighted network, a network cannot be generated. Next, when resampling the raw data that underlies the network, it is important to make sure that the resulting network is biologically feasible. For example, resampling of spatial locations could allow an individual to interact simultaneously with two individuals that are on opposite ends of the study site if not conducted with appropriate caution.

Finally, pitfalls of resampling-based approaches also include over- or under-sampling certain values and deviating from the observed network in unexpected ways. Such biased sampling is likely to be a particular issue for small networks in which the observed degree sequence represents a small sample from the degree distribution. For example, resampling from small degree sequences could lead to repeated sampling of a particular degree value that is an outlier in the observed degree sequence.

Alternatively, rare values of degree might be omitted in the reference model, leading to substantial changes in certain network measures. These biases could result in very broad or even multimodal reference distributions in some contexts, and potentially cause problems with inference. Test statistics that are based on edge strength could be highly impacted by resampling from the degree sequence, especially if the observed strength distribution is skewed. For example, resampling could alter the strength distribution of the reference network by omitting the tail of the distribution. The effects of resampling on different types of measures could become part of the research question if thought through carefully, otherwise it risks leading to erroneous inferences.

## 7. Distribution-based reference models

Reference models can emerge from general processes that shape a network rather than from the data itself. One can generalize the features of the observed network, as we detail in this section, or the processes that underlie the formation of the network, as we discuss in Section 8. In Section 6 we discussed resampling from the observed data; a further generalization of this approach is to create reference models based on inferences of the probabilistic description of the observed data, such as the degree distribution. Distribution-based approaches can result in reference datasets that diverge from some of the specific characteristics of observed networks that are often preserved in permutation-based reference model approaches (such as group size, or the number of interactions), making distribution-based approaches a method for generating reference datasets which are more abstracted from observed datasets (Fig. 1).

There are a number of technical approaches for implementing distribution-based randomization. To maintain the observed degree distribution in the reference models, researchers can either permute the network edges so that the reference network will have the exact same degree sequence as the observed but is otherwise random (as described in Section 5). Alternatively, researchers could create a reference network by resampling (with replacement) a new degree sequence from the observed degrees (as described in Section 6) or generate a network from the configuration model (as described in Section 8). Resampling from the degree sequences is equivalent to drawing random samples from an empirical degree sequence defined as

$$P_k = \frac{\text{number of nodes with degree } k}{\text{total number of nodes}}$$

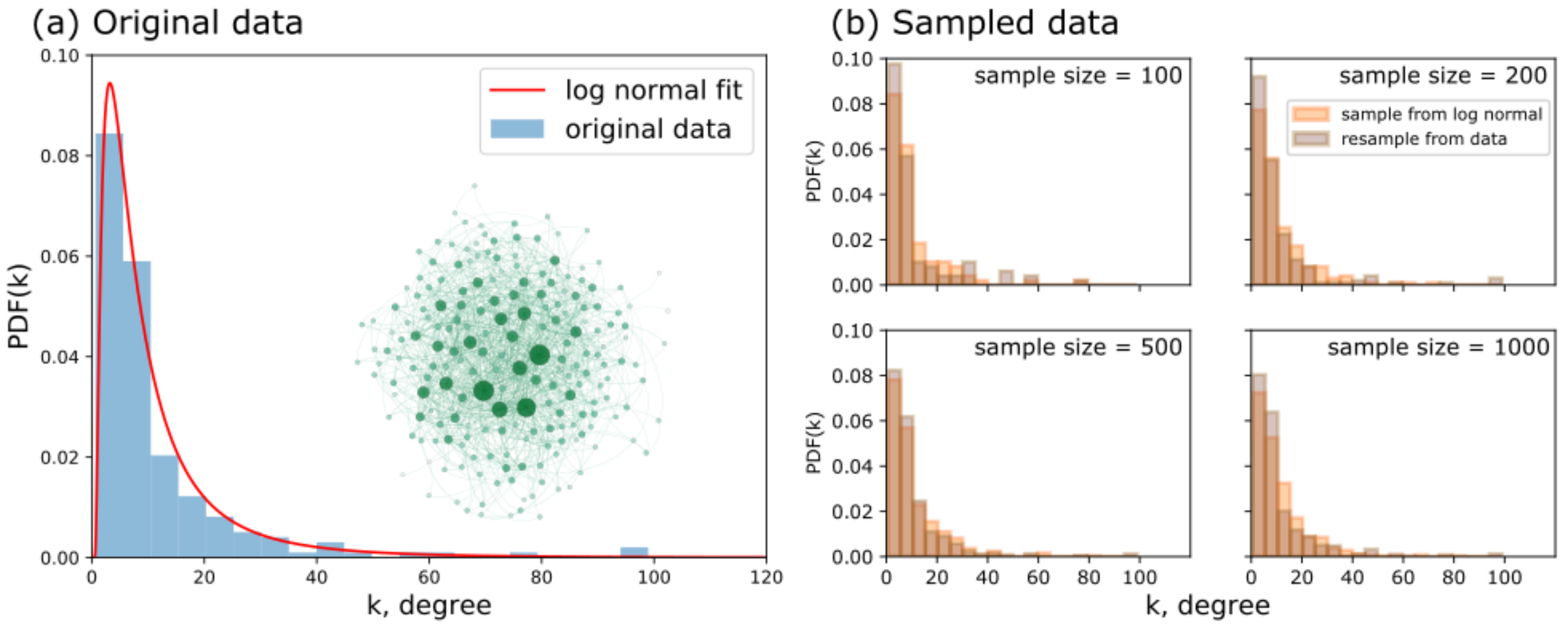

**Figure 4:** Drawing random degree sequences from the distribution-based model: (a) histogram of the degree sequence of the network shown in the inset and a fitted lognormal distribution (red line), (b) random samples of different sizes (100, 200, 500, 1000 randomization iterations) drawn from the fitted lognormal distribution (orange) and by resampling the original degree sequence (gray). Network visualization was done in Gephi (Bastian et al., 2009) with force atlas, a force directed layout. Node color and size correspond to degree.

However, if the functional form of the underlying degree distribution is unknown, it is possible to draw random samples from a fitted distribution to obtain a new degree sequence and subsequently generate a network (Fig. 4). For example, in many social networks there are right-skewed degree distributions in which most individuals have few interactions, and few individuals have many interactions. Such a degree distribution often fits a geometric distribution. Therefore, if researchers are interested in maintaining the shape of the distribution, but not necessarily the exact number of times each degree was observed, then reference models can be generated by resampling from a geometric distribution that has the same parameters as the observed data. Sampling from a fitted distribution can result in sampling nodes with degree $k$ that were not present in the observed network, unlike the resampling approach detailed in Section 6. Sampling from a fitted distribution imposes fewer restrictions on the reference model, which can have both statistical and computational advantages.

Drawing from a distribution can be thought of as sampling from a 'smoothed' version of the observed network. The biggest challenge is to find an appropriate statistical model for the fit. In many cases, finding an appropriate model can be done by fitting a parametric distribution to the data (for example, using maximum likelihood estimation) and drawing random samples from that distribution (e.g., Rozins et al., 2018). It is more convenient to fit continuous distributions, even when describing a discrete behavior, and one should be conscious of the implications of various rounding procedures to turn the sample into whole numbers (Clauset et al., 2009). In some cases, there are efficient stochastic processes that can be used for the distributions-based randomization approach. For example, to generate a network with the same degree distribution as the observed network, researchers can use the Chung-Lu model, which draws an edge between every pair of nodes $i$, $j$, with probability proportional to $k_i \cdot k_j$ where $k_i$ and $k_j$ are the degrees of node $i$ and $j$ respectively. Using this process would generate networks with degrees that were not present in the observed network, despite having similar degree distributions to the observed network.

Distribution-based models can offer flexibility and robustness. They are especially useful when other randomization procedures result in too few unique reference networks that satisfy all the randomization constraints, i.e., there are not enough unique random samples to compare the observed with (e.g., in small networks, see Section 5.3). Furthermore, the inferences from a distribution-based randomization approach emerge from the statistical features of the observed data and therefore may uncover inherent patterns in the underlying social processes. However, selecting appropriate distribution-based reference models can also come with challenges, which we outline below.

### 7.1. Key pitfalls for distribution-based reference models

An important potential pitfall when sampling from a distribution is failing to fit the correct distribution to the observed data and therefore simulating a reference dataset that differs from the observed one in key parameters. For example, a uniform random network has a Poisson degree distribution. However, many real-world social networks have overdispersed (right-skewed) degree distributions (e.g., Rozins et al. 2018) and failing to account for this overdispersion in a distribution-based reference model will lead to errors in inference.

A second potential pitfall arises when sampling independently from two distributions that co-vary. For example, consider a theoretical distribution-based reference model that preserves both the degree distribution and the distribution of clustering coefficients of an observed network. The clustering coefficient of a node measures the fraction of pairs of neighbors of that node which share a link. This quantity tends to co-vary with degree, often in a negative direction, especially in networks with an assortative community structure (e.g., Fig. 5). The negative relationship between degree and clustering coefficient emerges from the fact that high-degree nodes tend to connect different communities and therefore their friends are not tightly connected to each other because they belong to many different communities. In Fig. 5 we show a burbil association network with an assortative community structure in which node size corresponds to degree and node color corresponds to clustering coefficient (see Supplementary Material Section 3.3). If a researcher ignored correlations between degree and clustering coefficient and sampled two sequences of numbers independently from the distribution in (a) and (b) respectively, the resulting distributions would mimic the dataset individually but not jointly. We illustrate another example of failing to account for the correlation between two distributions (degree and mean edge weight) in the Supplementary Material (Section 3.3). For some correlations there may be easy solutions to this co-variance, for example if degree distributions differed between two sexes then they could be simulated separately for each sex. For other distributions (of network measures or in the raw data) it will be necessary to draw simulations from the appropriate multivariate distribution.

A third pitfall of using distribution-based reference models is that it is not (currently) possible to simulate networks with fixed distributions of many social network measures, one example being clustering coefficient (as per the example above). For a fixed number of nodes and edges, or for a fixed degree distribution, we know how to sample a network uniformly over all networks with such properties. However, conducting such uniform sampling can be done for very few other network properties.

Researchers often use reference models that do not sample uniformly from the space of all networks with a given property, but rather use reference models that happen to have properties that are close to the network in question (like the generative models in Section 8). It is important to understand the difference between sampling uniformly over all networks with a given property and sampling from a set of networks that tend to have the property while also having other constraints on their structure, because of the influence that these sampling methods will have on the inference process. These potential pitfalls of generating distribution-based reference models limits the contexts in which such randomization can be applied.

## 8. Generative reference models

Generative models produce a set of reference networks according to stochastic rules or processes which encode assumptions about how the network was formed. Thus, generative models are like recipes for creating networks from scratch. For instance, a researcher might know the behavioral rules that typically underlie the formation of interactions and might therefore create a network-forming generative model that instantiates those rules. However, care must be taken when modeling networks using such general rules about interaction formation because they have the potential to produce reference networks that are very different from those observed, in spite of sharing the same number of nodes, links, or other high-level features. In particular, when a generative process is fundamentally non-biological, that generative model may be a poor reference model because it differs too dramatically—and implausibly—from the observed network.

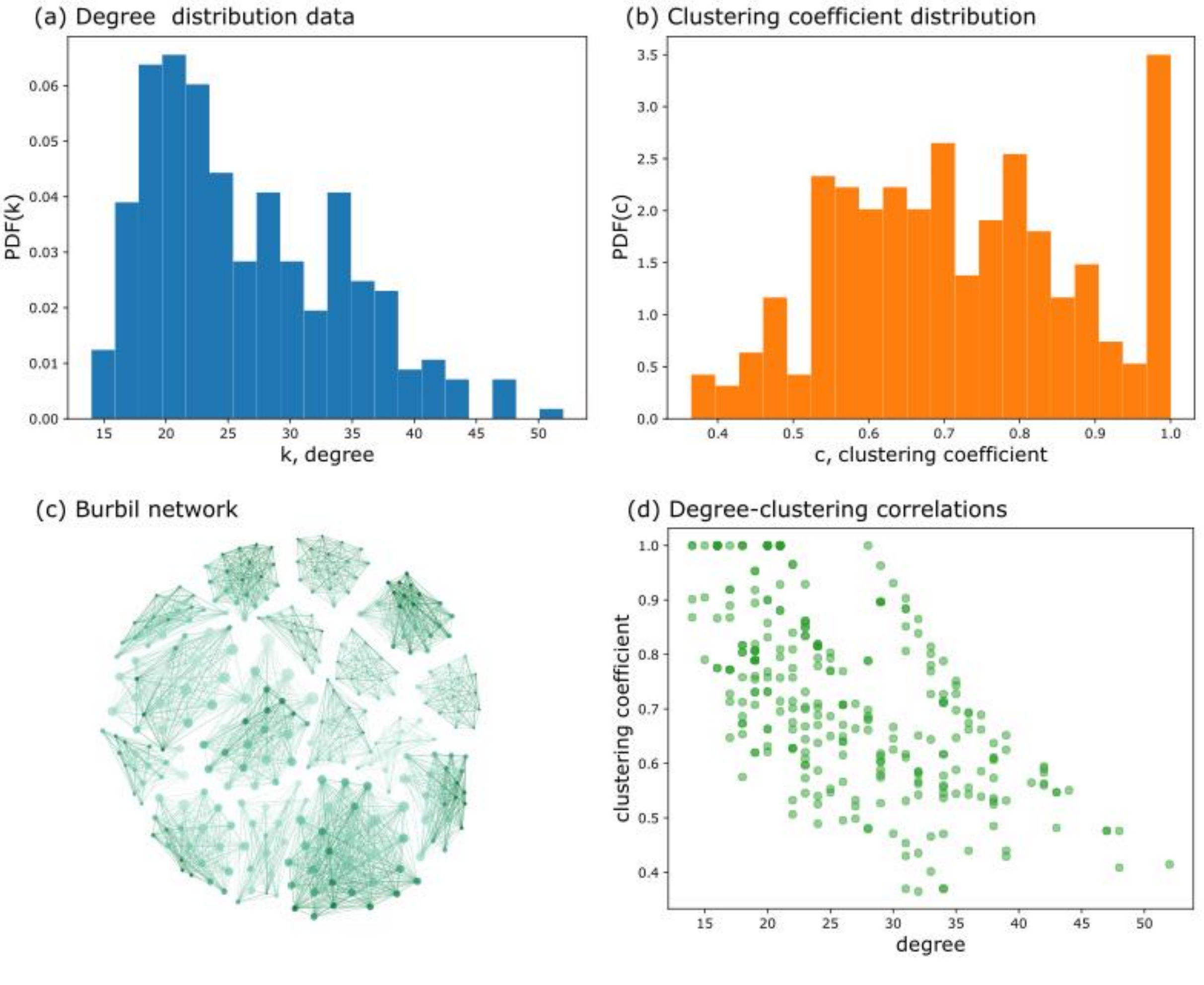

**Figure 5:** An illustration of covariance between two network properties in a burbil association network generated in Supplementary Material Section 3.3: (a) the degree distribution of the network, (b) the distribution of the clustering coefficient - the fraction of a node's friends that are friends with each other, (c) a visualization of the network where node size corresponds to degree and node color corresponds to clustering coefficient (network visualization was done in Gephi (Bastian et al., 2009) with force atlas, a force directed layout), (d) the correlation between clustering coefficient and degree in the network.

One example of a common but usually implausible reference model used in studies of animal behavior is the *uniform model G(n, p)* (Gilbert, 1959), also referred to as the Erdős–Rényi (ER) model. This model produces reference networks according to a simple recipe: begin with n nodes, and then place a link between each pair of nodes with probability p, independently of other pairs. While this model has the potential to create any simple network, i.e., a network without self loops or multiedges, it is designed to maximize entropy and uniformity, and is therefore unlikely to mimic any of the features of a network arising from animal behavior. Indeed, even animals following a Brownian motion in space will encounter each other in a way that is constrained by physical distance and barriers (Pinter-Wollman, 2015) meaning that even random encounters are poorly captured by the uniform reference model. Another example of a common reference model is the *configuration model*, introduced in Section 5. While the configuration model is commonly associated with the degree-preserving permutation of edges via rewiring, it is also simply a modified uniform model with more constraints: it chooses uniformly from all networks with a given degree sequence. The configuration model differs from *G(n,p)* in two key ways, by (i) having a fixed and non-random degree sequence and number of edges, and (ii) potentially containing self-loops and multi-edges. The configuration model is a generative reference model, for which there are a large number of different variations (Fosdick et al., 2018).

There is no shortage of generative models for networks. In fact, many common statistical models of networks, which we may usually think of as models to *fit to* data, are generative, including exponential random graph models (ERGMs: Lusher et al. 2013; Robins et al. 2007; Snijders et al. 2006) and stochastic block models (SBMs: Bollobás 1980; Snijders and Nowicki 1997). Just as with other classes of reference models, generative reference models require the careful consideration of the research question and hypothesis to inform the choice of the generative rules. For instance, ERGMs are dyadic models that can be used to test hypotheses about which features of dyads affect the presence or strength of edges. By including sex as an explanatory variable in an ERGM, it becomes possible for there to be differences between the likelihood of edges between female-female, female-male and male-male dyads. We illustrate some simple examples of the use of these models in our burbil case study. In the Supplementary Material (Section 3.4.1) we fit an ERGM to a within-group dominance network to simultaneously test hypotheses about the role of individual traits in explaining dominance relationships and an SBM to population-level association network to examine how well the community structure of the association network is explained by group membership.

A class of system-specific generative reference models are agent-based models (ABMs). In network analysis, ABMs can be spatially-explicit or socially-explicit. Spatially-explicit models can help reveal the role of spatial behaviour in explaining social network structure. For example, a generative model in which the movement of individuals is constrained by the spatial organization of the environment could be used to test whether spatial constraints are sufficient to explain social structure. Researchers could further include differences in spatial behaviour between individuals within such an ABM (Pinter-Wollman, 2015). In the Supplementary Material (Section 3.4.2) we use a spatially-explicit agent-based model to test whether the space use of burbils can explain patterns of between-group associations. Note that if we do not include any social component in the model then while our reference network is correlated with the observed network, it predicts far too many between-group associations.

Socially-explicit ABMs incorporate social behaviour (e.g., interaction preferences). One example of a socially-explicit ABM in the study of animal behavior is the *social inheritance model*, in which offspring are likely to form connections with friends of their parents while avoiding parents' enemies (Ilany and Akçay, 2016). While such a mechanism is highly likely, and has indeed been supported in some social systems, such as hyenas (Ilany et al., 2020), this model requires knowledge about relatedness and historical interactions, or long-term relationships, that are not available in all study systems. In our burbil case study in the Supplementary Material (Section 3.4.2) we develop two socially-explicit agent-based models that build on our spatially-explicit model. The first uses knowledge about burbil sub-group size to simulate burbils moving within groups rather than independently. The reference network generated is much more similar to the observed network than the previous version, which was only spatially-explicit. We then test the hypothesis that "clan" membership (burbil groups belong to three distinct clans) can help explain patterns of between-group associations. When we include clan membership in our ABM, the reference model produces a network that is very similar to the observed one, suggesting that clan membership can indeed explain the observed social interactions. In reality we would replicate these ABMs 1000 or more times to generate a full reference distribution rather than providing a single comparison, which we did to reduce computational run-time.

### 8.1. Key pitfalls for generative reference models

Comparing observed data with generative reference models provides insights about what processes might underlie observed interactions, and what processes might not. However, as a note of caution, it is possible to create the same types of networks with multiple generative processes—multiple recipes can generate similar patterns. Therefore, when observed data match a generative reference model, it does not necessarily mean that the modeled generative process is indeed the biological process that actually generated the observed network. Instead, it means that the modeled generative process is a plausible hypothesis that needs to be tested mechanistically.

Further, as generative models become more and more complicated, constraints on one property that is being modeled can have cascading effects on other properties. Complicated generative models with many parameters can result in one desirable property while other properties of the model remain poorly understood. Furthermore, complicated models require the specification of many parameters, which, if misspecified, can produce reference distributions that significantly differ from observations, leading to spurious conclusions. Uniform and configuration models have enjoyed much usage because their complete distributions, constraints, and correlations among their properties are well understood. However, these simple models might not encapsulate all the biological complexities a researcher might be interested in. As we experiment with more exotic and complex generative models, which capture more realistic aspects of observed behavior, it is increasingly important to carefully check for the unintentional creation of fundamentally *unrealistic* patterns and behaviors in our reference models. Such unrealistic patterns can be identified through an iterative approach, for example, by going back to the study system and asking if patterns observed in the reference models are feasible in real life.

## 9. Conclusions

1. We provided an overview of the process and caveats of using reference models when analyzing social networks. We detailed common approaches to generating reference distributions that increase in level of abstraction with respect to the observed dataset.

2. We highlighted the strengths and weaknesses of each approach, drawing attention to common pitfalls that can arise when using them.

3. Our goal is to provide a guide for researchers using social network analysis for hypothesis testing in diverse study systems. We anticipate that our overview will help researchers better appreciate the similarities and differences between different analytic approaches and also encourage greater confidence in designing appropriate reference models for their research questions.

4. Our key message is that the construction of reference models should depend closely on both the research question and study system and that the use of generic approaches applied without careful evaluation as to their suitability can lead to incorrect inference.

## Acknowledgements

This work was conducted as a part of the *Null Models in Social Behavior* Working Group at the National Institute for Mathematical and Biological Synthesis, supported by the National Science Foundation through NSF Award #DBI-1300426, with additional support from The University of Tennessee, Knoxville. We are grateful to Maureen Rombach for providing artwork used in the paper.

# Supplementary Material: A guide to choosing and implementing reference models for social network analysis

Matthew Silk

02/02/2021

Please direct any questions about the examples presented in this R script to Matthew Silk (matthewsilk@outlook.com)

# Section 1 - Preparation

First we are going to prepare the R environment and load the necessary packages for our case study. If you want to explore the variability in the networks and result possible you can change the number in the set.seed() function to produce different networks.

Note that throughout this script we use an edited version of the asnipe get_network2 function that doesn't print messages

```
#Set seed for reproducibility
#Can be changed to produce different networks and explore variability in results
set.seed(35)

##load packages
library(asnipe)
library(igraph)
library(boot)
library(prodlim)
library(sna)
library(assortnet)
library(blockmodels)
library(ergm)
library(ergm.count)
library(tnet)
library(vegan)
```

# Section 2 - Network Generation

## *Creating a population of burbils with social networks*

---

Burbils live in open habitats throughout the world. They form fission-fusion societies characterised by stable social groups that roost together but fission into smaller subgroups when foraging during the day. Foraging subgroups from different groups occasionally meet and intermingle creating opportunities for between-group interactions. Burbil groups vary in size and we are unsure whether groups of different sizes have similar social network structures. Groups also contain two unique colour morphs: burbils with red noses, and those with orange noses. As well as being able to identify individual burbils (which we use to construct their social networks!), we are also able to distinguish male and female burbils as well as those from three distinct age classes (adults, subadults and juveniles). We know that burbils are involved in both dominance interactions and affiliative interactions with group-mates. We suspect they may have a dominance hierarchy, but we don't know this for sure. We have a lot to find out!

---

Burbils form fission-fusion societies characterised by large groups that roost together at night but fission into smaller subgroups when foraging during the day. Foraging subgroups from different roosting groups occasionally meet and intermingle, creating opportunities for between-group associations. These between-group associations are more likely if the two Burbil groups belong to the same "clan".

---

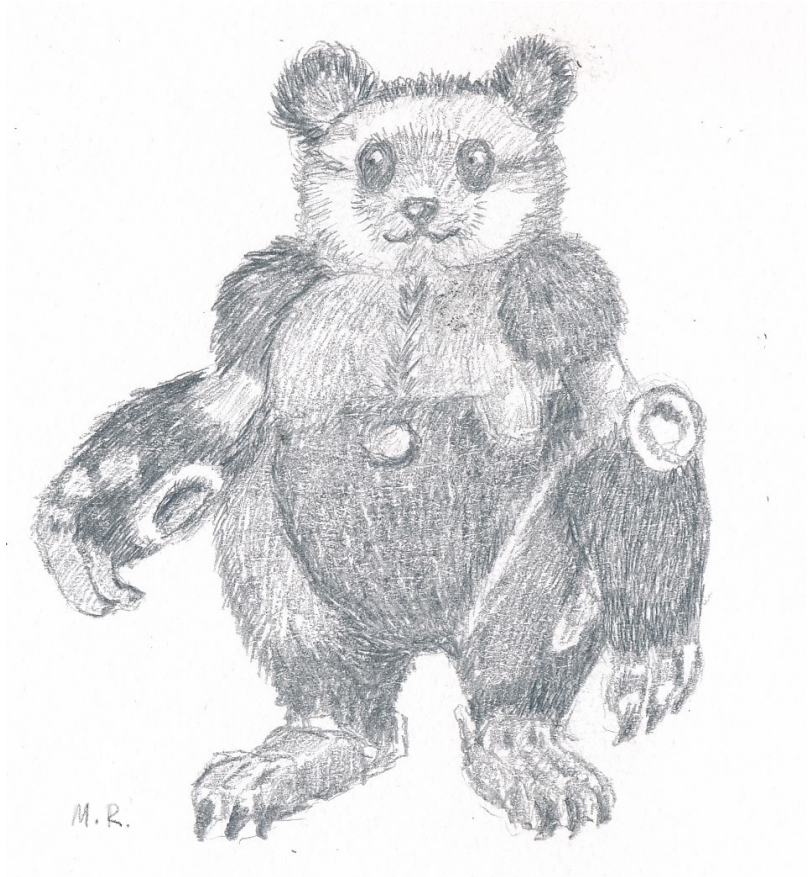

A burbil

---

---

# Section 2.1 - Generate population network

In this section of the code we create our burbil society (starting with the association network), explaining what we do as we go along. With practice it should be possible to change some of the numbers in this code to change the nature of social relationships in your burbil society.

---

```
#Set the mean group size
GS<-20

#Here we create a grid of locations for our observations
x<-seq(3,18,1)
y<-seq(3,18,1)
locs<-expand.grid(x,y)
names(locs)<-c("x","y")

#Here we assign coordinates to our groups. We create 9 groups in total.
group_locs<-locs[locs$x%%4==0&locs$y%%4==0,]

#Here we store the total number of groups
n_groups<-dim(group_locs)[1]

#Here we create three distinct clans of burbils. This will effect associations bet
ween members of different groups
group_clans<-sample(c("A","B","C"),n_groups,replace=TRUE)
```

```
#Set the probability of burbils from the same clan intermingling if they happen to
forage at the same location
p_wc<-1
#Set the probability of burbils from different clans intermingling if they happen
 to forage at the same location
p_bc<-0.4

#Create a list to store individual IDs
indss<-list()

#Create a list to store group sizes
gss<-list()

#Create a list to store the sex of each individual
sexes<-list()

#Create a list to store the age of each individual
ages<-list()

#Create a list to store the nose colour of each individual
noses<-list()

#Create a list to store information on which day a subgroup is observed on
daysl<-list()

#Create a list to store a group-by-individual matrix for each burbil group
gbis<-list()

#Set the mean number of subgroups observed for each group each day
sg_mn<-5

#Set the strength of assortativity based on nose colour
#Set a number between 0 and 1
sg_ass<-0.15

#Genereate association data within each burbil group!
for(j in 1:n_groups){

#individual identities
inds<-seq(1,rpois(1,GS),1)
indss[[j]]<-inds

#group size
gs<-length(inds)
gss[[j]]<-gs

#sex
sex<-sample(c("M","F"),gs,replace=TRUE)
sexes[[j]]<-sex

#age
age<-sample(c("AD","SUB","JUV"),gs,replace=TRUE,prob=c(0.6,0.2,0.2))
ages[[j]]<-age

#nose
nose<-sample(c("RED","ORANGE"),gs,replace=TRUE,prob=c(0.7,0.3))
noses[[j]]<-nose
```

```
#----------------------------------

#Define number of subgroups on the first day
n_sg<-rpois(1,sg_mn-1)+1

#find halfway point
max_red<-floor(n_sg/2)

#Sample subgroups on the first day
subgroups1<-sample(n_sg,sum(nose=="RED"),replace=TRUE,prob=c(rep(0.5+sg_ass,max_re
d),rep(0.5-sg_ass,n_sg-max_red)))
subgroups2<-sample(n_sg,sum(nose=="ORANGE"),replace=TRUE,prob=c(rep(0.5-sg_ass,max
_red),rep(0.5+sg_ass,n_sg-max_red)))

subgroups<-rep(NA,gs)
subgroups[nose=="RED"]<-subgroups1
subgroups[nose=="ORANGE"]<-subgroups2

#Store relevant information in the group-by-individual matrix and days vector
gbi<-matrix(0,nc=gs,nr=n_sg)
gbi[cbind(subgroups,seq(1,gs,1))]<-1
days<-rep(1,nrow(gbi))

#Repeat process over 100 days of observations
for(i in 2:100){

  n_sg<-rpois(1,sg_mn-1)+1

  #find halfway point
  max_red<-floor(n_sg/2)

  subgroups1<-sample(n_sg,sum(nose=="RED"),replace=TRUE,prob=c(rep(0.5+sg_ass,max_
red),rep(0.5-sg_ass,n_sg-max_red)))
  subgroups2<-sample(n_sg,sum(nose=="ORANGE"),replace=TRUE,prob=c(rep(0.5-sg_ass,m
ax_red),rep(0.5+sg_ass,n_sg-max_red)))

  subgroups<-rep(NA,gss[[j]])
  subgroups[nose=="RED"]<-subgroups1
  subgroups[nose=="ORANGE"]<-subgroups2

  tgbi<-matrix(0,nc=gs,nr=n_sg)
  tgbi[cbind(subgroups,seq(1,gs,1))]<-1
  days<-c(days,rep(i,nrow(tgbi)))
  gbi<-rbind(gbi,tgbi)
}

#We edit the group-by-individual matrix and days vector to delete any "empty" grou
ps
gbi2<-gbi[rowSums(gbi)>0,]
days<-days[rowSums(gbi)>0]
gbi<-gbi2

#We could create and plot the network for each burbil group
#(NOT RUN HERE)
#net<-get_network2(gbi)
#net2<-graph.adjacency(net,mode="undirected",weighted=TRUE)
```

```
#plot(net2,vertex.color=noses[[j]],edge.width=(edge_attr(net2)$weight*10)^2)

daysl[[j]]<-days
gbis[[j]]<-gbi

}

#We now go through and assign a location to every subgroup
sglocs<-list()
for(i in 1:n_groups){
  tx<-rep(NA,dim(gbis[[i]])[1])
  ty<-rep(NA,dim(gbis[[i]])[1])
  sglocs[[i]]<-data.frame(tx,ty)
  names(sglocs[[i]])<-c("x","y")
  sglocs[[i]]$x<-group_locs[i,1]+round(rnorm(dim(gbis[[i]])[1],0,2))
  sglocs[[i]]$y<-group_locs[i,2]+round(rnorm(dim(gbis[[i]])[1],0,2))
}

#Vector recording number of individuals in each group
n_inds<-numeric()
for(i in 1:n_groups){
  n_inds[i]<-dim(gbis[[i]])[2]
}

#Calculate total individuals in the population
n_tot<-sum(n_inds)

#Population-level individuals identities
inds_tot<-seq(1,n_tot,1)

#Information on each individual's group membership
g_tot<-rep(seq(1,n_groups,1),n_inds)

#Information on each individual's within-group identity
gi_tot<-seq(1,n_inds[1],1)
for(i in 2:n_groups){
  gi_tot<-c(gi_tot,seq(1,n_inds[i],1))
}

#We now calculate the full population association network
full_net<-matrix(0,nr=n_tot,nc=n_tot)

#Counts up between-group associations
for(i in 1:100){
  for(j in 1:(n_groups-1)){
    for(k in (j+1):n_groups){
      tA<-paste0(sglocs[[j]][,1],"-",sglocs[[j]][,2])
      tB<-paste0(sglocs[[k]][,1],"-",sglocs[[k]][,2])
      tA2<-tA[daysl[[j]]==i]
      tB2<-tB[daysl[[k]]==i]
      tt<-match(tA2,tB2)
      if(sum(is.na(tt))<length(tt)){
      if(group_clans[j]==group_clans[k]){same<-rbinom(1,1,p_wc)}
      if(group_clans[j]!=group_clans[k]){same<-rbinom(1,1,p_bc)}
      if(same==1){
        paste(i,j,k)
        for(m in length(tt)){
```

```
          if(is.na(tt[m])==FALSE){
            tsg1<-which(tA==tA2[m]&daysl[[j]]==i)
            tsg2<-which(tB==tB2[tt[m]]&daysl[[k]]==i)
            tid1<-which(gbis[[j]][tsg1,]==1)
            tid2<-which(gbis[[k]][tsg2,]==1)
            tid1a<-inds_tot[g_tot==j&gi_tot%in%tid1]
            tid2a<-inds_tot[g_tot==k&gi_tot%in%tid2]
            full_net[tid1a,tid2a]<-full_net[tid1a,tid2a]+1
            full_net[tid2a,tid1a]<-full_net[tid1a,tid2a]
          }
        }
      }
      }
    }
  }
}

#converts between group assocaitions to SRIs
for(i in 1:(nrow(full_net)-1)){
  for(j in (i+1):nrow(full_net)){
    full_net[i,j]<-full_net[i,j]/(200-full_net[i,j])
    full_net[j,i]<-full_net[i,j]
  }
}

#Adds within-group associations to the population network
for(i in 1:n_groups){
  full_net[inds_tot[g_tot==i],inds_tot[g_tot==i]]<-get_network2(gbis[[i]])
}

#Plots the population social network
full_net2<-graph.adjacency(full_net,mode="undirected",weighted=TRUE)
par(mar=c(0,0,2,0))
plot(full_net2,vertex.color=unlist(noses),vertex.label=NA,vertex.size=4,edge.width
=(edge_attr(full_net2)$weight*10)^2,main="Population association network")
```

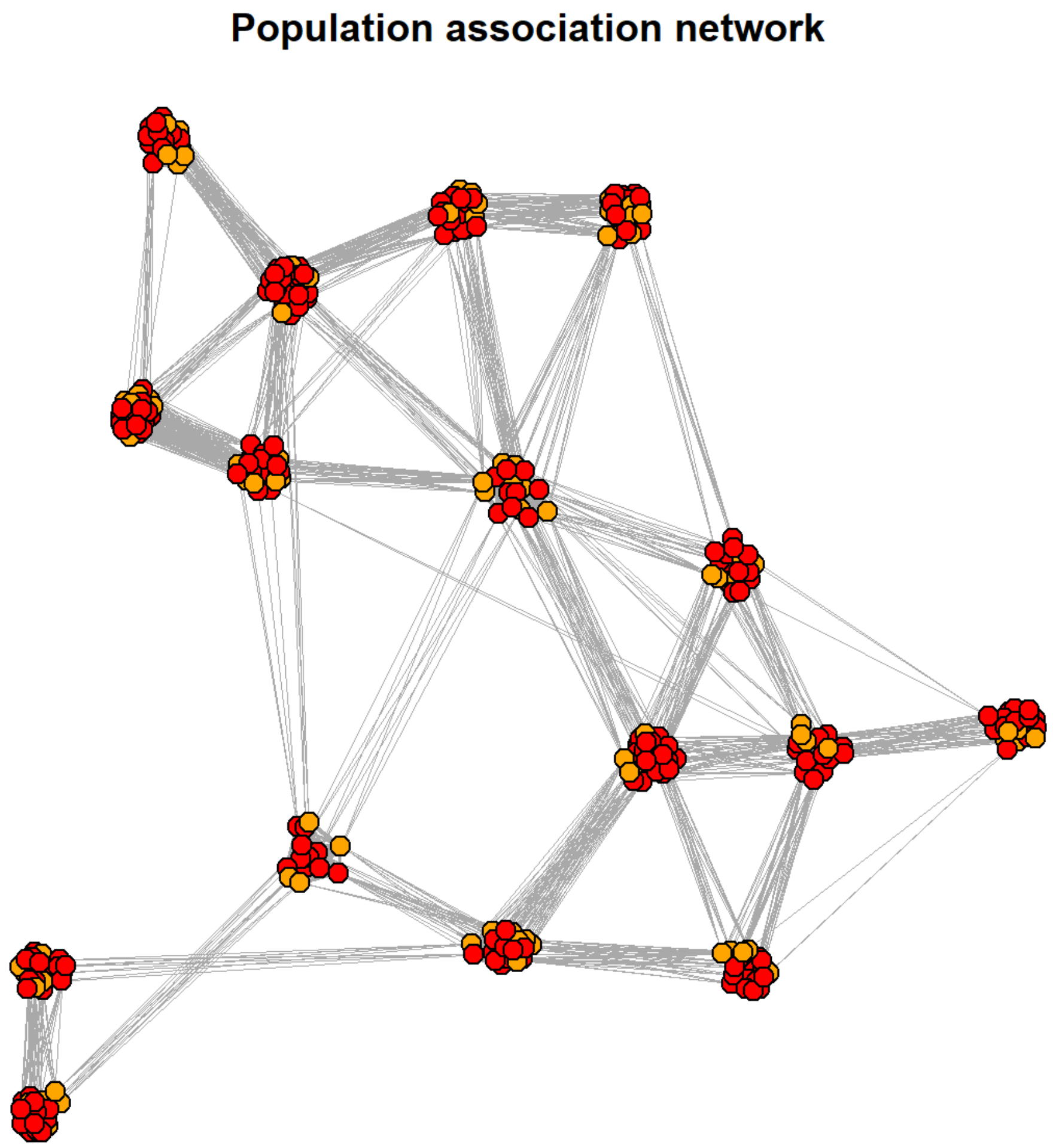

```
par(mar=c(5,6,2,2))
```

## Section 2.2 - Generate within-group networks of behavioural interactions

We now focus in on a single burbil group (group 1) and generate data on dominance interactions and affiliative behaviours. We are going to generate different structure in these interactions that we will attempt to uncover in subsequent analyses.

Another important feature is that dominance and affiliative interactions are only possible within subgroups. We use scan sampling of subgroups (as they are small) to record all interactions occurring. The number of interactions recorded depends on the size of the subgroup. As diligent researchers, we record the day and subgroup of all interactions.

### Create dominance interactions

```
#individual identities
inds1<-indss[[1]]

#group size
gs1<-gss[[1]]
```

```
#sex
sex1<-sexes[[1]]

#age
age1<-ages[[1]]

#nose
nose1<-noses[[1]]

gbi<-gbis[[1]]

#Set-up vectors to store results
GROUP<-numeric()
WINNER<-numeric()
LOSER<-numeric()

#Define the resource holding potential of different individuals
RHP_ad<-1
RHP_sub<-0
RHP_juv<- -1
RHP_M<--0.5
RHP_resid<-0.2
RHPs1<-rnorm(gs1,RHP_ad*(age1=="AD")+RHP_sub*(age1=="SUB")+RHP_juv*(age1=="JUV")+R
HP_M*(sex1=="M"),RHP_resid)

#Define the mean number of interactions observed per individual in a subgroup
m_nipi<-2

#record which group the interactions occur in
grD<-numeric()

#Generate dominance interaction data!
c<-1
for(g in 1:nrow(gbi)){
if(rowSums(gbi)[g]>1){
nipi<-rpois(1,m_nipi)
indivs<-which(gbi[g,]==1)
ni<-nipi*length(indivs)
for(n in 1:ni){
i1<-sample(indivs,1)
ifelse(rowSums(gbi)[g]==2,i2<-indivs[indivs!=i1],i2<-sample(indivs[indivs!=i1],1))
winner<-rbinom(1,1,inv.logit(RHPs1[i1]-RHPs1[i2]))
GROUP[c]<-g
if(winner==1){
  WINNER[c]<-i1
  LOSER[c]<-i2
}
if(winner==0){
  WINNER[c]<-i2
  LOSER[c]<-i1
}
grD[c]<-g
c<-c+1
}
}
}
```

```
#Create the dominance network in igraph format
dom_net<-graph_from_edgelist(cbind(WINNER,LOSER), directed = TRUE)
E(dom_net)$weight <- 1
dom_net<-simplify(dom_net, edge.attr.comb=list(weight="sum"))

#Plot the dominance network that results (it is densely connected and so the netwo
rk plot isn't especially informative)
plot(dom_net,edge.width=log(edge_attr(dom_net)$weight,10)^5,layout=layout_in_circl
e,main="Dominance network",edge.arrow.size=0.5)
```

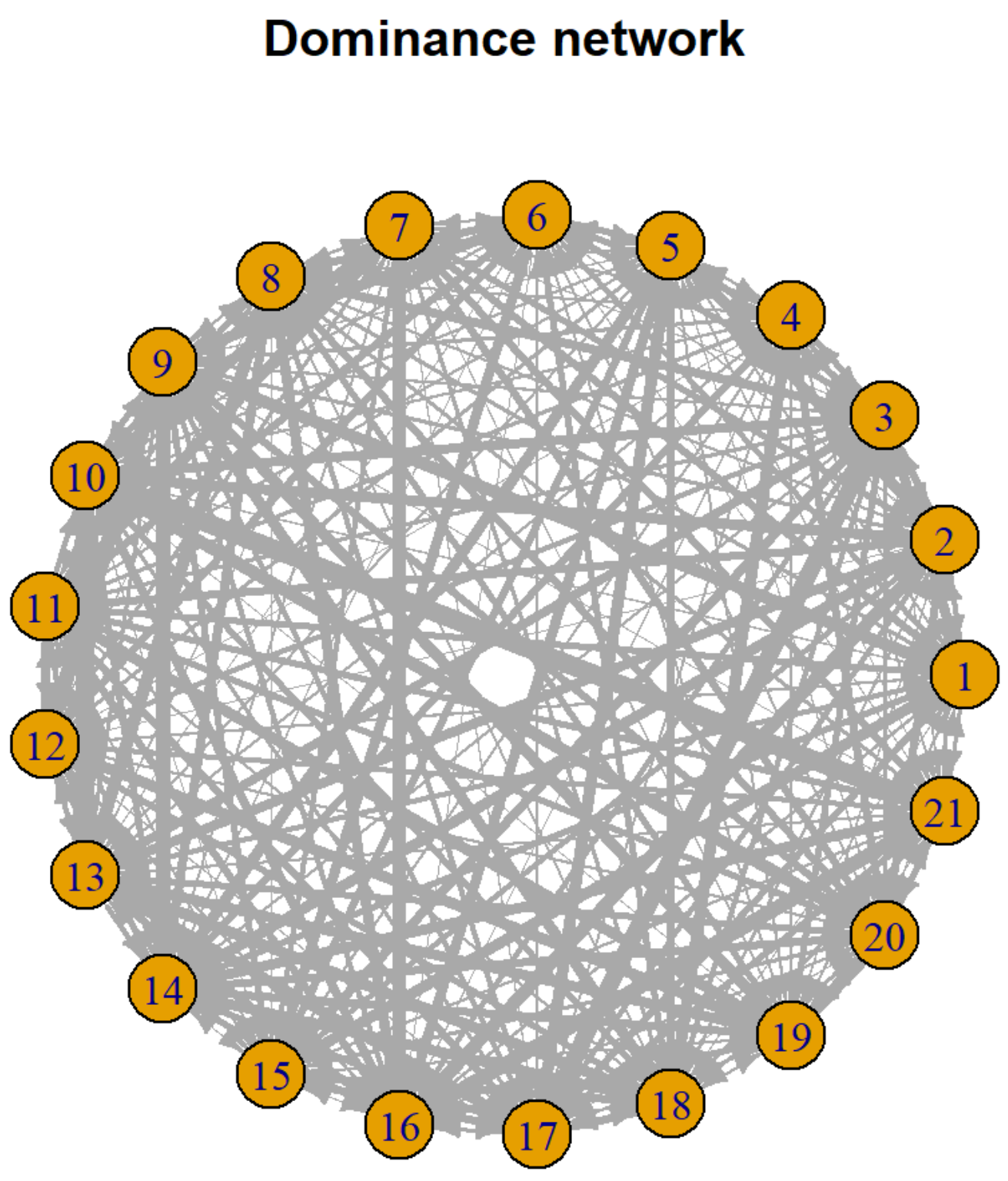

```
#To show that our code to generate the dominance network works we plot the relatio
nship between in-strength and out-strength and it is negatively correlated as woul
d be expected for a linear dominance hierarchy
plot(strength(dom_net,mode="out"),strength(dom_net,mode="in"),pch=16,xlab="Out-Deg
ree",ylab="In-Degree",cex.lab=1.5,cex.axis=1,main="Correlation between out- and in
-degree of nodes in the dominance network")
```

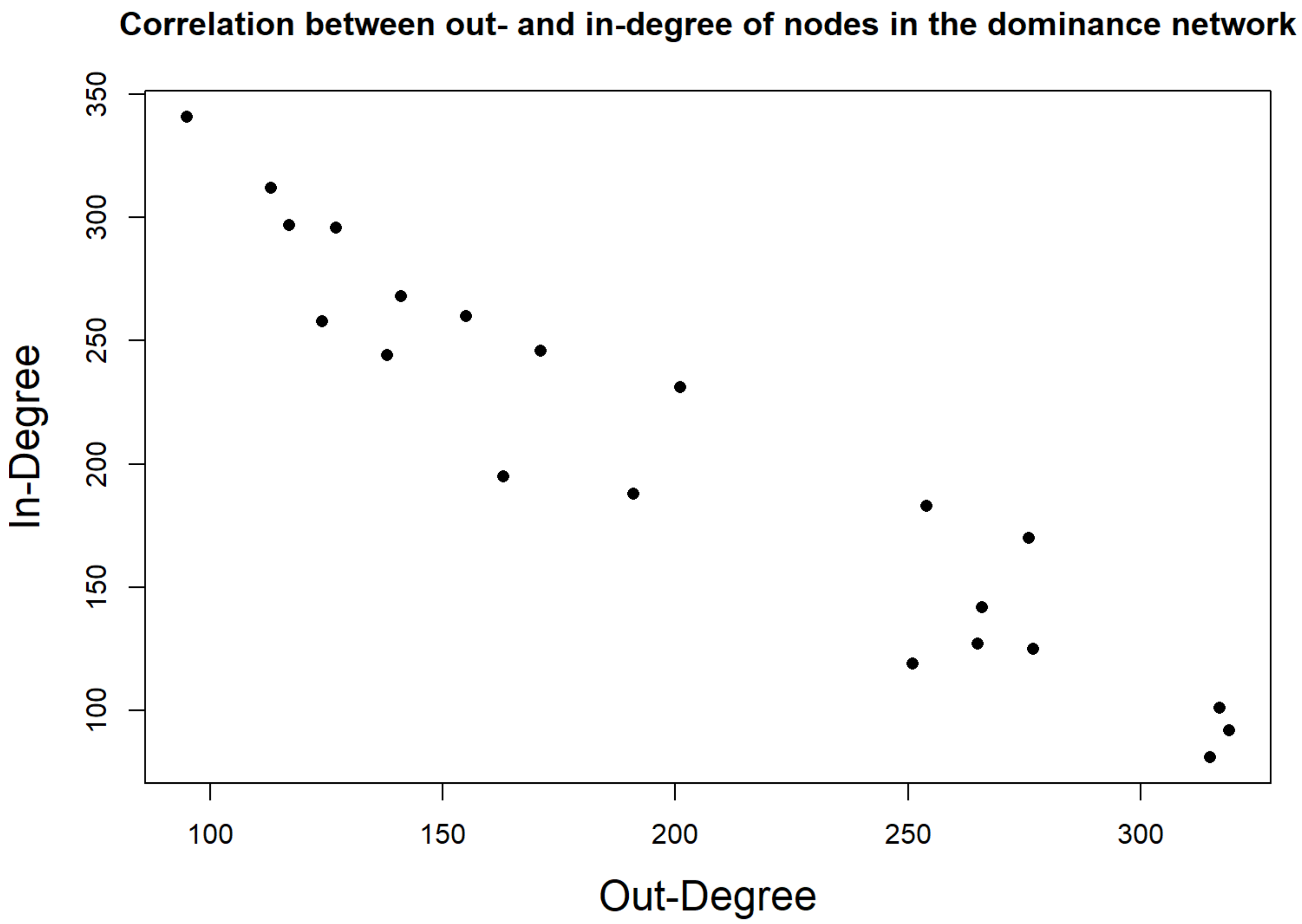

---

Create affiliative interactions

```
#We use an equivalent approach as for dominance networks so have kept much of the
 coding the same (hence the mismatch in names)

#Set-up vectors to store results
GROUP<-numeric()
GIV<-numeric()
REC<-numeric()

#Define the tendency of different individuals to initiate affiliative interactions
AHP_ad<- -1
AHP_sub<- -1
AHP_juv<-1
AHP_M<-0
AHP_nose<-1
AHP_resid<-0.2
AHPs1<-rnorm(gs1,AHP_ad*(age1=="AD")+AHP_sub*(age1=="SUB")+AHP_juv*(age1=="JUV")+A
HP_M*(sex1=="M"),AHP_resid)

#Define the mean number of interactions observed per individual in a subgroup
m_nipi<-0.5

#record which group interactions occur in
grA<-numeric()
```

```
#Generate affiliative interaction data!
c<-1
for(g in 1:nrow(gbi)){
  if(rowSums(gbi)[g]>1){
    nipi<-rpois(1,m_nipi)
    indivs<-which(gbi[g,]==1)
    ni<-nipi*length(indivs)
    for(n in 1:ni){
      i1<-sample(indivs,1)
      ifelse(rowSums(gbi)[g]==2,i2<-indivs[indivs!=i1],i2<-sample(indivs[indivs!=i
1],1))
      tn<-0
      if(nose1[i1]==nose1[i2]){tn<-1}
      winner<-rbinom(1,1,inv.logit(AHPs1[i1]-AHPs1[i2]+tn))
      GROUP[c]<-g
      if(winner==1){
        GIV[c]<-i1
        REC[c]<-i2
      }
      if(winner==0){
        GIV[c]<-i2
        REC[c]<-i1
      }
      grA[c]<-g
      c<-c+1
    }
  }
}

#Create the affiliative network in igraph format
aff_net<-graph_from_edgelist(cbind(GIV,REC), directed = TRUE)
E(aff_net)$weight <- 1
aff_net<-simplify(aff_net, edge.attr.comb=list(weight="sum"))

#Plot the affiliative network that results
plot(aff_net,edge.width=log(edge_attr(aff_net)$weight,6)^5,layout=layout_in_circl
e,main="Affiliative network",edge.arrow.size=0.5)
```

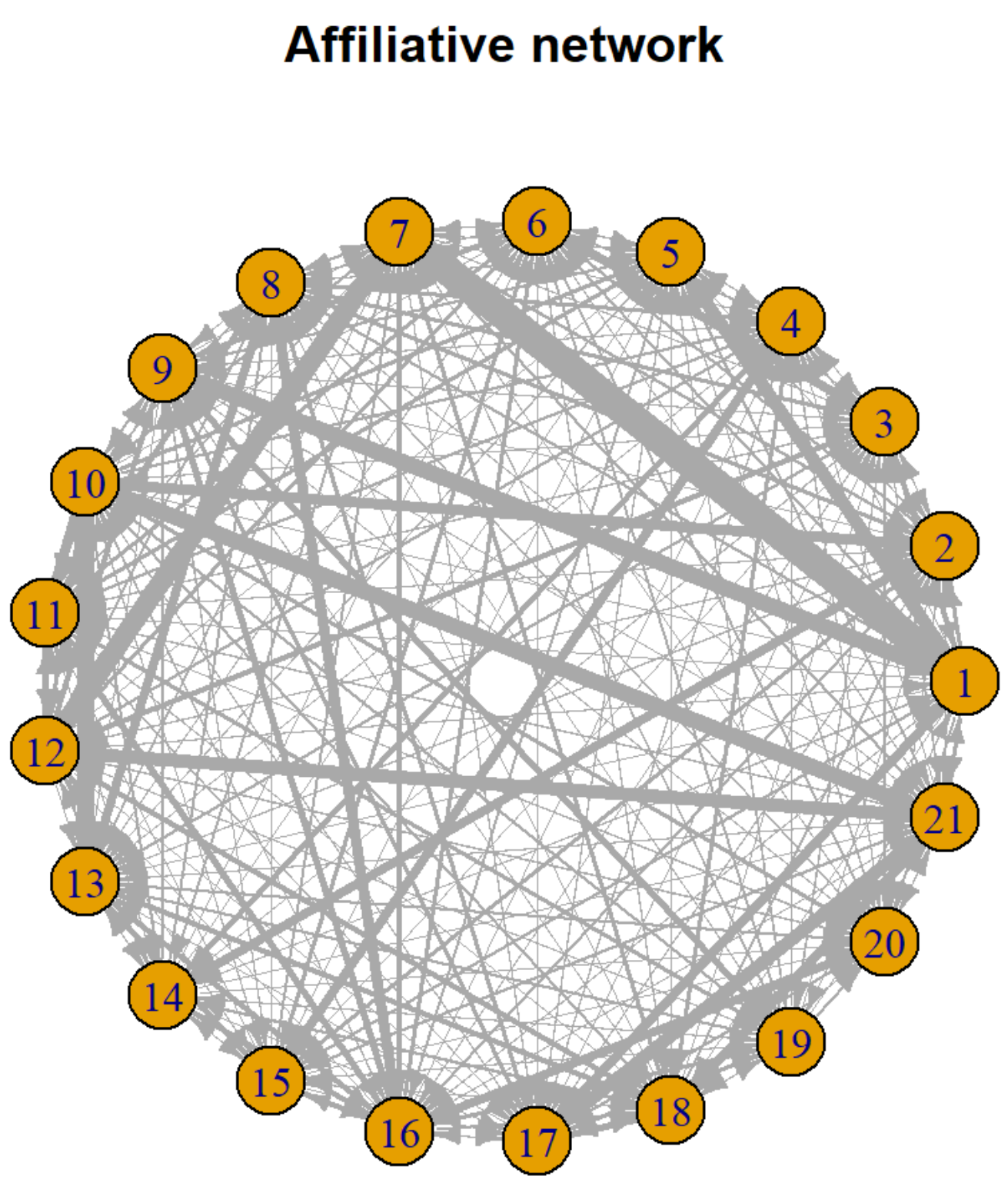

```
#Plot the same correlation used for dominance networks
plot(strength(aff_net,mode="out"),strength(aff_net,mode="in"),pch=16,xlab="Out-Deg
ree",ylab="In-Degree",cex.lab=1.5,cex.axis=1,main="Correlation between out- and in
-degree of nodes in the affiliative network")
```

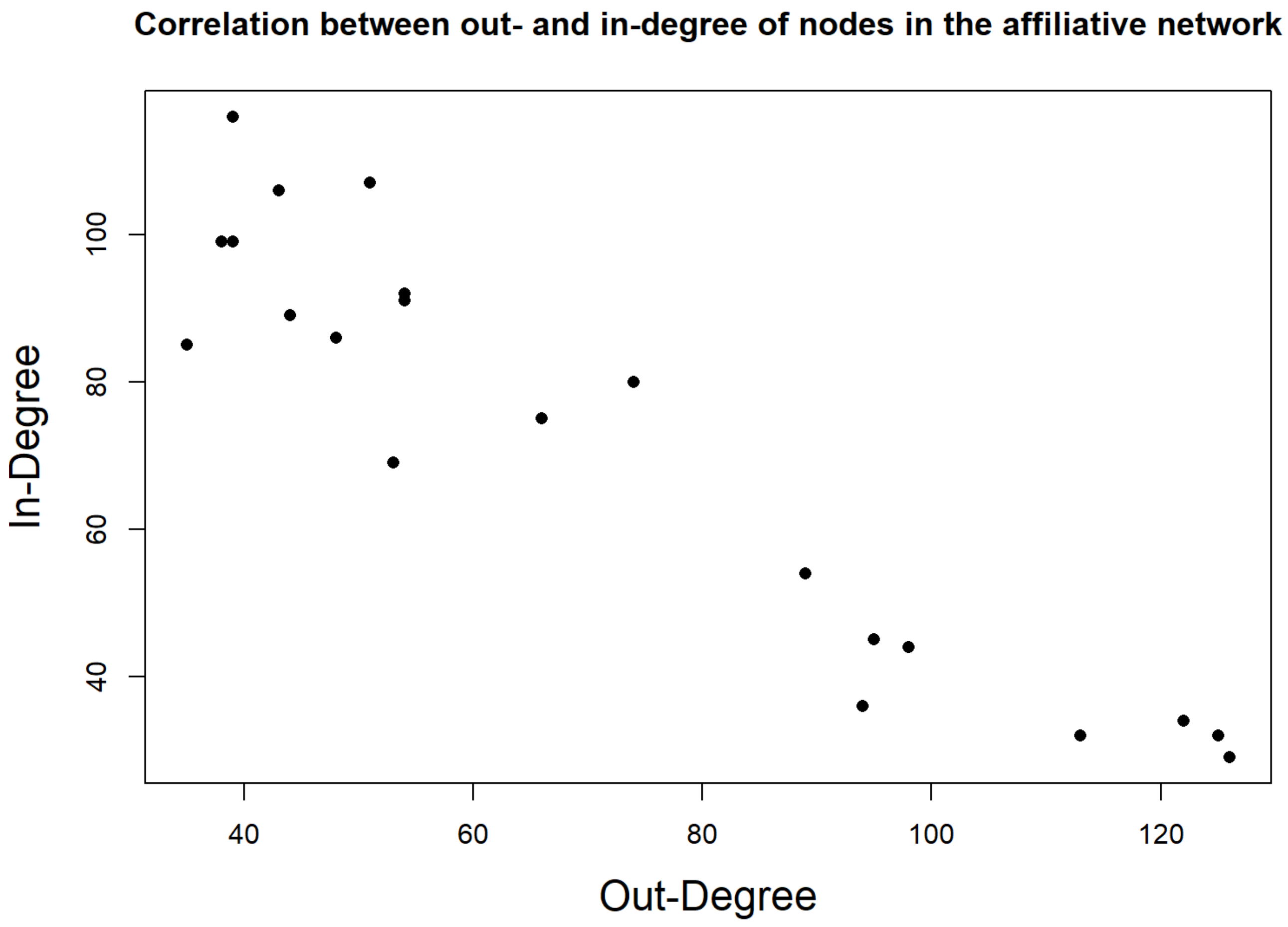

## Section 2.3 - Generate huddling networks

Data were also collected on the huddling networks of two burbil groups while they were roosting during summer and winter. This data can be used to test if the huddling networks differs between small and large groups. We simulate that data here.

```
sm_g<-which.min(n_inds)
bi_g<-which.max(n_inds)

#Generate "roosting/huddling network of burbils in the smallest group in the summe
r
hud_netSM<-sample_smallworld(dim=1, size=gss[[sm_g]], nei=3, p=0.05, loops = FALSE
, multiple = FALSE)

#Plot network
plot(hud_netSM,main="Huddling network in small group")
```

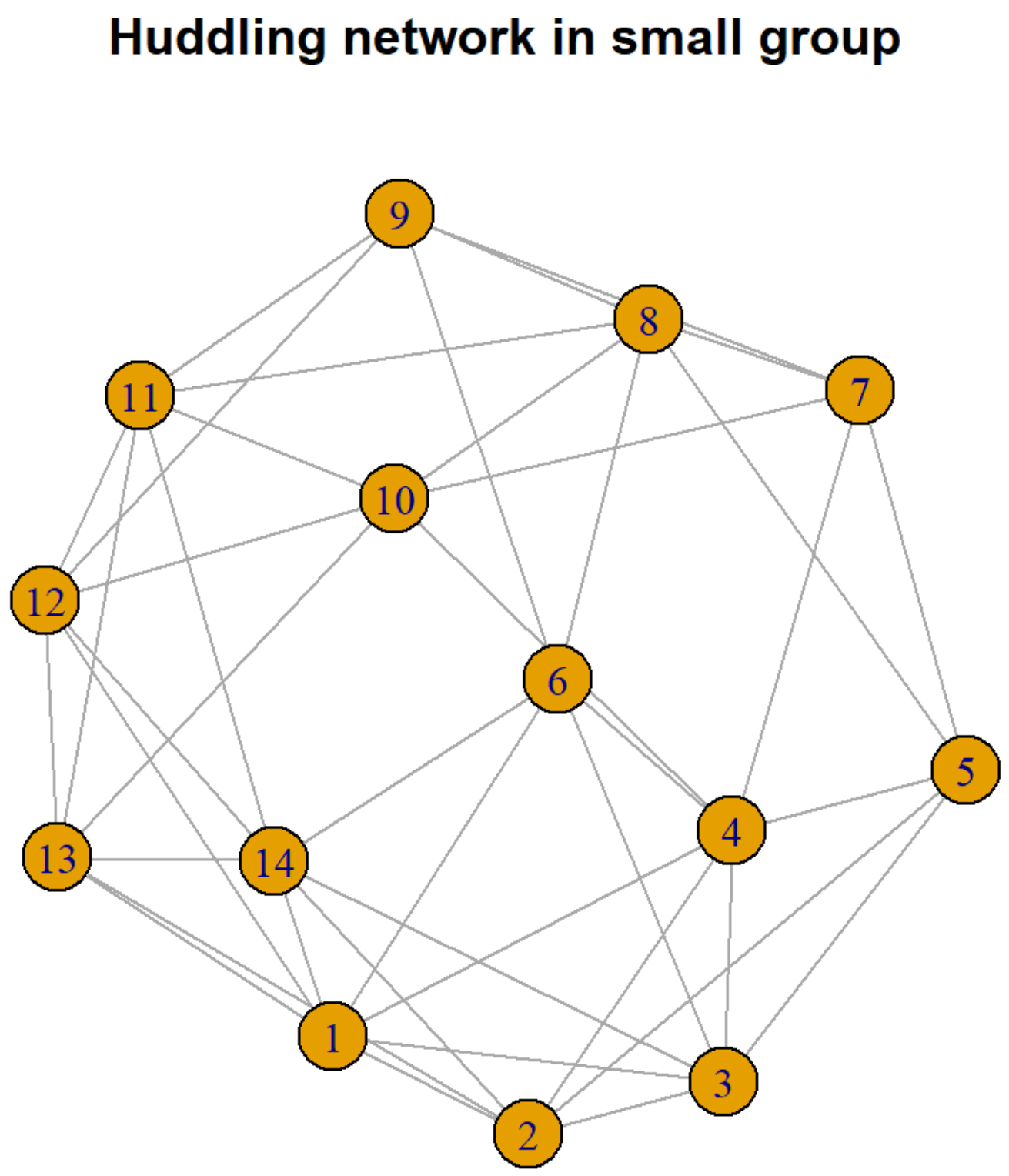

```
#Calculate betweenness of network
igraph::betweenness(hud_netSM)
```

```
##  [1] 3.821429 3.142857 2.059524 6.535714 3.059524 5.380952 2.166667
##  [8] 4.392857 2.547619 5.214286 3.214286 2.976190 2.654762 4.833333
```

```
##-----------------------------------------
##-----------------------------------------

#Generate "roosting/huddling network of burbils in the biggest group  in the summe
r
hud_netBI<-sample_smallworld(dim=1, size=gss[[bi_g]], nei=3, p=0.05, loops = FALSE
, multiple = FALSE)

#Plot network
plot(hud_netBI,main="Huddling network in big group")
```

## Huddling network in big group

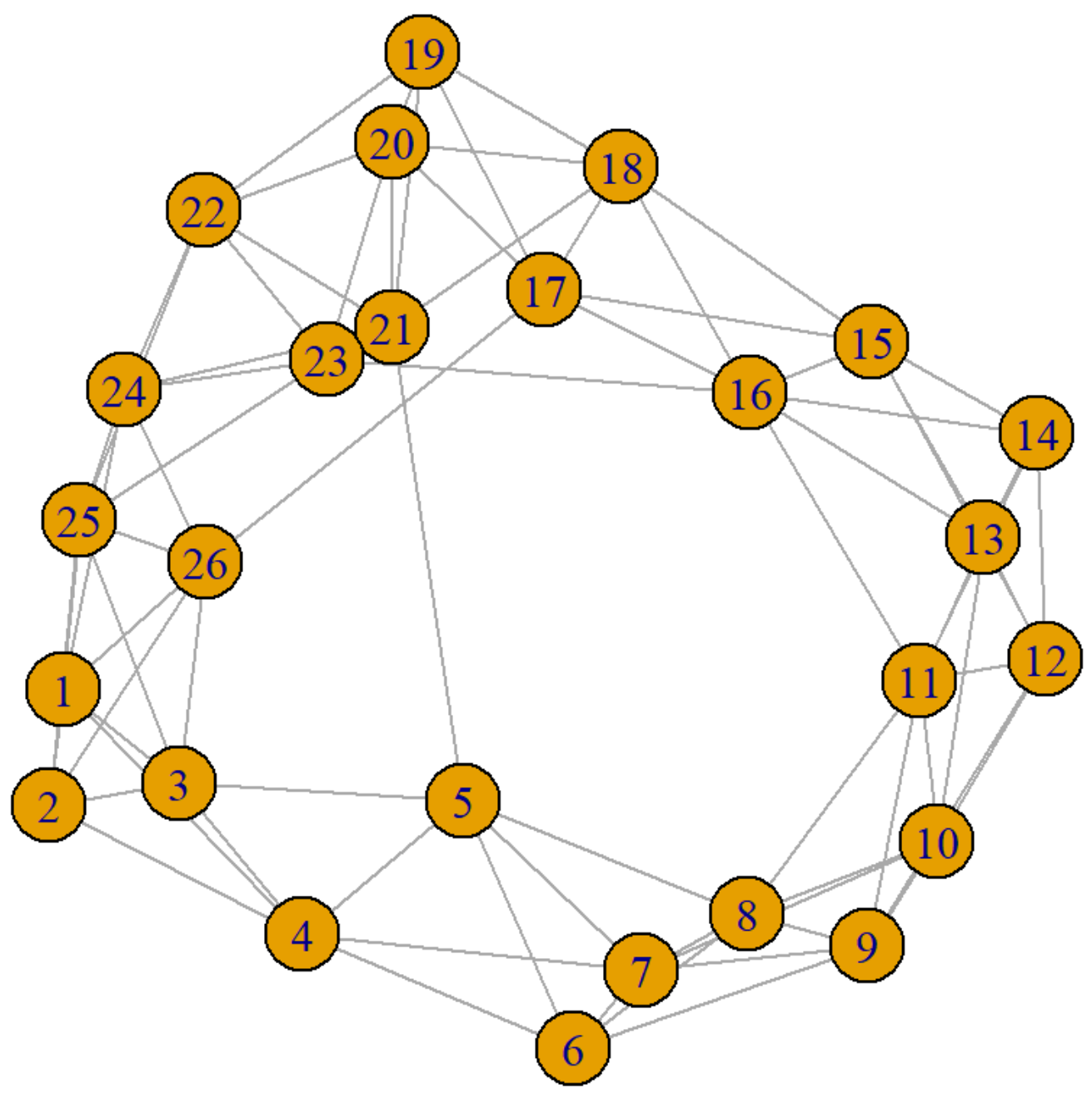

```r
#Calculate betweenness of network
igraph::betweenness(hud_netBI)
```

```
##  [1]  7.262572  3.457215 11.813276 22.990693 44.068957  7.520382 18.858081
##  [8] 14.385155 11.266270 10.802778 21.436941  9.367496  7.158929  1.370635
## [15] 19.455988 44.348846 30.635191 14.511310  2.268651  3.560317 37.887302
## [22]  5.767857 26.572817  7.732738 14.036706 21.462897
```

```r
#Examine differences in betweenness by inspecting histograms
hist(igraph::betweenness(hud_netSM),breaks=seq(0,200,1),col=rgb(1,0,0,0.3),border=
NA,xlab="Betweenness",cex.lab=1.5,cex.axis=1,main="Betweenness centrality distribu
tion in\n small group (red) and big group (blue) networks")
hist(igraph::betweenness(hud_netBI),breaks=seq(0,200,1),col=rgb(0,0,1,0.3),border=
NA,add=TRUE,cex.lab=1.5,cex.axis=1,main="")
```

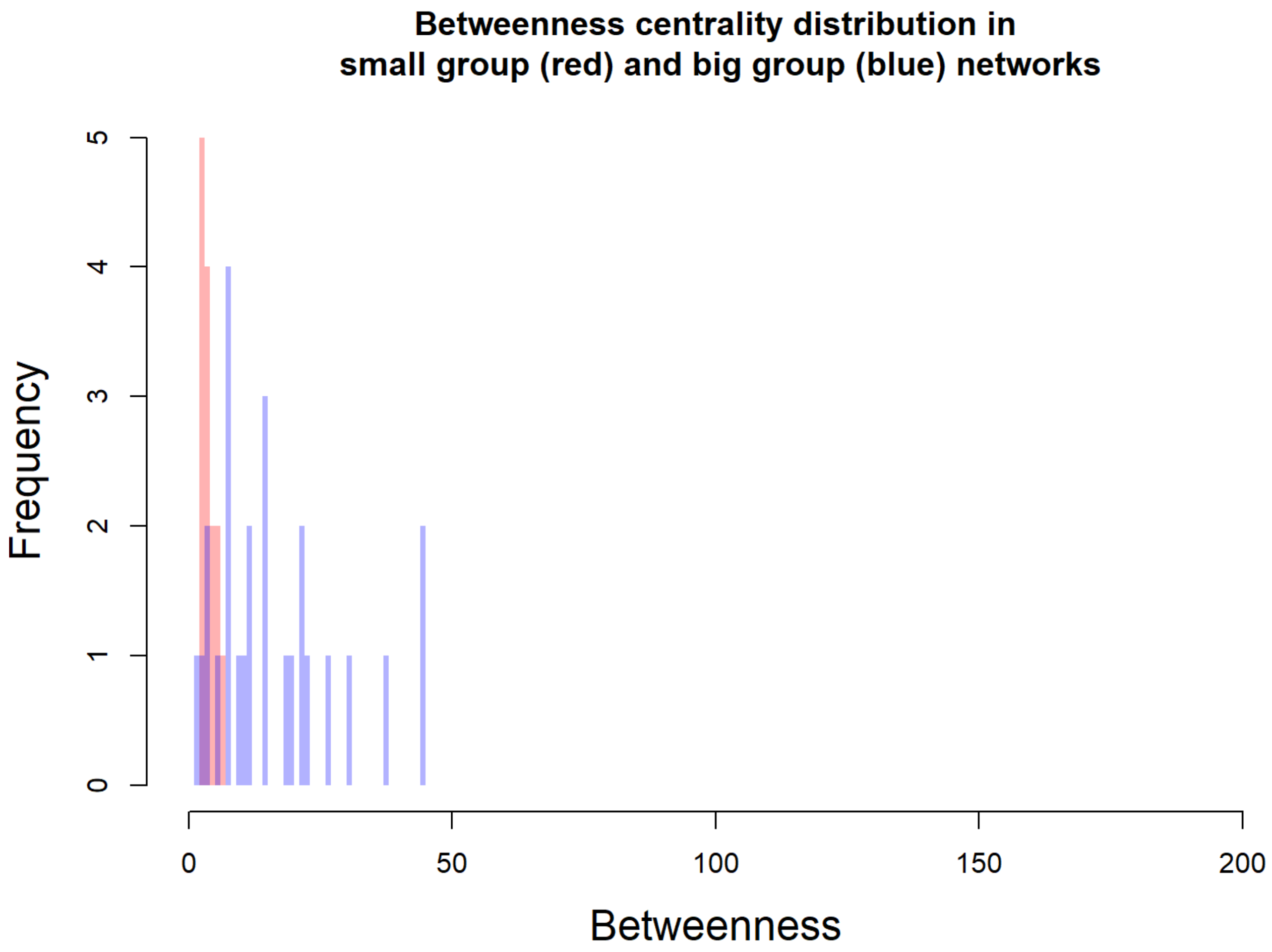

```
##-----------------------------------------
##-----------------------------------------

#Generate "roosting/huddling network of burbils in the smallest group in the winte
r
hud_netSM_w<-erdos.renyi.game(n=gss[[sm_g]], p=0.3, loops = FALSE, multiple = FALS
E)
hud_netBI_w<-erdos.renyi.game(n=gss[[bi_g]], p=0.3, loops = FALSE, multiple = FALS
E)
```

## Section 2.4 - A second population network

We have also been sent association data from a similar but smaller burbil population by a colleague. They want to know whether their burbil population has a similar network structure to ours.

```
#Set the mean group size
GS_B<-20

#Here we create a grid of locations for our observations
x_B<-seq(3,13,1)
y_B<-seq(3,9,1)
locs_B<-expand.grid(x_B,y_B)
names(locs_B)<-c("x","y")
```

```
#Here we assign coordinates to our groups. We create 9 groups in total.
group_locs_B<-locs_B[locs_B$x%%4==0&locs_B$y%%4==0,]

#Here we store the total number of groups
n_groups_B<-dim(group_locs_B)[1]

#Here we create three distinct clans of burbils. This will effect associations bet
ween members of different groups
group_clans_B<-sample(c("A","B","C"),n_groups_B,replace=TRUE)

#Set the probability of burbils from the same clan intermingling if they happen to
forage at the same location
p_wc_B<-1
#Set the probability of burbils from different clans intermingling if they happen
 to forage at the same location
p_bc_B<-0.4

#Create a list to store individual IDs
indss_B<-list()

#Create a list to store group sizes
gss_B<-list()

#Create a list to store the sex of each individual
sexes_B<-list()

#Create a list to store the age of each individual
ages_B<-list()

#Create a list to store the nose colour of each individual
noses_B<-list()

#Create a list to store information on which day a subgroup is observed on
daysl_B<-list()

#Create a list to store a group-by-individual matrix for each burbil group
gbis_B<-list()

#Set the mean number of subgroups observed for each group each day
sg_mn_B<-5

#Set the strength of assortativity based on nose colour
#Set a number between 0 and 1
sg_ass_B<-0.1

#Genereate association data within each burbil group!
for(j in 1:n_groups_B){

#individual identities
inds_B<-seq(1,rpois(1,GS_B),1)
indss_B[[j]]<-inds_B

#group size
gs_B<-length(inds_B)
gss_B[[j]]<-gs_B

#sex
```

```
sex_B<-sample(c("M","F"),gs_B,replace=TRUE)
sexes_B[[j]]<-sex_B

#age
age_B<-sample(c("AD","SUB","JUV"),gs_B,replace=TRUE,prob=c(0.6,0.2,0.2))
ages_B[[j]]<-age_B

#nose
nose_B<-sample(c("RED","ORANGE"),gs_B,replace=TRUE,prob=c(0.7,0.3))
noses_B[[j]]<-nose_B

#---------------------------------

#Define number of subgroups on the first day
n_sg_B<-rpois(1,sg_mn_B-1)+1

#find halfway point
max_red_B<-floor(n_sg_B/2)

#Sample subgroups on the first day
subgroups1_B<-sample(n_sg_B,sum(nose_B=="RED"),replace=TRUE,prob=c(rep(0.5+sg_ass_
B,max_red_B),rep(0.5-sg_ass_B,n_sg_B-max_red_B)))
subgroups2_B<-sample(n_sg_B,sum(nose_B=="ORANGE"),replace=TRUE,prob=c(rep(0.5-sg_a
ss_B,max_red_B),rep(0.5+sg_ass_B,n_sg_B-max_red_B)))

subgroups_B<-rep(NA,gs_B)
subgroups_B[nose_B=="RED"]<-subgroups1_B
subgroups_B[nose_B=="ORANGE"]<-subgroups2_B

#Store relevant information in the group-by-individual matrix and days vector
gbi_B<-matrix(0,nc=gs_B,nr=n_sg_B)
gbi_B[cbind(subgroups_B,seq(1,gs_B,1))]<-1
days_B<-rep(1,nrow(gbi_B))

#Repeat process over 100 days of observations
for(i in 2:100){

  n_sg_B<-rpois(1,sg_mn_B-1)+1

  #find halfway point
  max_red_B<-floor(n_sg_B/2)

  subgroups1_B<-sample(n_sg_B,sum(nose_B=="RED"),replace=TRUE,prob=c(rep(0.5+sg_as
s_B,max_red_B),rep(0.5-sg_ass_B,n_sg_B-max_red_B)))
  subgroups2_B<-sample(n_sg_B,sum(nose_B=="ORANGE"),replace=TRUE,prob=c(rep(0.5-sg
_ass_B,max_red_B),rep(0.5+sg_ass_B,n_sg_B-max_red_B)))

  subgroups_B<-rep(NA,gss_B[[j]])
  subgroups_B[nose_B=="RED"]<-subgroups1_B
  subgroups_B[nose_B=="ORANGE"]<-subgroups2_B

  tgbi_B<-matrix(0,nc=gs_B,nr=n_sg_B)
  tgbi_B[cbind(subgroups_B,seq(1,gs_B,1))]<-1
  days_B<-c(days_B,rep(i,nrow(tgbi_B)))
  gbi_B<-rbind(gbi_B,tgbi_B)
}
```

```
#We edit the group-by-individual matrix and days vector to delete any "empty" grou
ps
gbi2_B<-gbi_B[rowSums(gbi_B)>0,]
days_B<-days_B[rowSums(gbi_B)>0]
gbi_B<-gbi2_B

daysl_B[[j]]<-days_B
gbis_B[[j]]<-gbi_B

}

#We now go through and assign a location to every subgroup
sglocs_B<-list()
for(i in 1:n_groups_B){
  tx_B<-rep(NA,dim(gbis_B[[i]])[1])
  ty_B<-rep(NA,dim(gbis_B[[i]])[1])
  sglocs_B[[i]]<-data.frame(tx_B,ty_B)
  names(sglocs_B[[i]])<-c("x","y")
  sglocs_B[[i]]$x<-group_locs_B[i,1]+round(rnorm(dim(gbis_B[[i]])[1],0,2))
  sglocs_B[[i]]$y<-group_locs_B[i,2]+round(rnorm(dim(gbis_B[[i]])[1],0,2))
}

#Vector recording number of individuals in each group
n_inds_B<-numeric()
for(i in 1:n_groups_B){
  n_inds_B[i]<-dim(gbis_B[[i]])[2]
}

#Calculate total individuals in the population
n_tot_B<-sum(n_inds_B)

#Population-level individuals identities
inds_tot_B<-seq(1,n_tot_B,1)

#Information on each individual's group membership
g_tot_B<-rep(seq(1,n_groups_B,1),n_inds_B)

#Information on each individual's within-group identity
gi_tot_B<-seq(1,n_inds_B[1],1)
for(i in 2:n_groups_B){
  gi_tot_B<-c(gi_tot_B,seq(1,n_inds_B[i],1))
}

#We now calculate the full population association network
full_net_B<-matrix(0,nr=n_tot_B,nc=n_tot_B)

#Counts up between-group associations
for(i in 1:100){
  for(j in 1:(n_groups_B-1)){
    for(k in (j+1):n_groups_B){
      tA_B<-paste0(sglocs_B[[j]][,1],"-",sglocs_B[[j]][,2])
      tB_B<-paste0(sglocs_B[[k]][,1],"-",sglocs_B[[k]][,2])
      tA2_B<-tA_B[daysl_B[[j]]==i]
      tB2_B<-tB_B[daysl_B[[k]]==i]
      tt_B<-match(tA2_B,tB2_B)
      if(sum(is.na(tt_B))<length(tt_B)){
      if(group_clans_B[j]==group_clans_B[k]){same<-rbinom(1,1,p_wc_B)}
```

```
      if(group_clans_B[j]!=group_clans_B[k]){same<-rbinom(1,1,p_bc_B)}
      if(same==1){
        paste(i,j,k)
        for(m in length(tt_B)){
          if(is.na(tt_B[m])==FALSE){
            tsg1_B<-which(tA_B==tA2_B[m]&daysl_B[[j]]==i)
            tsg2_B<-which(tB_B==tB2_B[tt_B[m]]&daysl_B[[k]]==i)
            tid1_B<-which(gbis_B[[j]][tsg1_B,]==1)
            tid2_B<-which(gbis_B[[k]][tsg2_B,]==1)
            tid1a_B<-inds_tot_B[g_tot_B==j&gi_tot_B%in%tid1_B]
            tid2a_B<-inds_tot_B[g_tot_B==k&gi_tot_B%in%tid2_B]
            full_net_B[tid1a_B,tid2a_B]<-full_net_B[tid1a_B,tid2a_B]+1
            full_net_B[tid2a_B,tid1a_B]<-full_net_B[tid1a_B,tid2a_B]
          }
        }
      }
      }
    }
  }
}

#converts between group assocaitions to SRIs
for(i in 1:(nrow(full_net_B)-1)){
  for(j in (i+1):nrow(full_net_B)){
    full_net_B[i,j]<-full_net_B[i,j]/(200-full_net_B[i,j])
    full_net_B[j,i]<-full_net_B[i,j]
  }
}

#Adds within-group associations to the population network
for(i in 1:n_groups_B){
  full_net_B[inds_tot_B[g_tot_B==i],inds_tot_B[g_tot_B==i]]<-get_network2(gbis_B
[[i]])
}

#Plots the population social network
full_net2_B<-graph.adjacency(full_net_B,mode="undirected",weighted=TRUE)
plot(full_net2_B,vertex.color=unlist(noses_B),vertex.label=NA,vertex.size=4,edge.w
idth=(edge_attr(full_net2_B)$weight*8)^2,main="Population association network for
 collaborator")
```

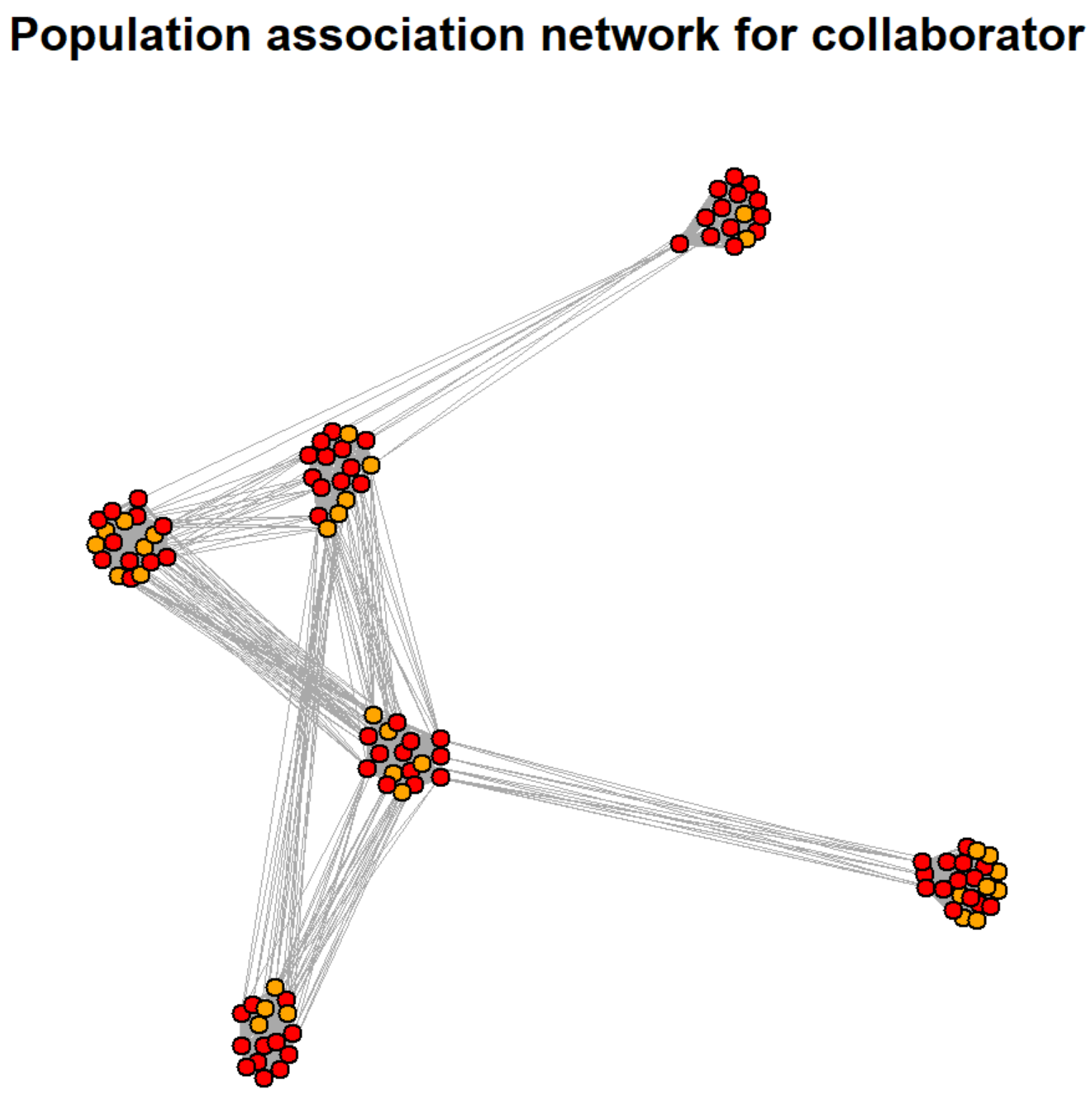

## Section 2.5 - Summary for Section 2

*Burbils then (coincidentally, of course) are a study system in which social network analysis offers a perfect tool to answer key questions about social behaviour and ecology*

# Section 3

## *Social network analysis examples*

## Section 3.1 - PERMUTATION-BASED REFERENCE MODELS

### Section 3.1.1 - Comparing two approaches to using reference models

Our first analyses are for the examples presented in Box 2, with two research groups asking questions about the associations of Burbils in group one.

The first group ask a specific research question

```
#First we extract the association network for group 1 from the group-by-individual
matrix using the asnipe package
MAT1<-get_network2(gbis[[1]])

#We can the plot the network
NET1<-graph.adjacency(MAT1,mode="undirected",weighted=TRUE)
plot(NET1,vertex.label=NA,vertex.color=noses[[1]],edge.width=(edge_attr(NET1)$weig
ht*8)^2)
```

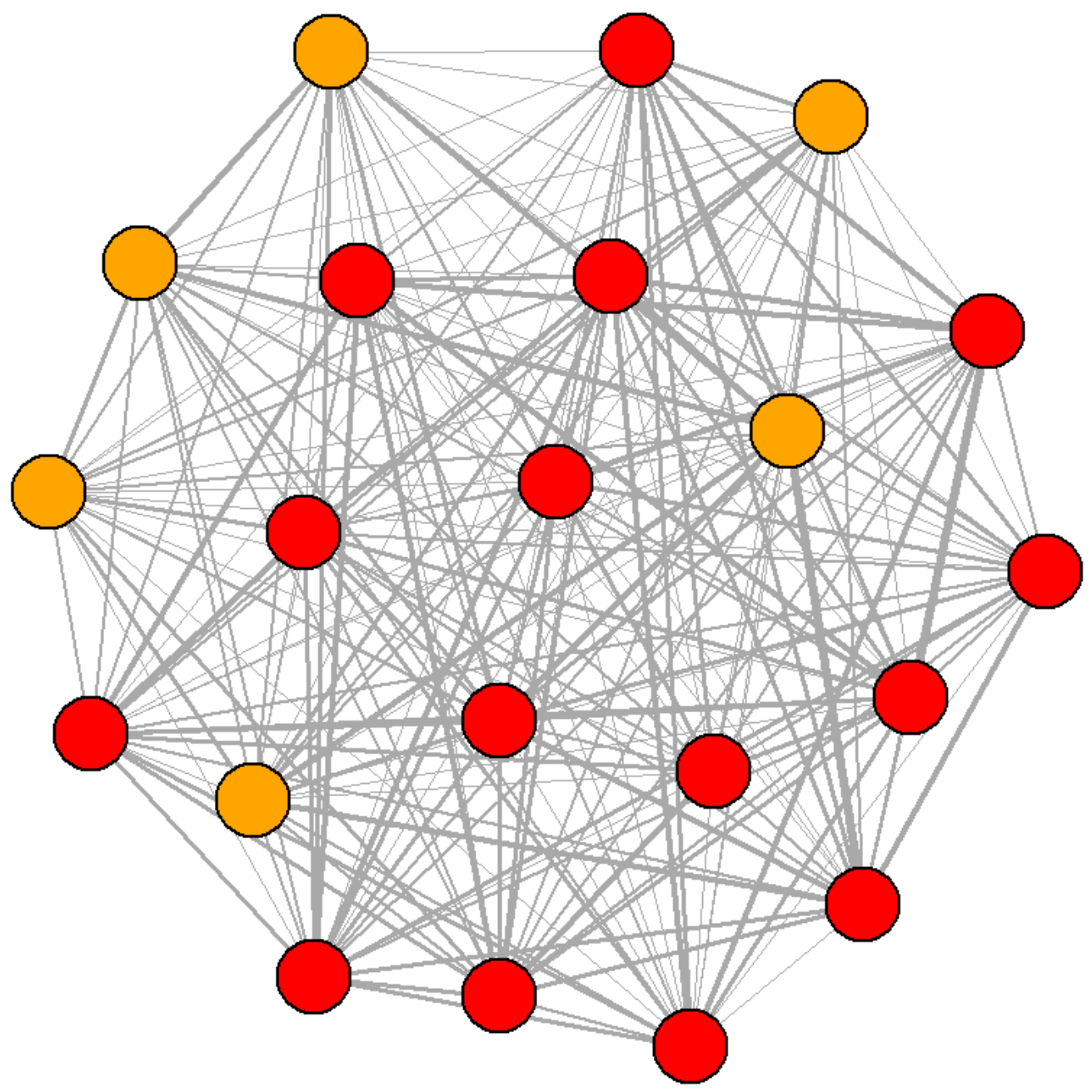

```
#We now calculate assortativity by nose colour in the real network
obs<-assortnet::assortment.discrete(MAT1, types=noses[[1]], weighted = TRUE, SE =
FALSE, M = 1)

#We then use node swap permutations to generate a reference distribution for assor
tativity
#We choose this type of permutation to break the correlation between nose colour a
nd network position
reference<-numeric()
MAT_T<-sna::rmperm(MAT1)
for(i in 1:9999){
reference[i]<-assortnet::assortment.discrete(MAT_T,types=noses[[1]], weighted = TR
UE, SE = FALSE, M = 1)$r
MAT_T<-sna::rmperm(MAT_T)
}
```

```
#We then add the observed assortativity to the reference distribution
reference2<-c(obs$r,reference)

#We can then calculate a p value by comparing the observed assortativity to the re
ference dataset. We are using a two-tailed test. Therefore, if we assume alpha=0.0
5 then assortativity is different to that expected by chance when p<0.025 (greater
than chance) or p>0.975 (less than chance)
sum(obs$r<reference2)/length(reference2)
```

```
## [1] 0
```

```
#Here we produce a plot to show this result. The grey histogram is the reference d
ataset, the blue dashed lines the 2.5% and 97.5% quantiles of the reference datase
t and the red line is the observed assortativity
par(xpd=FALSE)
hist(reference,las=1,xlim=c(-0.2,0.1),col="grey",border=NA,main="Reference Distrib
ution",xlab="Test statistic values",cex.lab=1.5,cex.axis=1)
lines(x=c(obs$r,obs$r),y=c(0,5000),col="red",lwd=4)
lines(x=rep(quantile(reference2,0.025),2),y=c(0,5000),col="darkblue",lwd=2,lty=2)
lines(x=rep(quantile(reference2,0.975),2),y=c(0,5000),col="darkblue",lwd=2,lty=2)
```

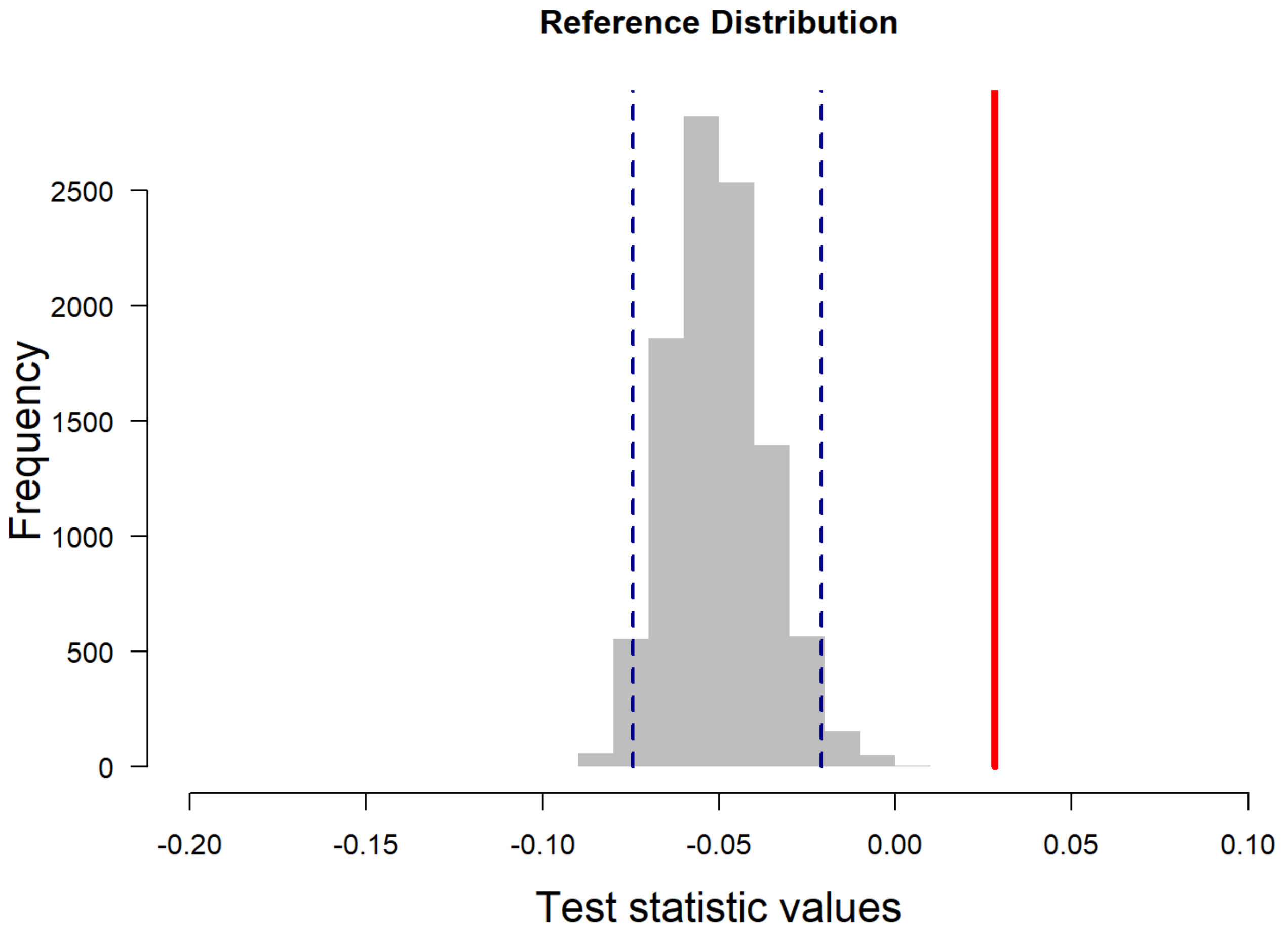

The second group ask a rather vague question and are less careful when designing and implementing their reference model

```r
#First we extract the association network for group 1 from the group-by-individual
matrix using the asnipe package
MAT1<-get_network2(gbis[[1]])

#Calculate the coefficient of variation in weighted degree (naively)
obs<-sd(rowSums(MAT1))/mean(rowSums(MAT1))

#Generate reference model using a random graph and edge weight distribution
reference<-numeric()
for(i in 1:9999){
  net_r<-igraph::erdos.renyi.game(n=nrow(MAT1),p.or.m=sum(sign(MAT1))/2,type="gnm"
)
net_r<-set_edge_attr(net_r,"weight",value=rnorm(n=sum(sign(MAT1))/2,mean=mean(MAT
1),sd=sd(MAT1)))
mat_r<-as_adjacency_matrix(net_r,type="both",attr="weight",sparse=FALSE)
diag(mat_r)<-0
  reference[i]<-sd(rowSums(mat_r))/mean(rowSums(MAT1))
}

#We then add the observed coefficient of variation to the reference distribution
reference2<-c(obs,reference)

#Calculate p value
sum(obs<reference2)/length(reference2)
```

```
## [1] 0.656
```

```r
#Plot randomisation result
par(xpd=FALSE)
hist(reference,las=1,xlim=c(0,0.2),col="grey",border=NA,main="Reference Distributi
on",xlab="Test statistic values",cex.lab=1.5,cex.axis=1)
lines(x=c(obs,obs),y=c(0,5000),col="red",lwd=4)
lines(x=rep(quantile(reference2,0.025),2),y=c(0,5000),col="darkblue",lwd=2,lty=2)
lines(x=rep(quantile(reference2,0.975),2),y=c(0,5000),col="darkblue",lwd=2,lty=2)
```

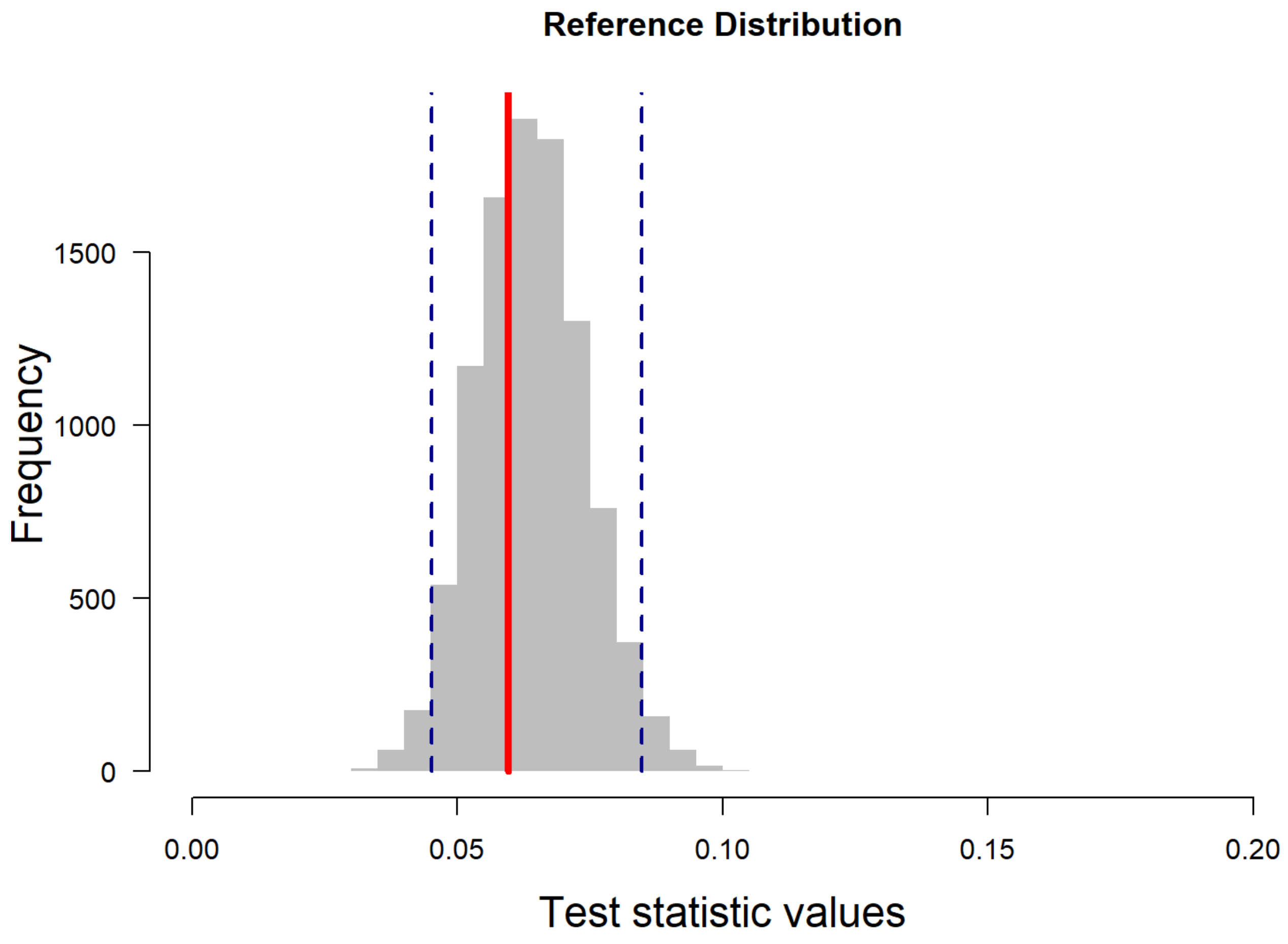

## Section 3.1.2 - Node feature swaps

Node swaps can be used to test a range of hypotheses and network types, for example they are also appropriate to test statistical significance of regression models when a network measure is the response variable.

Here we test the relationship between sex (female versus male) and weighted degree.

```
#First we plot the relationship
boxplot(strength(dom_net,mode="out")~sexes[[1]],xlab="Sex",ylab="Out-strength",ce
x.lab=1.5)
```

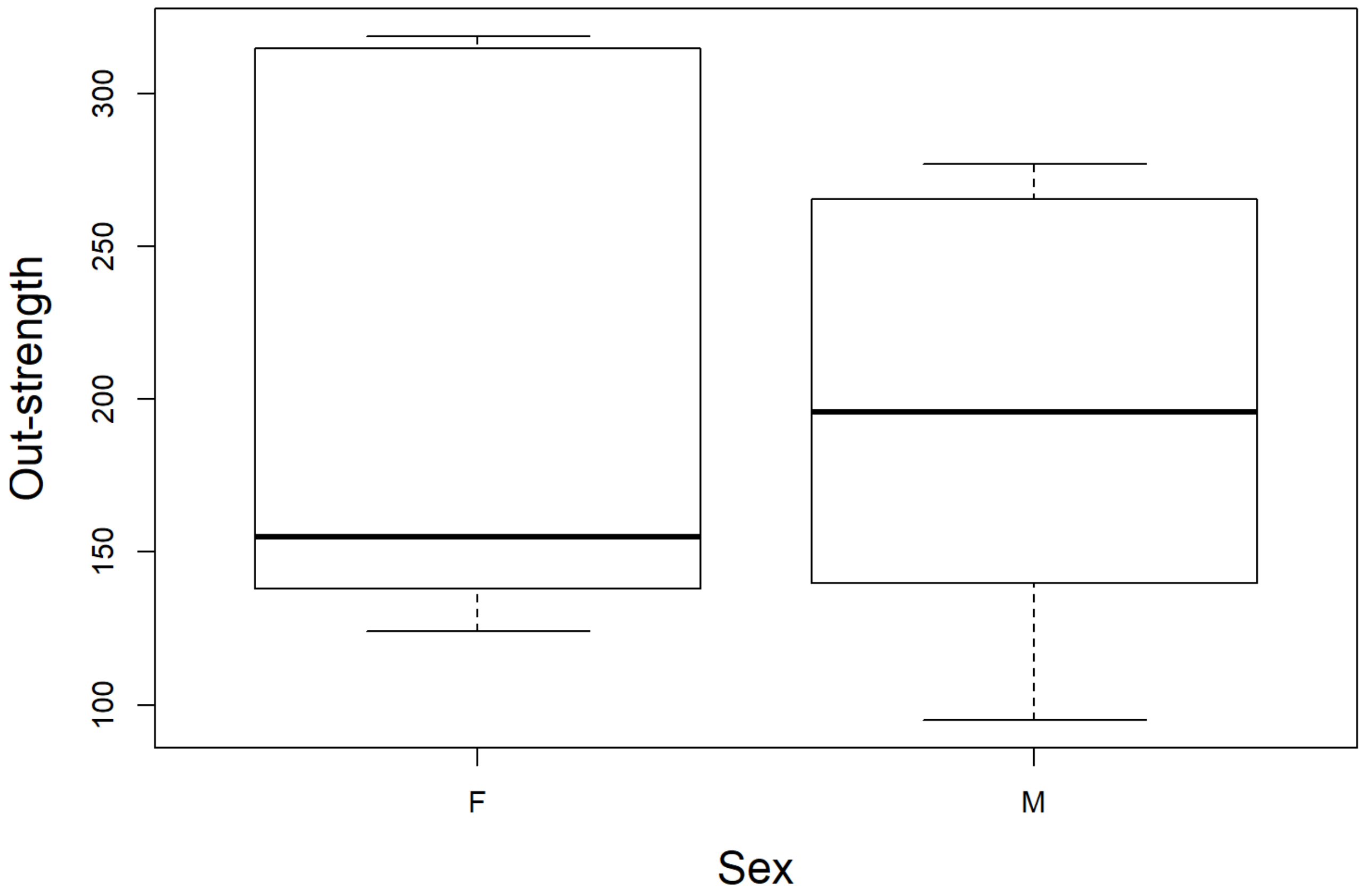

```
#We then convert our dominance network into an adjacency matrix form
MAT_DOM<-as_adjacency_matrix(dom_net,sparse=FALSE,attr="weight")

#We can the calculate the weighted out-degree for all individuals in the dominance
network
str_obs<-rowSums(MAT_DOM)

#We then choose our test statistic. We select the coefficient for effect of sex on
weighted degree estimated from a linear model. We calculate this for the observed
 network here
obs<-coef(lm(str_obs~sexes[[1]]))[2]

#We then use node swap permutations to generate a reference distribution for the n
ull relationship. Our node swaps (as above) using the rmperm function in the R pac
kage sna.
reference<-numeric()
MAT_T<-sna::rmperm(MAT_DOM)
for(i in 1:9999){
str_perm<-rowSums(MAT_T)
reference[i]<-coef(lm(str_perm~sexes[[1]]))[2]
MAT_T<-sna::rmperm(MAT_T)
}

#We then add the observed coefficient to the reference distribution
reference2<-c(obs,reference)

#We can then calculate a p value by comparing the observed linear relationship to
 those in the reference dataset. We are using a two-tailed test. Therefore, if we
```

```
assume alpha=0.05 then assortativity is different to that expected by chance when
p<0.025 (weighted out-degree of males higher than females) or p>0.975 (weighted ou
t-degree of males less than females)
sum(obs<reference2)/length(reference2)
```

```
## [1] 0.638
```

```
#Here we produce a plot to show this result. The grey histogram is the reference d
ataset, the blue dashed lines the 2.5% and 97.5% quantiles of the reference datase
t and the red line is the observed relationship
par(xpd=FALSE)
hist(reference,las=1,xlim=c(-200,200),col="grey",border=NA,main="Reference Distrib
ution",xlab="Test statistic values",cex.lab=1.5)
lines(x=c(obs,obs),y=c(0,5000),col="red",lwd=4)
lines(x=rep(quantile(reference2,0.025),2),y=c(0,5000),col="darkblue",lwd=2,lty=2)
lines(x=rep(quantile(reference2,0.975),2),y=c(0,5000),col="darkblue",lwd=2,lty=2)
```

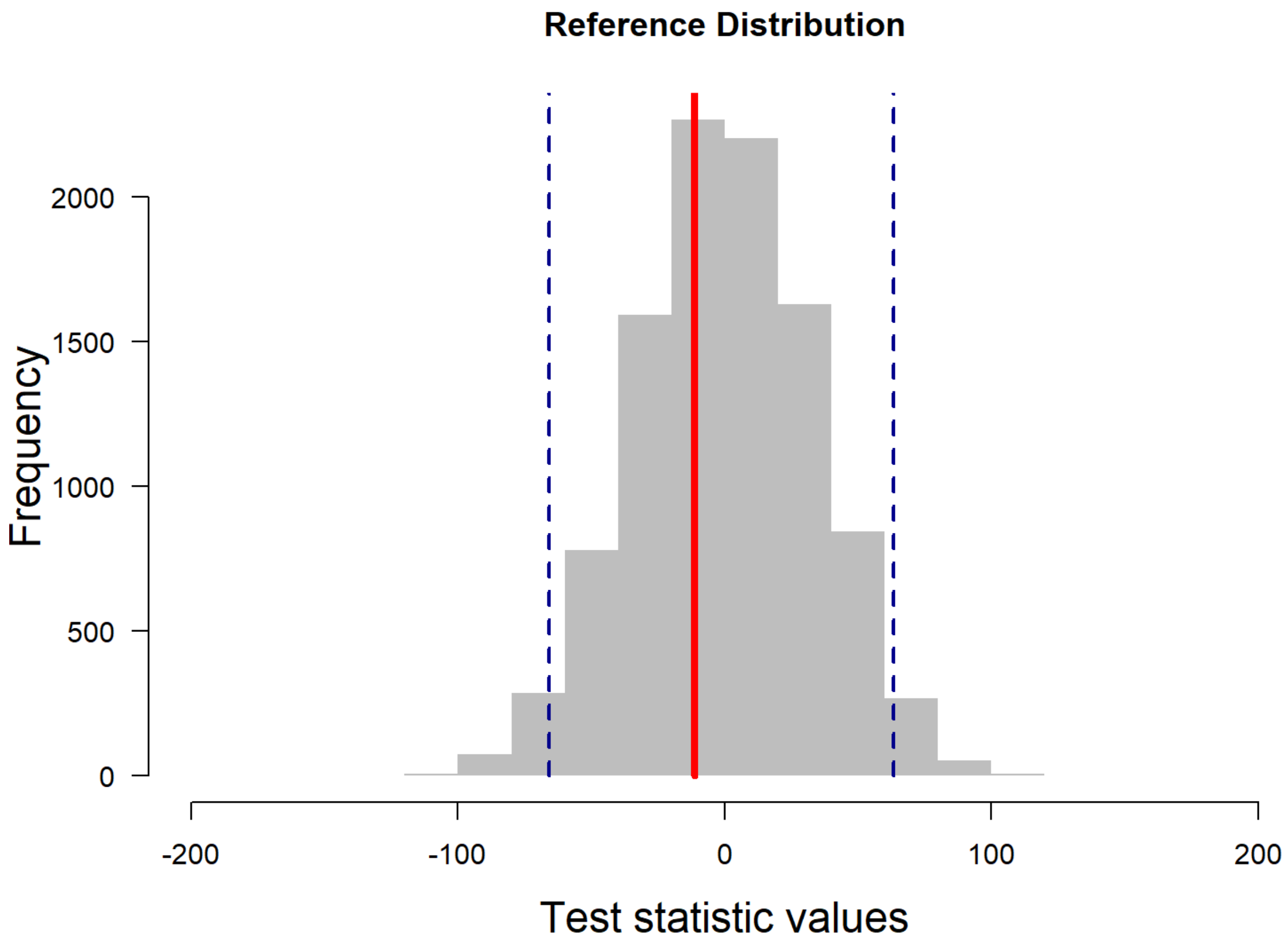

An important thing to bear in mind with node swaps, especially unconstrained node swaps as we have used above, is that they can't control for structure in the network. This could lead to potentially misleading conclusions being drawn when: a) there are biological processes operating at different scales that might be important in driving observed patterns; or b) there is variation in sampling intensity. For example, in the example above we know (because we simulated the data!) that there is an equal probability of males and females being observed. However, if one sex was more likely to be observed than the other, it would be expected to have a higher weighted degree, but this would be driven by sampling bias and not biology. It is hard to deal with this directly using node feature permutations, and they would need to be combined with other methodologies.

---

We provide an example of a) here. We know that the burbil association network is assorted by nose colour. Therefore, we want to know if the network of affiliative interactions is too.

```
#Convert affiliative network into an adjacency matrix
MAT_AFF<-as_adjacency_matrix(aff_net,sparse=FALSE,attr="weight")

#Calculate the observed assortativity of the affiliative network
obs<-assortnet::assortment.discrete(MAT_AFF, types=noses[[1]], weighted = TRUE, SE
= FALSE, M = 1)$r

#Generate the reference distribution
reference<-numeric()
MAT_T<-sna::rmperm(MAT_AFF)
for(i in 1:9999){
reference[i]<-assortnet::assortment.discrete(MAT_T,types=noses[[1]], weighted = TR
UE, SE = FALSE, M = 1)$r
MAT_T<-sna::rmperm(MAT_T)
}

#Add the observed assortativity to the reference dataset
reference2<-c(obs,reference)

#Calculate the p value. (p<0.025 would equate to the network being positively asso
rted by nose colour and p>0.975 to the network being negatively assorted by nose c
olour)
sum(obs<reference2)/length(reference2)
```

```
## [1] 0.0048
```

```
#We can then plot the result as we have done above
par(xpd=FALSE)
hist(reference,las=1,xlim=c(-0.2,0.2),col="grey",border=NA,main="Reference Distrib
ution",xlab="Test statistic values",cex.lab=1.5)
lines(x=c(obs,obs),y=c(0,5000),col="red",lwd=4)
lines(x=rep(quantile(reference2,0.025),2),y=c(0,5000),col="darkblue",lwd=2,lty=2)
lines(x=rep(quantile(reference2,0.975),2),y=c(0,5000),col="darkblue",lwd=2,lty=2)
```

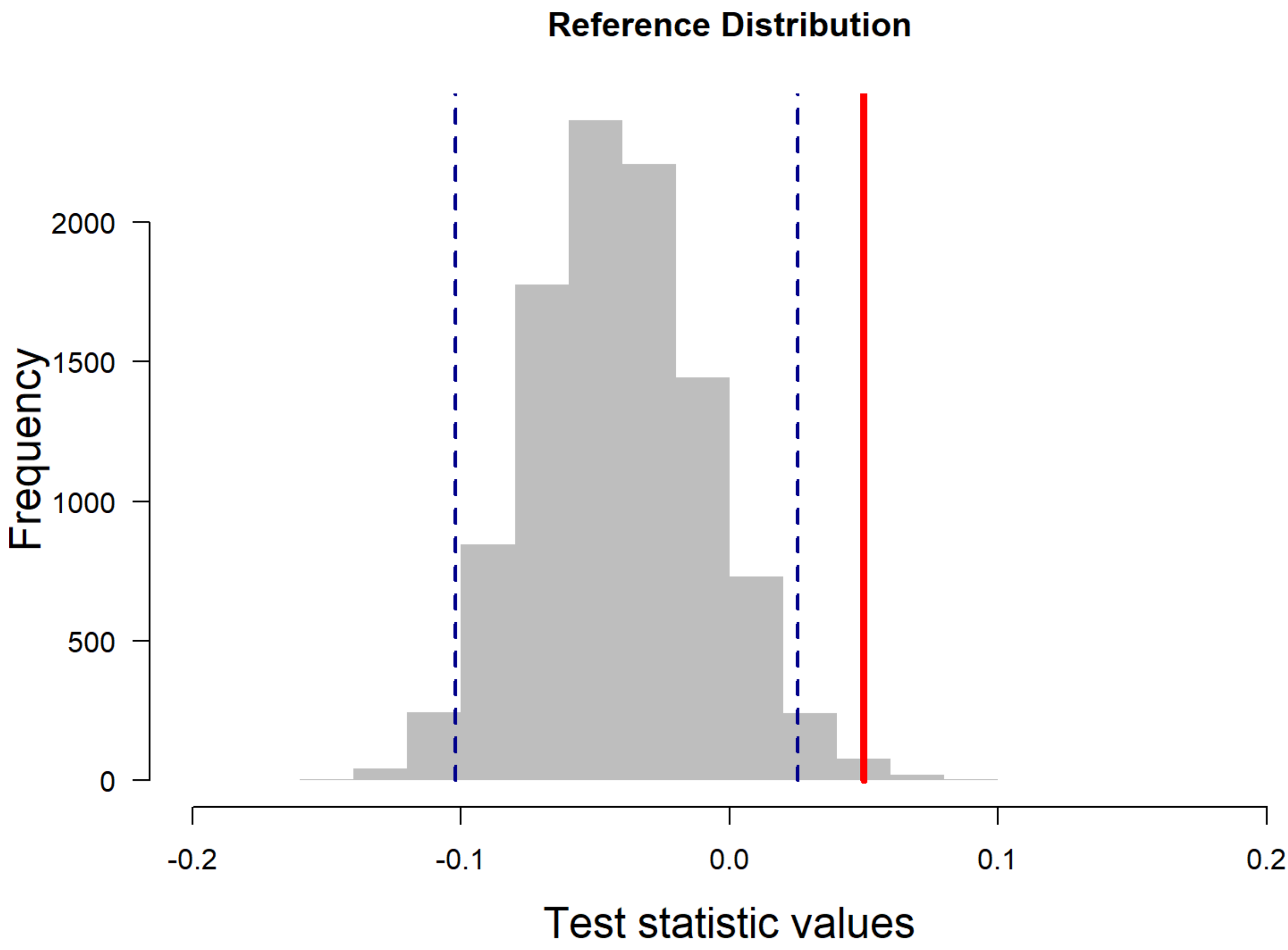

We find that the affiliative network is assorted by nose colour. However, we haven't controlled for association network structure in our reference model and we know that this places important constraints on the opportunities to interact. This is shows a key consideration when interpreting the results of simple reference models like this. We revisit this example later on.

---

## Section 3.1.3 - Edge feature swaps

As described in the main text, we don't just have to swap nodes. To test some hypotheses in directed networks, permuting the direction of edges can be a useful way to generate a reference distribution.

Here we provide an example of swapping edge directions. We test the hypothesis that adults tend to initiate more dominance interactions than younger individuals

*So far all of our permutations have randomised the whole network in one go. Now we move on to a type of permutation we make a single swap at a time (the direction of one edge) and these swaps occur successively causing the "permutedness" of the network to increase until it is a uniform sample of the reference distribution. We are generating what is known as a Markov Chain, and sampling from it*

```
#First we check the relationship
boxplot(strength(dom_net,mode="out")~ages[[1]],ylab="Out-strength",xlab="Age",cex.
lab=1.5)
```

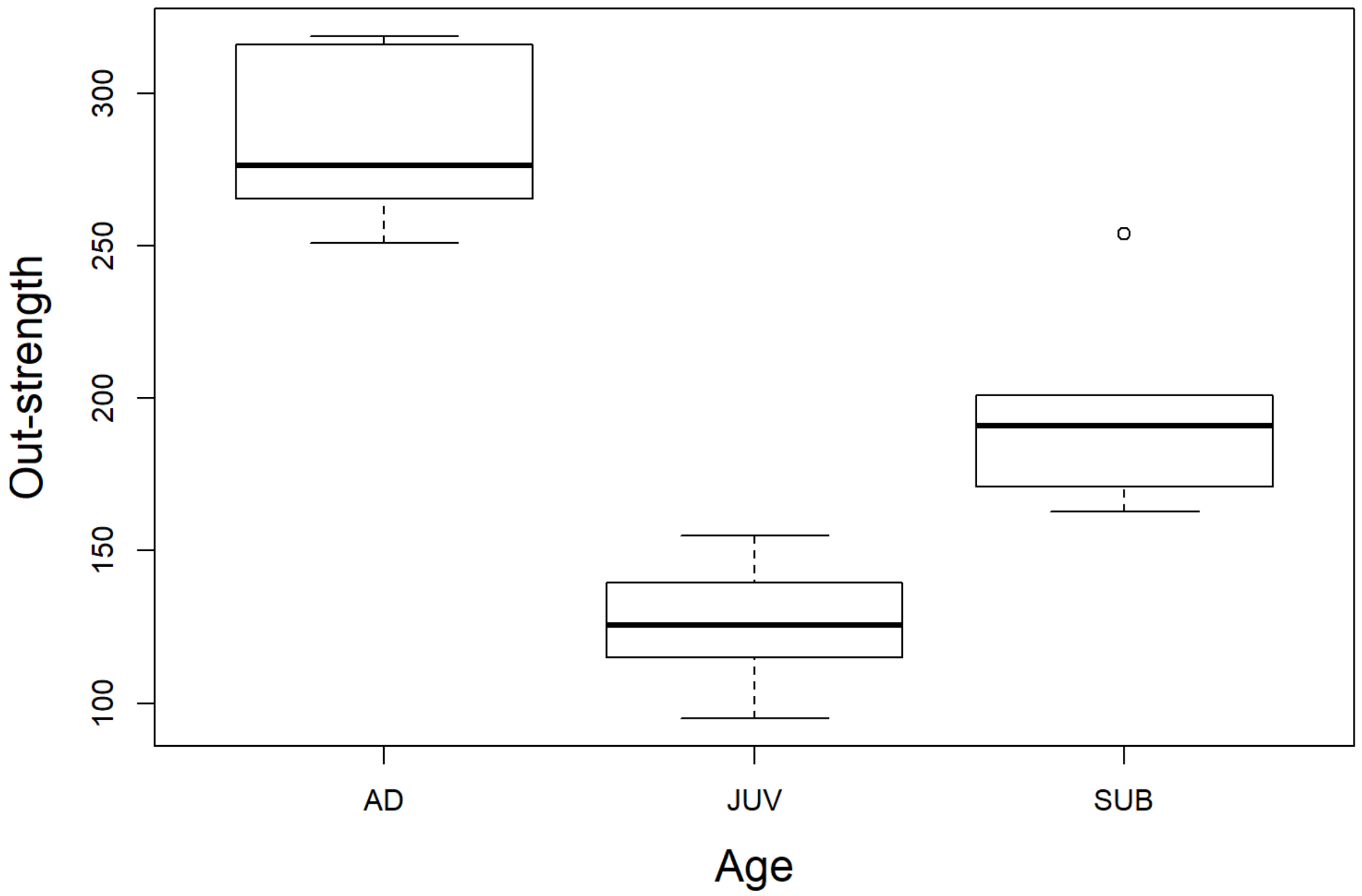

```
#Now we convert the network into an adjacency matrix
MAT_DOM<-as_adjacency_matrix(dom_net,sparse=FALSE,attr="weight")

#We now calculated the weighted out-degree (or out-strength) for all individuals i
n the dominance network
str_obs<-strength(dom_net,mode="out")

#Here we collapse the "age" variable into a simple version with just adults (A) an
d youngsters (Y)
ageT<-ages[[1]]
ageT[ageT=="AD"]<-"A"
ageT[ageT=="SUB"|ageT=="JUV"]<-"Y"

#As per the previous example we choose a test statistic which is the coefficient o
f the linear model between weighted degree (response variable) and our simplified
 measure of age (explanatory variable)
obs<-coef(lm(str_obs~ageT))[2]

#We now generate the reference distribution
burnin<-numeric()
reference<-numeric()
MAT_T<-MAT_DOM

#Unlike for the node swaps we first conduct a period known as the burn-in, during
 which time the "permutedness" of the data gradually increases towards the referen
ce distribution. We plot this to show it
for(i in 1:500){
  tid1<-sample(which(ageT=="A"),1)
```

```
  tid2<-sample(which(ageT=="Y"),1)
  MAT_T2<-MAT_T
  MAT_T2[tid1,tid2]<-MAT_T[tid2,tid1]
  MAT_T2[tid2,tid1]<-MAT_T[tid1,tid2]
  MAT_T<-MAT_T2
  dn_r<-graph_from_adjacency_matrix(MAT_T,weighted=TRUE,mode="directed")
  str_ref<-strength(dn_r,mode="out")
  burnin[i]<-coef(lm(str_ref~ageT))[2]
}

#When we plot how the value of the test statistic changes as the number of swaps increases, we can see that it moves quickly away from the observed value and then after ~100 swaps becomes relatively stable, or stationary.
plot(burnin,type="l",las=1,ylab="Test statistic value",xlab="Position in Markov Chain",cex.lab=1.5)
```

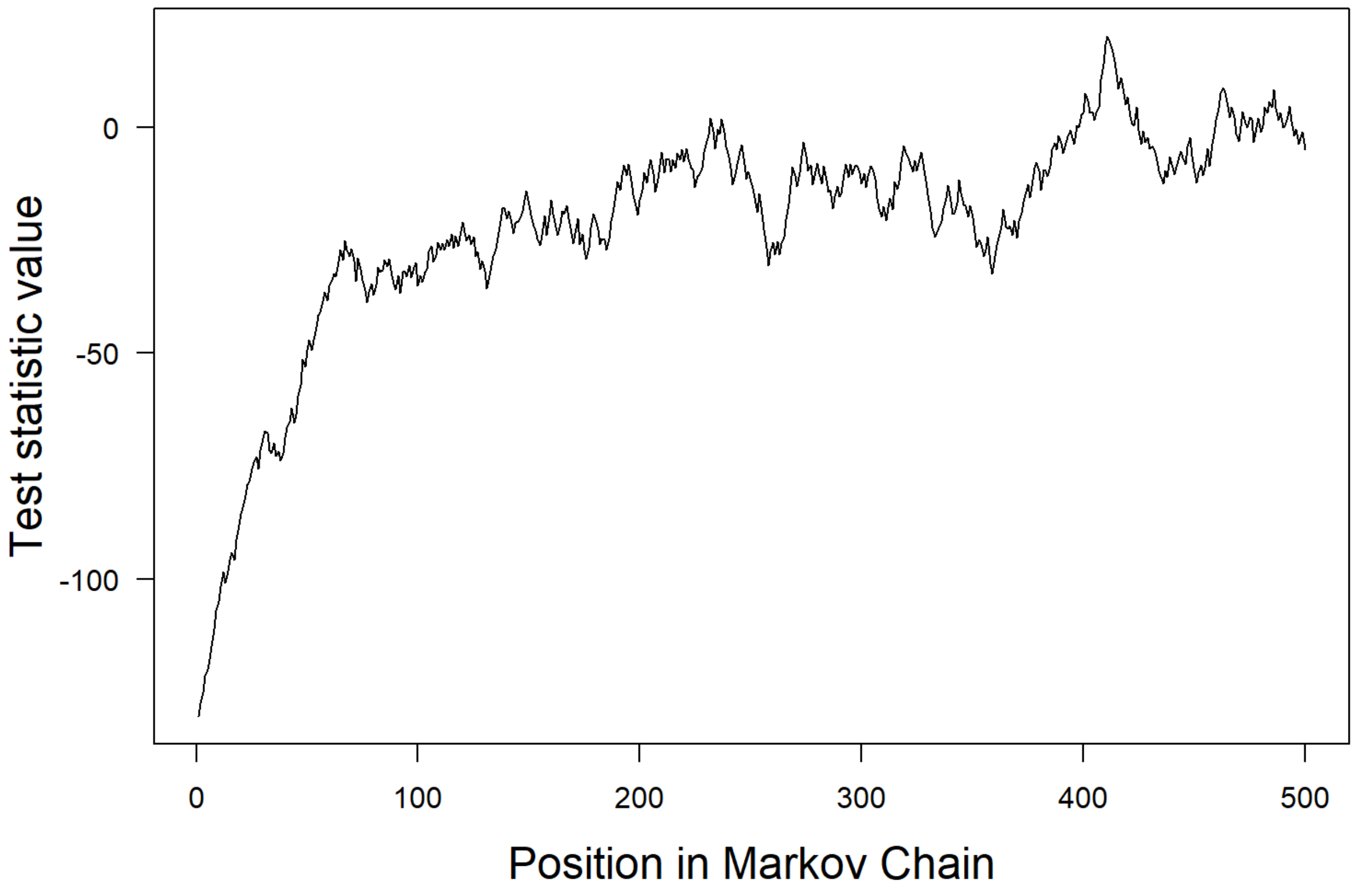

```
#We can then continue the Markov Chain and sample from it after each swap to generate our reference distribution of test statistics.
for(i in 1:999){
  tid1<-sample(which(ageT=="A"),1)
  tid2<-sample(which(ageT=="Y"),1)
  MAT_T2<-MAT_T
  MAT_T2[tid1,tid2]<-MAT_T[tid2,tid1]
  MAT_T2[tid2,tid1]<-MAT_T[tid1,tid2]
  MAT_T<-MAT_T2
  dn_r<-graph_from_adjacency_matrix(MAT_T,weighted=TRUE,mode="directed")
  str_ref<-strength(dn_r,mode="out")
```

```
  reference[i]<-coef(lm(str_ref~ageT))[2]
}

#We then add the observed value to the reference distribution
reference2<-c(obs,reference)

#And calculate the p value (p<0.025 would equate to the youngsters having higher o
ut-strength in the dominance network and p>0.975 to youngsters having lower out-st
rength)
sum(obs<reference2)/length(reference2)
```

```
## [1] 0.999
```

```
#We can then plot our results in the same way we have previously
par(xpd=FALSE)
hist(reference,las=1,xlim=c(-200,200),col="grey",border=NA,main="Reference Distrib
ution",xlab="Test statistic value",cex.lab=1.5)
lines(x=c(obs,obs),y=c(0,5000),col="red",lwd=4)
lines(x=rep(quantile(reference2,0.025),2),y=c(0,5000),col="darkblue",lwd=2,lty=2)
lines(x=rep(quantile(reference2,0.975),2),y=c(0,5000),col="darkblue",lwd=2,lty=2)
```

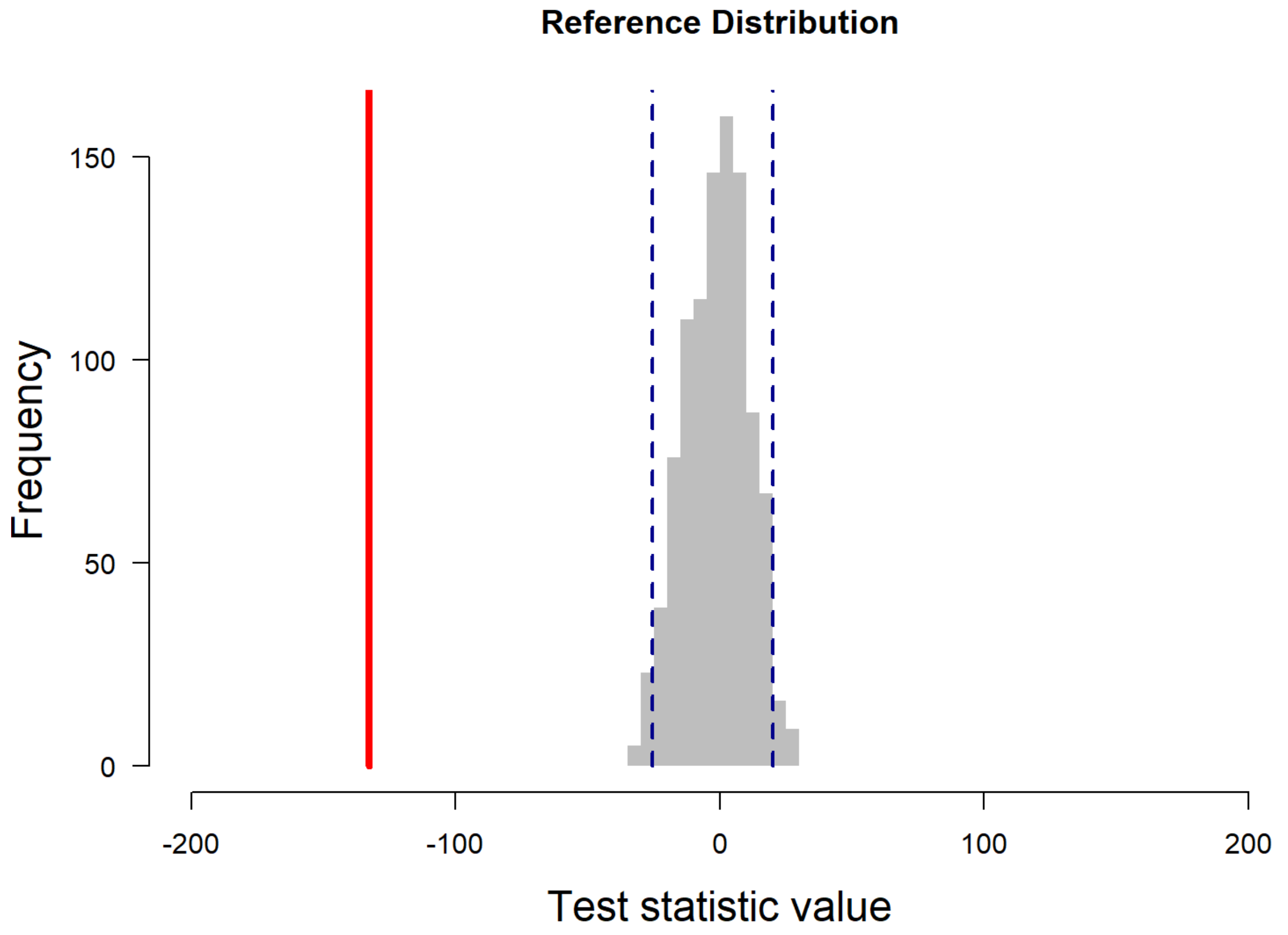

As well as swapping the directions of edges, you could also swap edge weights to address some research questions, especially for well-connected networks. We provide a quick demonstration of how you could use edge weight swaps using our burbil affiliative networks here.

We test the hypothesis that sexes differ in their weighted degree in burbil networks of affiliative interactions. While we could create a reference model in other ways, in this case we choose to permute edge weights to randomise them with respect to sex.

```
#Convert affiliative network into an adjacency matrix
MAT_AFF<-as_adjacency_matrix(aff_net,sparse=FALSE,attr="weight")

#We now calculated the weighted out-degree (or out-strength) for all individuals i
n the affiliative network
str_obs<-strength(aff_net,mode="out")

#Here we record the sexes of the individuals in the network
sexT<-sexes[[1]]

#Calculate the association between strength and sex in the observed affiliative ne
twork
obs<-coef(lm(str_obs~sexT))[2]

#First plot the relationship to look to see if any differences are apparent
boxplot(strength(aff_net,mode="out")~sexT)
```

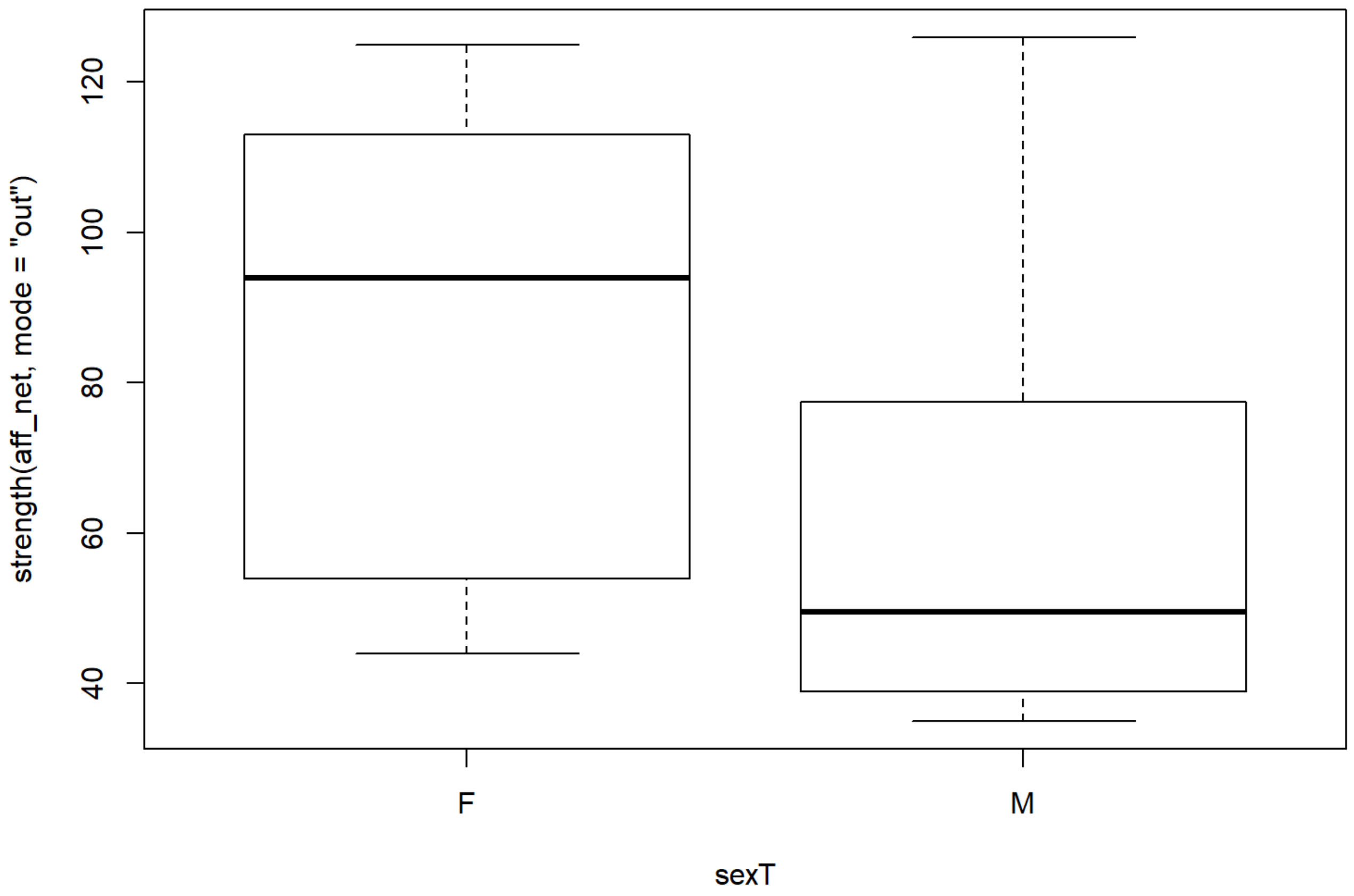

```
#The plot shows that, when we ignore other factors, males do initiate slightly mor
e affiliative interactions

#We now generate the reference distribution
burnin<-numeric()
reference<-numeric()
```

```
MAT_T<-MAT_AFF

#Unlike for the node swaps we first conduct a period known as the burn-in, during
 which time the "permutedness" of the data gradually increases towards the referen
ce distribution. We plot this to show it
for(i in 1:500){
  tid1<-sample(1:nrow(MAT_AFF),2,replace=F)
  tid2<-sample(1:nrow(MAT_AFF),2,replace=F)
  if(sum(tid1%in%tid2)!=2){
    MAT_T2<-MAT_T
    MAT_T2[tid1[1],tid1[2]]<-MAT_T[tid2[1],tid2[2]]
    MAT_T2[tid2[1],tid2[2]]<-MAT_T[tid1[1],tid1[2]]
    MAT_T<-MAT_T2
  }
  an_r<-graph_from_adjacency_matrix(MAT_T,weighted=TRUE,mode="directed")
  str_ref<-strength(an_r,mode="out")
  burnin[i]<-coef(lm(str_ref~sexT))[2]
}

#When we plot how the value of the test statistic changes as the number of swaps i
ncreases, we can see that it moves quickly away from the observed value and then a
fter ~100 swaps becomes relatively stable, or stationary.
plot(burnin,type="l",las=1,ylab="Test statistic value",xlab="Position in Markov Ch
ain",cex.lab=1.5)
```

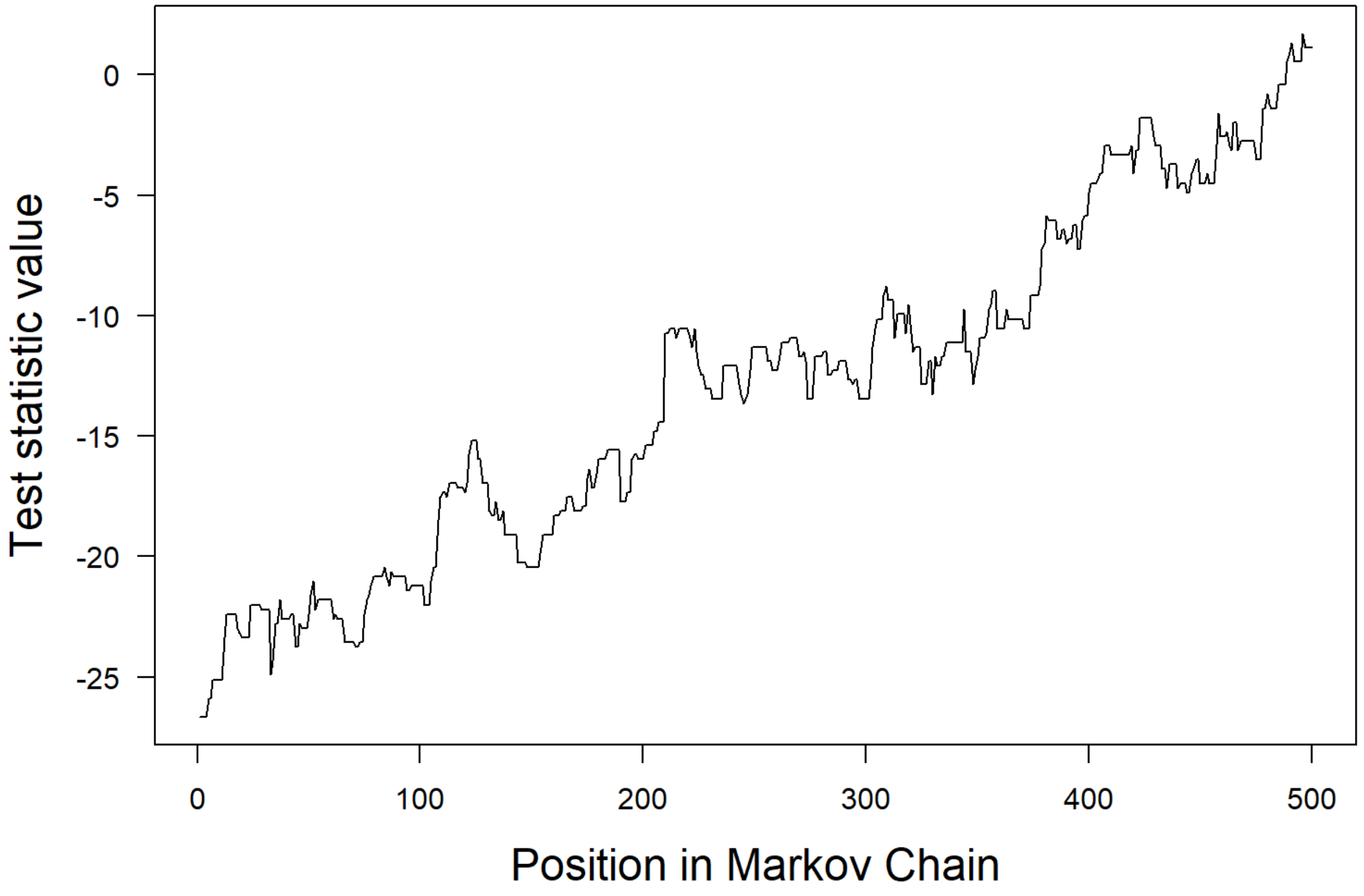

```
#We can then continue the Markov Chain and sample from it after each swap to gener
ate our reference distribution of test statistics.
```

```
for(i in 1:9999){
  tid1<-sample(1:nrow(MAT_AFF),2,replace=F)
  tid2<-sample(1:nrow(MAT_AFF),2,replace=F)
  if(sum(tid1%in%tid2)!=2){
    MAT_T2<-MAT_T
    MAT_T2[tid1[1],tid1[2]]<-MAT_T[tid2[1],tid2[2]]
    MAT_T2[tid2[1],tid2[2]]<-MAT_T[tid1[1],tid1[2]]
    MAT_T<-MAT_T2
  }
  an_r<-graph_from_adjacency_matrix(MAT_T,weighted=TRUE,mode="directed")
  str_ref<-strength(an_r,mode="out")
  reference[i]<-coef(lm(str_ref~sexT))[2]
}

#We then add the observed value to the reference distribution
reference2<-c(obs,reference)

#And calculate the p value (p<0.025 would equate to the youngsters having higher o
ut-strength in the dominance network and p>0.975 to youngsters having lower out-st
rength)
sum(obs<reference2)/length(reference2)
```

```
## [1] 0.9999
```

```
#We can then plot our results in the same way we have previously
par(xpd=FALSE)
hist(reference,las=1,xlim=c(-25,25),col="grey",border=NA,main="Reference Distribut
ion",xlab="Test statistic value",cex.lab=1.5)
lines(x=c(obs,obs),y=c(0,5000),col="red",lwd=4)
lines(x=rep(quantile(reference2,0.025),2),y=c(0,5000),col="darkblue",lwd=2,lty=2)
lines(x=rep(quantile(reference2,0.975),2),y=c(0,5000),col="darkblue",lwd=2,lty=2)
```

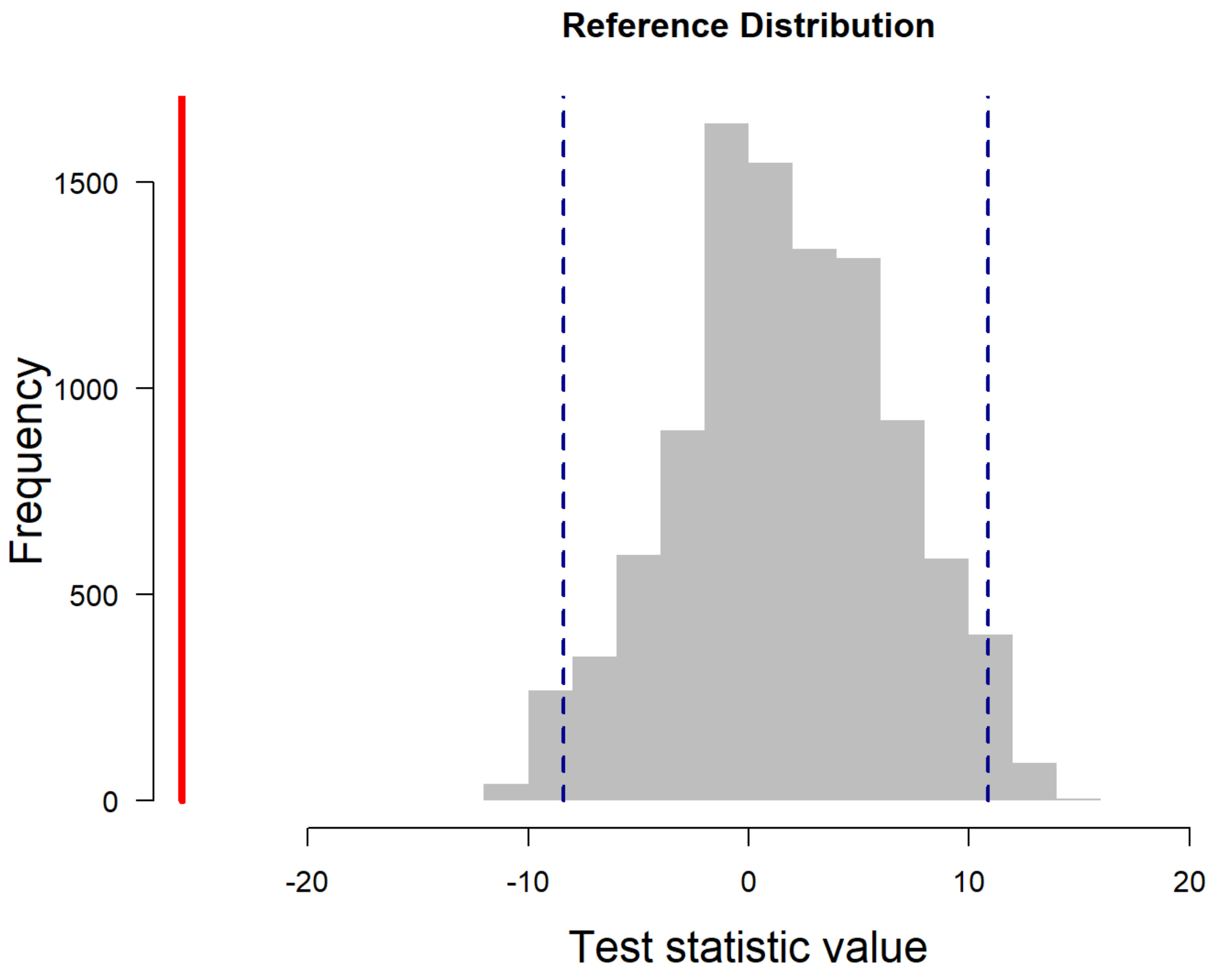

```
#We interpret this result as showing that females tend to have higher out-strength
than males.

#However, when we evaluate our analysis we consider the role of other variables, e
specially as our group of burbils is relatively small. So we inspect the data we h
ave on individual traits and realise that more juveniles in this group happen to b
e female

sum(sexes[[1]]=="F"&ages[[1]]=="JUV")
```

```
## [1] 5
```

```
sum(sexes[[1]]=="M"&ages[[1]]=="JUV")
```

```
## [1] 3
```

```
#A better approach would be to include age in our reference model (more tightly co
ntrol the permutations). This example also shows the challenge of using network an
alyses in small groups like this, even with plentiful data on the interactions the
mselves.
```

## Section 3.1.4 - Raw data swaps

We can also use permutations to make swaps in the raw data used to generate the network. It can be helpful to think of these as networks in themselves. The raw data used to construct animal social networks tends to come in two forms:

- Group-by-individual (GBI) matrices: these are effectively bipartite networks in which individuals are connected to particular grouping events. GBI matrices are used to generate association networks by collapsing this bipartite network using the assumption that individuals within each grouping event are connected
- Edge lists: these list the initiator and receiver of a set of behavioural interactions (or can also be used for contacts detected using proximity loggers). They are edge lists for a multigraph (i.e. a network with multiple rather than weighted edges) comprising the same set of individuals.

### Permutations such as this are often called **datastream permutations**

---

---

### Section 3.1.4.1 - Permutations applied to group-by-individual matrices

We provide an example of datastream permutations for a GBI here. We are going to test the hypothesis that our association network within a single group is different from random. We first compare it to a completely randomised network but then realise that this isn't especially interesting as we already know that the network is assorted by nose colour. Therefore, we test a second hypothesis that associations within a group are random once we have accounted for the assortativity by nose colour.

This helps us show how additional constraints can be added to these datastream permutation, and also highlights the value of constructing multiple reference models to have a good understanding of your data.

```
#Define some functions that we will use to calculate our test statistic
CoV<-function(a){
  a2<-a
  diag(a2)<-NA
  return(sd(a2,na.rm=T)/mean(a2,na.rm=T))
}
CoV2<-function(a){
  return(sd(a[a!=0])/mean(a[a!=0]))
}

#We choose the coefficient of variation in edge weight to be our test statistic. T
his is often used to test whether networks are different to the random expectation
within particular constraints
obs_cv<-CoV(get_network2(gbi))


#We first store the GBI and relevant information on which days groups were observe
d. We use the latter to constrain our permutations
gbi<-gbis[[1]]
day<-daysl[[1]]

## We then generate our reference distribution
#We constrain our swaps so that they have to occur between two groups occurring on
the same day.
#We have additional constraints that stop individuals being swapped into a group t
hey already occur in (this shouldn't matter here, but does happen if an individual
```

```
can be recorded multiple times within the time period that swaps are constrained t
o occur within)
#Note that if we try and swap an individual into a group that it already occurs in
and reject the swap then we keep the current version of the permuted GBI for the n
ext step of the Markov Chain rather than simply trying again. This is important to
ensure uniform sampling of the reference distribution.
gbi_t<-gbi
rgbis<-list()

#As above we have a burn-in period for the Markov Chain
for(i in 1:500){
  #sample an individual/grouping-event
  pind<-which(gbi_t>0,arr.ind=TRUE)
  tind1<-pind[sample(1:nrow(pind),1),]
  #record the day on which that individual/grouping-event occurred
  td<-which(day==day[tind1[1]])
  #sample a second individual/grouping-event that occurs on the same day
  pind2<-pind[which(pind[,1]%in%td),]
  tind2<-pind2[sample(1:nrow(pind2),1),]
  #If additional constraints are met then conduct swap
  if(tind1[1]!=tind2[1]&tind1[2]!=tind2[2]){
    if(gbi_t[tind1[1],tind2[2]]==0&gbi_t[tind2[1],tind1[2]]==0){
        gbi_t2<-gbi_t
        gbi_t2[tind2[1],tind1[2]]<-gbi_t[tind1[1],tind1[2]]
        gbi_t2[tind1[1],tind1[2]]<-gbi_t[tind2[1],tind1[2]]
        gbi_t2[tind1[1],tind2[2]]<-gbi_t[tind2[1],tind2[2]]
        gbi_t2[tind2[1],tind2[2]]<-gbi_t[tind1[1],tind2[2]]
        gbi_t<-gbi_t2
    }
  }
}

#We can then continue the Markov Chain and sample from it to generate our referenc
e distribution of test statistics. Here we conduct 10000 swaps but we only save ev
ery 10 iterations (known as a thinning interval) to avoid auto-correlation that ma
y occur because of rejected swaps
c<-1
for(i in 1:10000){
  pind<-which(gbi_t>0,arr.ind=TRUE)
  tind1<-pind[sample(1:nrow(pind),1),]
  td<-which(day==day[tind1[1]])
  pind2<-pind[which(pind[,1]%in%td),]
  tind2<-pind2[sample(1:nrow(pind2),1),]
  if(tind1[1]!=tind2[1]&tind1[2]!=tind2[2]){
    if(gbi_t[tind1[1],tind2[2]]==0&gbi_t[tind2[1],tind1[2]]==0){
        gbi_t2<-gbi_t
        gbi_t2[tind2[1],tind1[2]]<-gbi_t[tind1[1],tind1[2]]
        gbi_t2[tind1[1],tind1[2]]<-gbi_t[tind2[1],tind1[2]]
        gbi_t2[tind1[1],tind2[2]]<-gbi_t[tind2[1],tind2[2]]
        gbi_t2[tind2[1],tind2[2]]<-gbi_t[tind1[1],tind2[2]]
        gbi_t<-gbi_t2
    }
  }
  #This is where we save the swaps. Notice we only save every 10th swap
  if(i%%10==0){
    rgbis[[c]]<-gbi_t
    c<-c+1
```

```
  }
}

#Here we convert our permuted GBIs to networks
rnets<-lapply(rgbis,get_network2)

#We can then calculate our reference distribution
ref_cvs<-unlist(lapply(rnets,CoV))

#We now are going to compare our Markov Chain with the observed coefficient of var
iation
#Unsurprisingly, our observed coefficient of variation lies outside the reference
 distribution, but then we knew our networks were non-random already
par(xpd=FALSE)
plot(ref_cvs,type="l",ylim=c(0,0.4),las=1,ylab="Value of test statistic",cex.lab=
1.5)
lines(x=c(-100,100000),y=c(obs_cv,obs_cv),col="red",lwd=2)
```

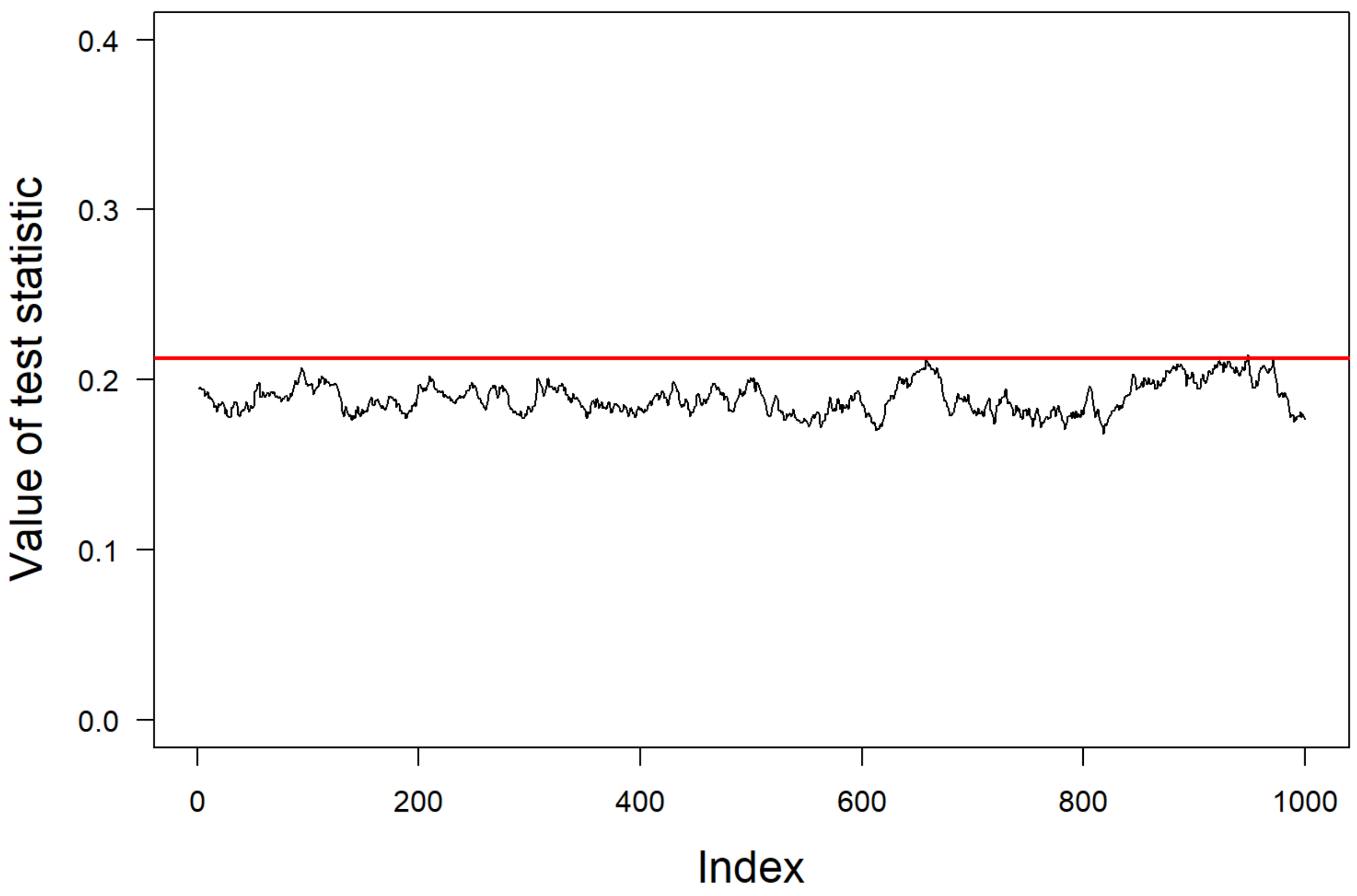

```
#Check p value from permutations
ref_cvs2<-c(obs_cv,ref_cvs)
sum(ref_cvs2<obs_cv)/length(ref_cvs2)
```

```
## [1] 0.998002
```

```
############################################################################
```

```
#Therefore we generate a new reference model where we additionally constrain swaps
to be between individuals with the same nose colour (given that we have already es
tablished this to be important)

gbi_t<-gbi
rgbis<-list()

#Burn-in period
for(i in 1:1000){
  pind<-which(gbi_t>0,arr.ind=TRUE)
  tind1<-pind[sample(1:nrow(pind),1),]
  td<-which(day==day[tind1[1]])
  #This is where we work out the nose colour of the individual sampled first
  tn1<-noses[[1]][tind1[2]]
  pind2<-pind[which(pind[,1]%in%td),]
  tind2<-pind2[sample(1:nrow(pind2),1),]
  #This is where we work out the nose colour of the individual sampled second
  tn2<-noses[[1]][tind2[2]]
  if(tind1[1]!=tind2[1]&tind1[2]!=tind2[2]){
    if(gbi_t[tind1[1],tind2[2]]==0&gbi_t[tind2[1],tind1[2]]==0){
      #We only conduct a swap if they have the same nose colour. If not then the c
urrent permuted GBI is resampled in       #the Markov Chain
      if(tn1==tn2){
        gbi_t2<-gbi_t
        gbi_t2[tind2[1],tind1[2]]<-gbi_t[tind1[1],tind1[2]]
        gbi_t2[tind1[1],tind1[2]]<-gbi_t[tind2[1],tind1[2]]
        gbi_t2[tind1[1],tind2[2]]<-gbi_t[tind2[1],tind2[2]]
        gbi_t2[tind2[1],tind2[2]]<-gbi_t[tind1[1],tind2[2]]
        gbi_t<-gbi_t2
      }
    }
  }
}

#Sampling period
#100000 swaps with every 100th swap saved
c<-1
for(i in 1:100000){
  pind<-which(gbi_t>0,arr.ind=TRUE)
  tind1<-pind[sample(1:nrow(pind),1),]
  td<-which(day==day[tind1[1]])
  tn1<-noses[[1]][tind1[2]]
  pind2<-pind[which(pind[,1]%in%td),]
  tind2<-pind2[sample(1:nrow(pind2),1),]
  tn2<-noses[[1]][tind2[2]]
  if(tind1[1]!=tind2[1]&tind1[2]!=tind2[2]){
    if(gbi_t[tind1[1],tind2[2]]==0&gbi_t[tind2[1],tind1[2]]==0){
      if(tn1==tn2){
        gbi_t2<-gbi_t
        gbi_t2[tind2[1],tind1[2]]<-gbi_t[tind1[1],tind1[2]]
        gbi_t2[tind1[1],tind1[2]]<-gbi_t[tind2[1],tind1[2]]
        gbi_t2[tind1[1],tind2[2]]<-gbi_t[tind2[1],tind2[2]]
        gbi_t2[tind2[1],tind2[2]]<-gbi_t[tind1[1],tind2[2]]
        gbi_t<-gbi_t2
      }
    }
  }
```

```
  if(i%%100==0){
    rgbis[[c]]<-gbi_t
    c<-c+1
  }
}

#Here we convert our permuted GBIs to networks
rnets<-lapply(rgbis,get_network2)

#We can then calculate our reference distribution
ref_cvs<-unlist(lapply(rnets,CoV))

#If we produce the same plot as before, we can see that now the observed coefficie
nt of variation lies within the reference distribution, suggesting that individual
s interact at random aside from assorting by nose colour (this is the result we ex
pect from how we simulated the data)
plot(ref_cvs,type="l",ylim=c(0,0.4),ylab="Value of test statistic",las=1,cex.lab=
1.5)
lines(x=c(-100,100000),y=c(obs_cv,obs_cv),col="red",lwd=2)
```

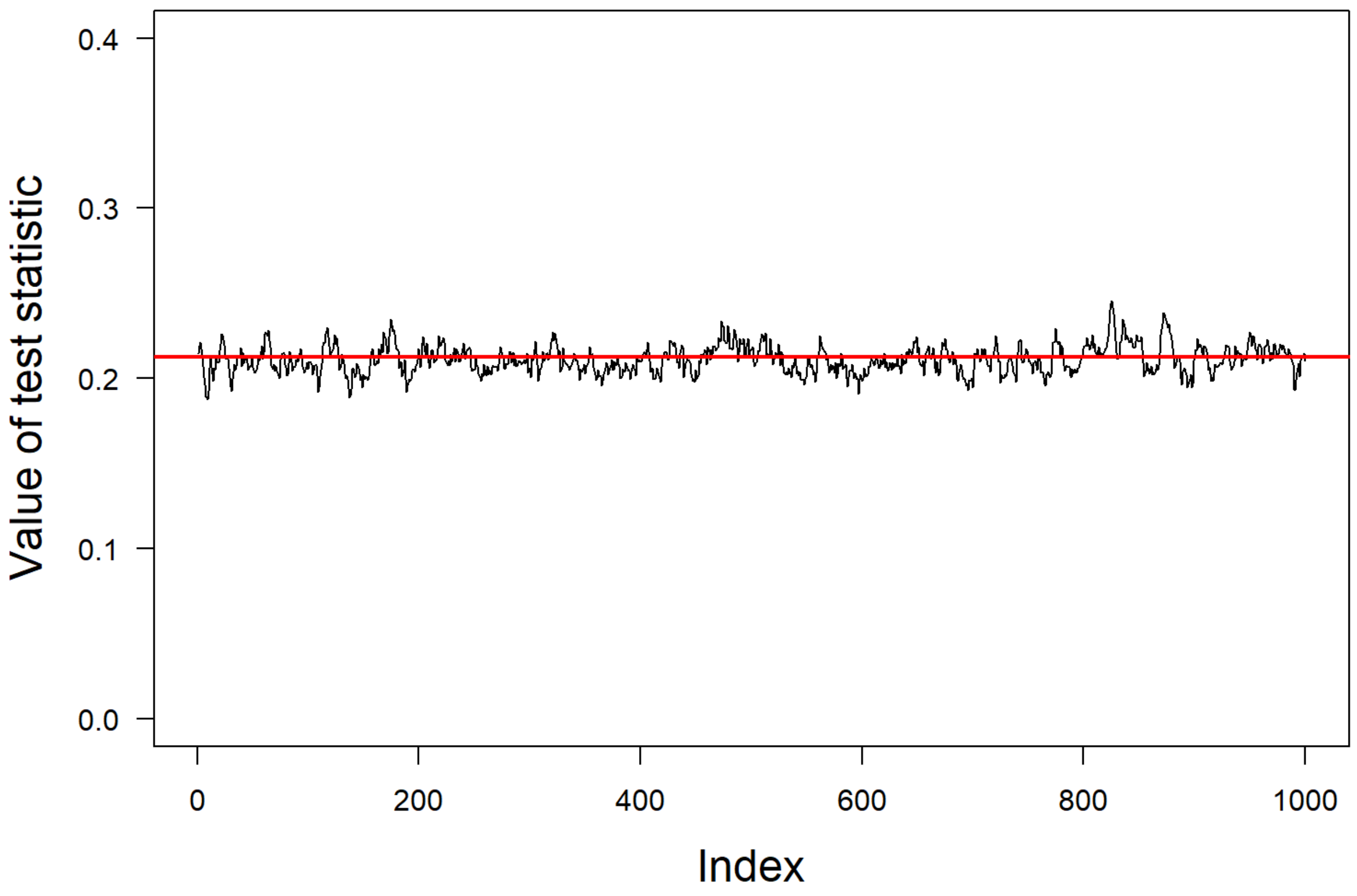

```
#Check p value from permutations
ref_cvs2<-c(obs_cv,ref_cvs)
sum(ref_cvs2<obs_cv)/length(ref_cvs2)
```

```
## [1] 0.5994006
```

```
#We could also produce histograms as we have previously
```

An important caveat to using datastream permutations of association data is that choosing the right constraints on swaps can be very important and have major effects on results. Without using very constraints then datastream permutations will swap edges at random and the reference distribution generated will be drawn from the configuration model. If it's unreasonable to expect relationships to be random within the constraints imposed then this can lead to errors in statistical inference.

Datastream permutations of are also not appropriate for testing the statistical significance of linear regressions (why we used node swaps) as they change the distribution of the response variable. This is discussed at length here: https://besjournals.onlinelibrary.wiley.com/doi/full/10.1111/2041-210X.13508

---

## Section 3.1.4.2 - Permutations applied to edge lists

We can also use datastream permutations in edge lists of interaction data

Before we were unable to test whether affiliative interactions were assorted by nose colour when you controlled for the structure of the association network. Using datastream permutations we can now test this hypothesis.

```
#Define function to convert permuted edge lists into adjacency matrices
matrix_gen<-function(a){
  Taff_net<-graph_from_edgelist(cbind(a,REC), directed = TRUE)
  E(Taff_net)$weight <- 1
  Taff_net<-simplify(Taff_net, edge.attr.comb=list(weight="sum"))
  b<-as_adjacency_matrix(Taff_net,sparse=FALSE,attr="weight")
  return(b)
}

#Define function to calculated assortativity in permuted networks
assortment2<-function(a){
  b<-assortnet::assortment.discrete(a, types=noses[[1]], weighted = TRUE, SE = FAL
SE, M = 1)
  return(b$r)
}

#We are going to swap the individual initiating affiliative interactions within su
b-groups to demonstrate that there are no more dominance interactions between indi
viduals of the same nose colour than expected by chance

#Calculate test statistic (assortativity) in observed network
obs<-assortnet::assortment.discrete(MAT_AFF, types=noses[[1]], weighted = TRUE, SE
= FALSE, M = 1)$r

#Generate reference distribution

T_W<-GIV
rGIV<-list()

#Burn-in period
for(i in 1:500){
  t1<-sample(1:length(T_W),1)
  tgr<-grA[t1]
  tw<-which(grA==tgr)
```

```
  t2<-sample(tw,1)
  T_W2<-T_W
  T_W2[t1]<-T_W[t2]
  T_W2[t2]<-T_W[t1]
  T_W<-T_W2
}

#Sampling period
c<-1
for(i in 1:100000){
  t1<-sample(1:length(T_W),1)
  tgr<-grA[t1]
  tw<-which(grA==tgr)
  t2<-sample(tw,1)
  T_W2<-T_W
  T_W2[t1]<-T_W[t2]
  T_W2[t2]<-T_W[t1]
  T_W<-T_W2
  if(i%%10==0){
    rGIV[[c]]<-T_W
    c<-c+1
  }
}

#convert permuted edgelists into adjacency matrices
r_affnets<-lapply(rGIV,matrix_gen)

#Calculate assortativity in permuted networks to generate reference distribution
refs<-unlist(lapply(r_affnets,assortment2))

#Add observed value to the reference distribution
refs2<-c(obs,refs)

#Calculate p value (0.025<p<0.975 indicates that the null hypothesis is accepted a
s expected)
sum(obs<refs)/length(refs2)
```

```
## [1] 0.1561844
```

```
#Plot the Markov Chain (could use histograms instead)
plot(refs,type="l",las=1,ylab="Value of test statistic",cex.lab=1.5)
lines(x=c(-100,100000),y=c(obs,obs),col="red",lwd=2)
```

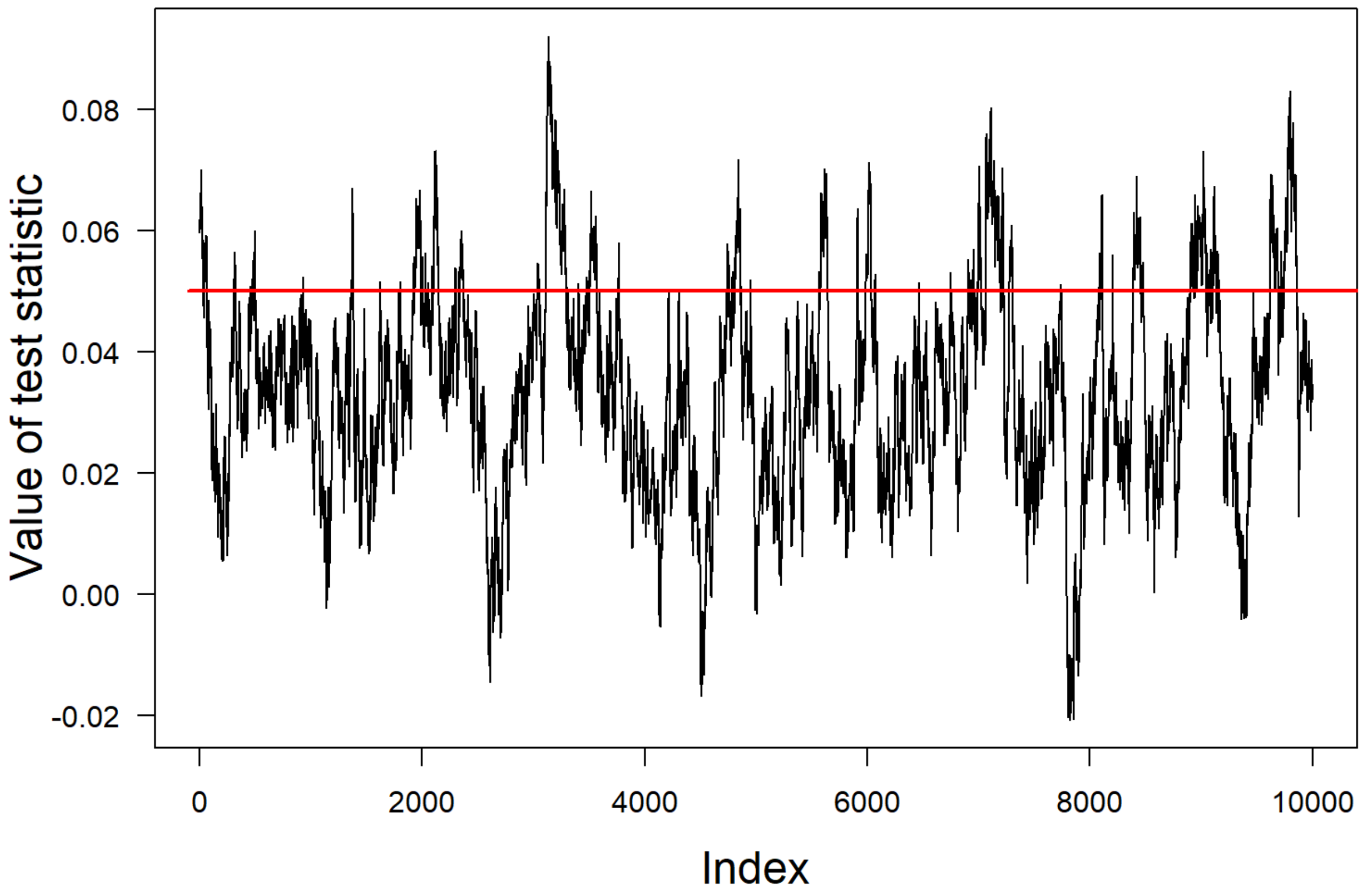

```
#As expected affiliative interactions aren't assorted by nose colour once you cont
rol for the structure of the association network
```

---

---

---

# Section 3.2 - RESAMPLING-BASED REFERENCE MODELS

Resampling-based reference models can be used to test a range of hypotheses, that may overlap or differ from the type of hypotheses tested using permutation-based reference models.

Resampling of either the network or the raw data can be used.

## Section 3.2.1 - Network resampling

First we demonstrate how resampling of the network might be used. We show how it might be applied to test the hypothesis that two networks of different sizes are generated by the same underlying process. However, we provide two examples to demonstrate how challenging this can be. We also briefly discuss the relationship between network size and network properties.

---

Here we test whether the underlying structure of the huddling network in our smallest and largest groups are the same. We do this for winter when huddling networks are random graphs and summer when huddling networks are small-world graphs

```
#Convert the winter huddling networks into adjacency matrices
hns2_m<-as_adjacency_matrix(hud_netSM_w,sparse=FALSE)
hnb2_m<-as_adjacency_matrix(hud_netBI_w,sparse=FALSE)

#Look at the degree distribution of the huddling network in the smallest (red) and
biggest (blue) groups
hist(colSums(hns2_m),col=adjustcolor("firebrick",0.2),border=NA,breaks=seq(0,20,1
),ylim=c(0,15),las=1,las=1,xlab="Betweenness",main="",cex.lab=1.5)
hist(colSums(hnb2_m),col=adjustcolor("dodgerblue",0.2),border=NA,breaks=seq(0,20,1
),add=TRUE)
```

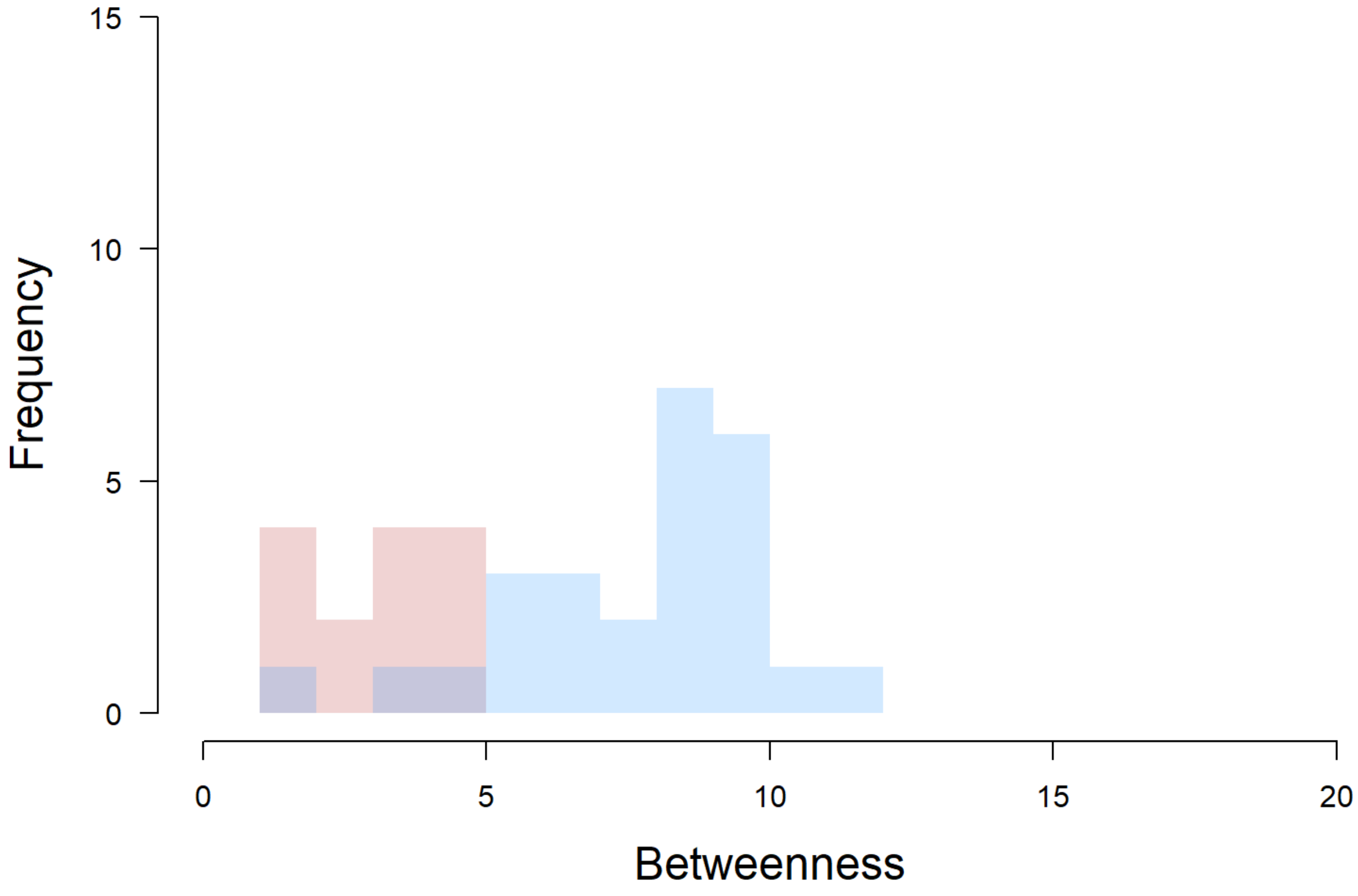

```
#We then decide that the mean degree of the network is our test statistic. We calc
ulate the mean degree for each network
smeanw<-mean(colSums(hns2_m))
bmeanw<-mean(colSums(hnb2_m))

#Print mean degree of each network
print(smeanw)
```

```
## [1] 3.571429
```

```
print(bmeanw)
```

```
## [1] 8.153846
```

```
#Note that the mean degree of the huddling network in the biggest group is much la
rger

#We then generate our reference distribution by sampling from our larger network t
o produce a network equivalent in size to the smaller network and then recalculati
ng the mean degree
bmeansw<-numeric()
for(i in 1:1000){
  samp<-sample(1:nrow(hnb2_m),nrow(hns2_m),replace=FALSE)
  tm<-hnb2_m[samp,samp]
  bmeansw[i]<-mean(colSums(tm))
}

#Calculate the p value. Assuming alpha=0.05 then p<0.025 indicates the small netwo
rk has a larger mean degree than the sampled larger network and p>0.975 indicates
 it has a smaller degree
sum(smeanw<bmeansw)/length(bmeansw)
```

```
## [1] 0.878
```

```
#0.025<p<0.975 indicating the networks have similar mean degree, which is unsurpri
sing given they are generated by the same process

#An alternative comparison might instead to be to correct the degree measure by th
e number of individuals minus one (proportion of group connected with)
smeanC<-mean(colSums(hns2_m))/(nrow(hns2_m)-1)
bmeanC<-mean(colSums(hnb2_m))/(nrow(hnb2_m)-1)

#These values are now much more similar to each other. In both groups are connecte
d to about 30% of others
print(smeanC)
```

```
## [1] 0.2747253
```

```
print(bmeanC)
```

```
## [1] 0.3261538
```

```
##However, the success of using a process like this is dependent on the structure
 of the network (see main text) and so would need to be done with great care.

#When we do the same with the summer huddling networks which have small-world prop
erties then the degree distribution of the two networks is very similar even thoug
h they are different sizes and when we do the same resampling procedure it indicat
es a difference between the small and large networks which we know are generated b
y the same process.
hns_m<-as_adjacency_matrix(hud_netSM,sparse=FALSE)
hnb_m<-as_adjacency_matrix(hud_netBI,sparse=FALSE)

#Plot histogram comparing the degree distribution in the large and small groups
hist(colSums(hns_m),col=adjustcolor("firebrick",0.2),border=NA,breaks=seq(0,10,1),
ylim=c(0,15),las=1,xlab="Betweenness",main="",cex.lab=1.5)
```

```
hist(colSums(hnb_m),col=adjustcolor("dodgerblue",0.2),border=NA,breaks=seq(0,10,1
),add=TRUE)
```

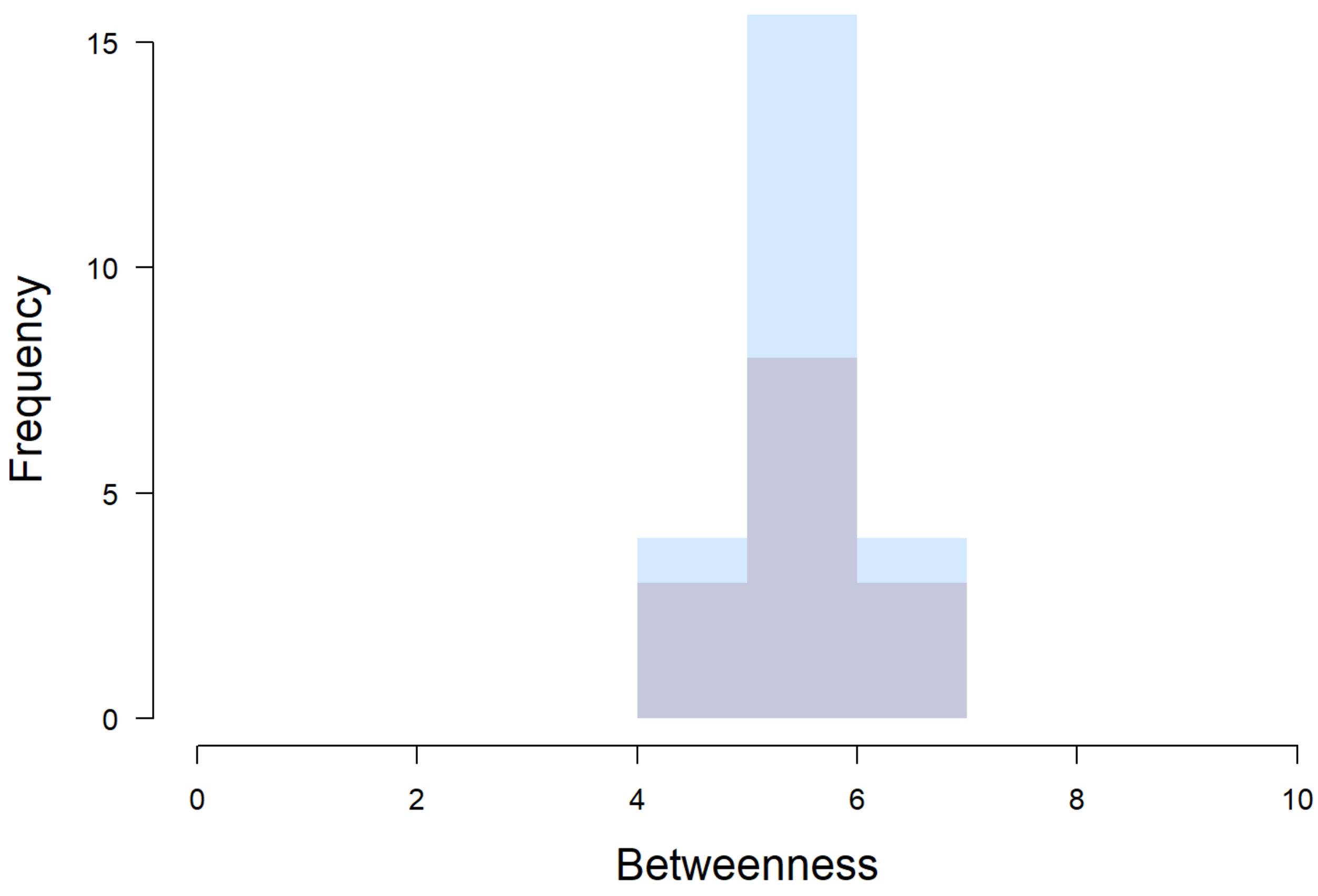

```
#Calculate mean degree of each network as a test statistic
smean<-mean(colSums(hns_m))
bmean<-mean(colSums(hnb_m))

#As above we then generate our reference distribution by sampling from our larger
 network to produce a network equivalent in size to the smaller network and then r
ecalculating the mean degree
bmeans<-numeric()
for(i in 1:1000){
  samp<-sample(1:nrow(hnb_m),nrow(hns_m),replace=FALSE)
  tm<-hnb_m[samp,samp]
  bmeans[i]<-mean(colSums(tm))
}

#Again we could correct degree measure by the number of individuals minus one (pro
portion of group connected with)
smeanC2<-mean(colSums(hns_m))/(nrow(hns_m)-1)
bmeanC2<-mean(colSums(hnb_m))/(nrow(hnb_m)-1)

#This time the corrected values are very different from each other, similar to the
result detected by the resampling approach
print(smeanC2)
```

```
## [1] 0.4615385
```

```
print(bmeanC2)
```

```
## [1] 0.24
```

```
#If we looked at the uncorrected degrees then they are the same
print(mean(colSums(hns_m)))
```

```
## [1] 6
```

```
print(mean(colSums(hns_m)))
```

```
## [1] 6
```

```
#This is an example of where evaluation is important. If we think about the proces
ses structuring the two networks then we can see that the resampling approach and
 the test statistic used are not appropriate in this particular case. This is then
made clearer still when we use corrections for network size instead of resampling.
```

Comparing networks of different sizes is challenging (see main text of the paper), but resampling can be useful in other contexts too. Below we provide an example of resampling the raw data used to construct the social networks.

In the main text of the paper we provide an example of betweenness centrality calculated from our main study population and from study population B (which has similar social structure but is smaller). We also illustrate that briefly here.

Population A has higher “raw” betweenness values as the population is larger and therefore there are more potential paths between individuals. However, if we normalise the betweenness calculated by the number of potential dyads (pairs of nodes) in the network we see that (while more similar) the betweenness estimated in Population B is higher because there are fewer options of shortest paths between pairs of nodes in the smaller network

```
#Calculate betweenness of the two population association networks
betA<-igraph::betweenness(full_net2,weights = 1/edge_attr(full_net2)$weight)
betB<-igraph::betweenness(full_net2_B,weights = 1/edge_attr(full_net2_B)$weight)

#Betweenness values are much higher (both mean and max) in the larger network
summary(betA)
```

```
##    Min. 1st Qu.  Median    Mean 3rd Qu.    Max.
##     0.0     0.0    42.0   511.9   371.0  7693.0
```

```
summary(betB)
```

```
##    Min. 1st Qu.  Median    Mean 3rd Qu.    Max.
##     0.0     0.0     0.0   106.7    59.5  1562.0
```

```
#However, when we normalise betweenness centrality values by the number of node pa
irs this difference disappears
betA2<-betA/(nrow(full_net)^2-nrow(full_net))
betB2<-betB/(nrow(full_net_B)^2-nrow(full_net_B))

#The distribution of betweenness values is now much more similar but normalised/co
rrected betweenness estimated in Population B is now higher as explained above
summary(betA2)
```

```
##      Min.   1st Qu.    Median      Mean   3rd Qu.      Max.
## 0.0000000 0.0000000 0.0003822 0.0046578 0.0033762 0.0700051
```

```
summary(betB2)
```

```
##     Min.  1st Qu.   Median     Mean  3rd Qu.     Max.
## 0.000000 0.000000 0.000000 0.009774 0.005449 0.143040
```

```
#Finding the correct approach to normalising measures is central to making appropr
iate comparisons between networks but also very challenging.
```

---

Another form of resampling-based reference model is to sample from a metric distribution, most commonly the degree distribution. For example, using igraph we can calculate the degree sequence of our full burbil association network and generate otherwise randomised graphs with the same degree sequence.

We can also resample edge weights from the observed network in a similar way.

```
#Calculate degree of the full population association network
deg<-igraph::degree(full_net2)

#Generate a reference network by resampling degree distribution
rdsn<-igraph::sample_degseq(deg,method="simple.no.multiple")

#Plot (randomised) reference network
plot(rdsn,vertex.label=NA,vertex.size=5)
```

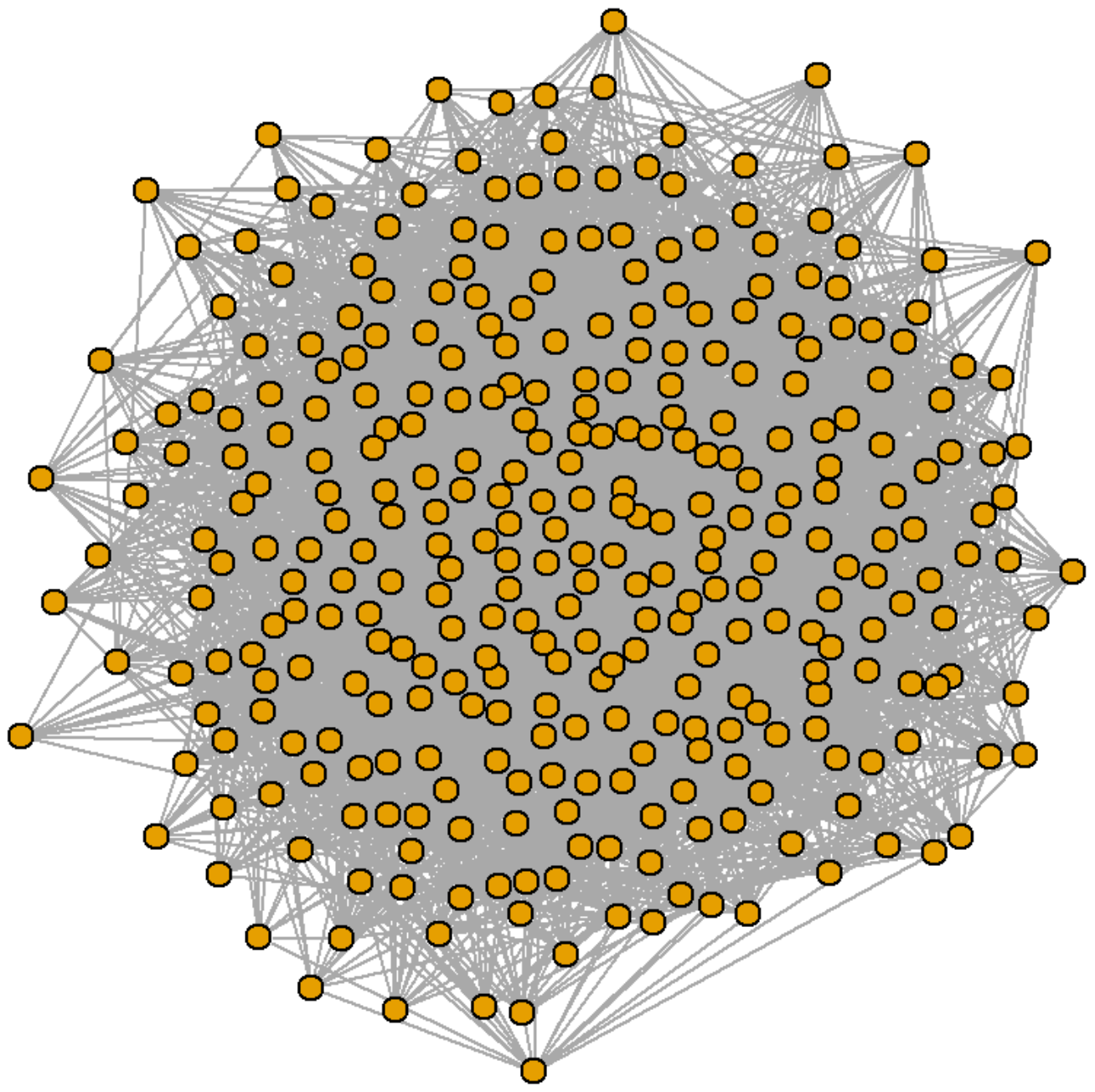

```
#Note that this new network no longer has the discernible community structure of t
he observed association network

#We could then use resampling to reassign edge weights to this same graph (with or
without replacement). For example

#Resample edge weights with replacement
edge_attr(rdsn)$weight<-sample(edge_attr(full_net2)$weight,gsize(rdsn),replace=TRU
E)

#Plot rewired, weighted network
#You can see from these plots that we lose the strong grouping/community structure
of our original network without additional rules when we resample.
plot(rdsn,vertex.label=NA,vertex.size=5,edge.width=(edge_attr(rdsn)$weight*8)^2.5)
```

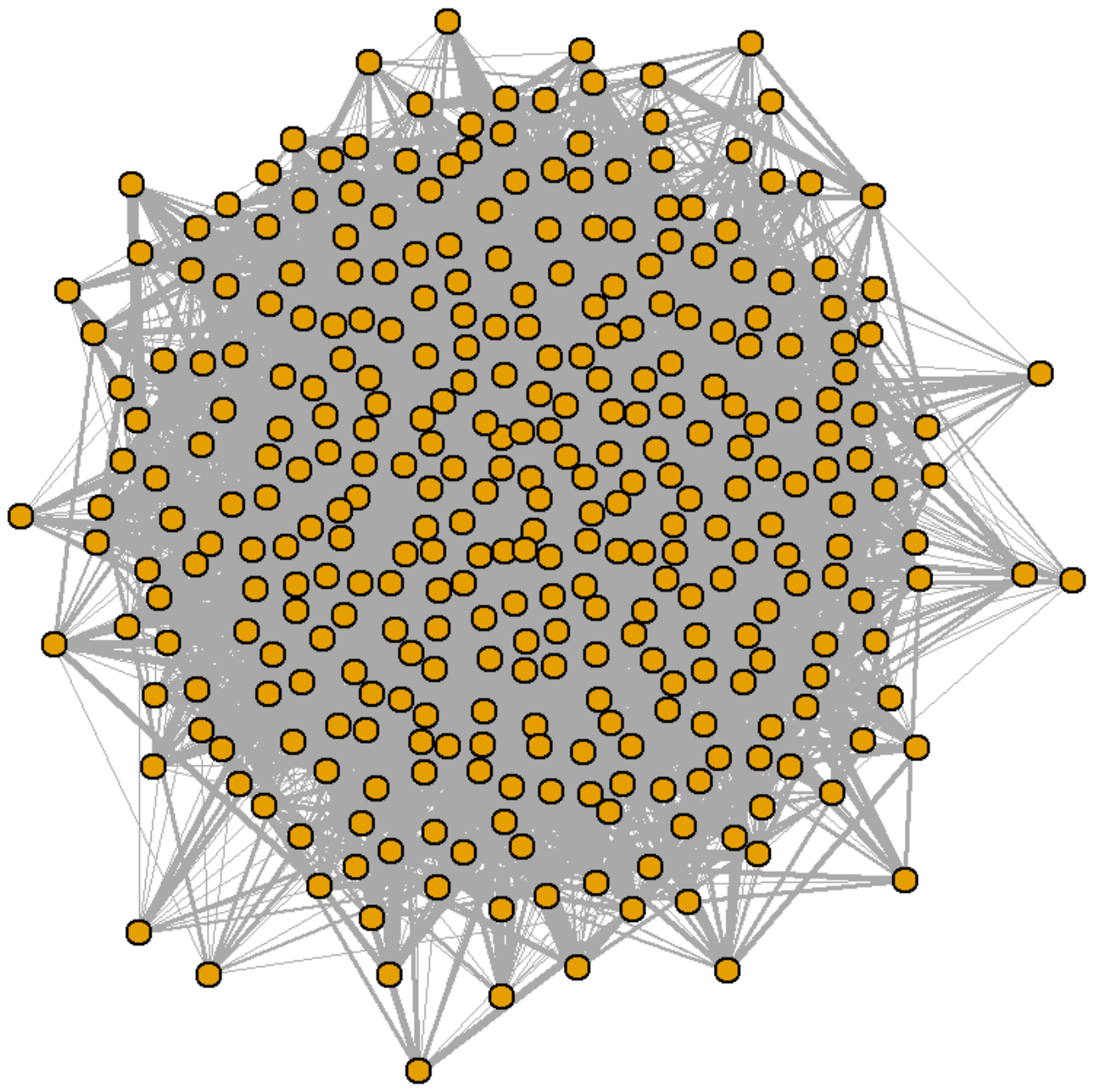

These types of reference model can be very useful when studying transmission through social networks. Degree distributions can have profound implications for spreading processes, and this type of approach provides a way to maintain the degree distribution to study the importance of other aspects of network structure.

---

## Section 3.2.1 - Resampling the raw data

A more common (and frequently more useful) way to use resampling to generate a reference model is to resample the raw data itself.

We know that space use is important in structuring the overall (between-group) burbil association network. Therefore we could construct a reference model by resampling the spatial locations that subgroups were observed at and re-constructing the social network.

We use this example to demonstrate bootstrapping as a technique in generating reference models. This is sampling **with** replacement.

Here we test the hypothesis that there is no preferential associations between members of different groups when they meet.

*N.B. Here we calculate only one instance of the reference distribution (a single reference network) but there is no reason this operation can't be repeated to generate a full reference distribution*

```
#Generate adjacency matrix to store reference model
R_fn_rs<-matrix(0,nr=nrow(full_net),nc=ncol(full_net))

#We would probably make the assumption that these two probabilities (for within an
```

```
d between clan associations) were both 1 in the absence of any other information
R_p_wc<-1
R_p_bc<-1

#Resample within-group GBIs
R_gbis<-list()
for(i in 1:length(gbis)){
  nsg<-sample(1:nrow(gbis[[i]]),nrow(gbis[[i]]),replace=TRUE)
  R_gbis[[i]]<-gbis[[i]][nsg,]
}

#Resample between-group GBIs
R_sglocs<-list()
for(i in 1:length(sglocs)){
  nsg<-sample(1:nrow(sglocs[[i]]),nrow(sglocs[[i]]),replace=TRUE)
  R_sglocs[[i]]<-sglocs[[i]][nsg,]
}

#Recalculate new between-group associations using resampled spatial data
for(i in 1:100){
  for(j in 1:(n_groups-1)){
    for(k in (j+1):n_groups){
      tA<-paste0(R_sglocs[[j]][,1],"-",R_sglocs[[j]][,2])
      tB<-paste0(R_sglocs[[k]][,1],"-",R_sglocs[[k]][,2])
      tA2<-tA[daysl[[j]]==i]
      tB2<-tB[daysl[[k]]==i]
      tt<-match(tA2,tB2)
      if(sum(is.na(tt))<length(tt)){
      if(group_clans[j]==group_clans[k]){same<-rbinom(1,1,R_p_wc)}
      if(group_clans[j]!=group_clans[k]){same<-rbinom(1,1,R_p_bc)}
      if(same==1){
        paste(i,j,k)
        for(m in length(tt)){
          if(is.na(tt[m])==FALSE){
            tsg1<-which(tA==tA2[m]&daysl[[j]]==i)
            tsg2<-which(tB==tB2[tt[m]]&daysl[[k]]==i)
            tid1<-which(gbis[[j]][tsg1,]==1)
            tid2<-which(gbis[[k]][tsg2,]==1)
            tid1a<-inds_tot[g_tot==j&gi_tot%in%tid1]
            tid2a<-inds_tot[g_tot==k&gi_tot%in%tid2]
            R_fn_rs[tid1a,tid2a]<-R_fn_rs[tid1a,tid2a]+1
            R_fn_rs[tid2a,tid1a]<-R_fn_rs[tid1a,tid2a]
          }
        }
      }
      }
    }
  }
}

#Convert between-group associations to SRIs
for(i in 1:(nrow(full_net)-1)){
  for(j in (i+1):nrow(full_net)){
    R_fn_rs[i,j]<-R_fn_rs[i,j]/(200-full_net[i,j])
    R_fn_rs[j,i]<-R_fn_rs[i,j]
  }
}
```

```
#Add within-group associations to the population network
for(i in 1:n_groups){
  R_fn_rs[inds_tot[g_tot==i],inds_tot[g_tot==i]]<-get_network2(R_gbis[[i]])
}

####################################

#We now calculate our test statistics. We choose three different test statistics.
 We do this in a different way to how we have used our test statistic before; this
time our test statistic is a comparison to the observed dataset. We can use this t
o quantify how well different reference models do in recreating the observed data.

#The first test statistic is to calculate the correlation between the network gene
rated using resampled data and the observed association network (a Mantel test)
vegan::mantel(R_fn_rs,full_net)
```

```
##
## Mantel statistic based on Pearson's product-moment correlation
##
## Call:
## vegan::mantel(xdis = R_fn_rs, ydis = full_net)
##
## Mantel statistic r: 0.9828
##       Significance: 0.001
##
## Upper quantiles of permutations (null model):
##     90%     95%   97.5%     99%
## 0.00572 0.00742 0.00911 0.01085
## Permutation: free
## Number of permutations: 999
```

```
#The second test statistic is the summed difference in values between the referenc
e network and observed network, which can highlight any bias in the edge weights o
f the reference network.
sum(R_fn_rs-full_net)
```

```
## [1] 14.00534
```

```
#The third test statistic is the summed absolute difference in values between the
 reference network and observed network, which shows how similar the reference net
work is to the observed network (smaller value is a better fit).
sum(abs(R_fn_rs-full_net))
```

```
## [1] 170.1541
```

What we see here is that the network generated from resampled GBIs and spatial locations is very similar to the observed association network. However, we may want to evaluate why this is the case. The very strong community structure to the network means that the use of the matrix correlation for the overall network is always likely to find a very strong correlation. The other two test statistics also suggest reasonable similarity, and this is likely to be that by simply resampling the GBIs and spatial locations within the groups we don't break down the social community structure and this is the overriding process explaining network structure. Therefore, because of the way we have constructed the reference model we have not learned much about our network.

We revisit this example in section 3.4 below.

---

---

---

# Section 3.3 - DISTRIBUTION-BASED REFERENCE MODELS

Instead of resampling from existing measures we can also fit statistical models to distributions of network measures and then re-generate networks accordingly.

This can be a relatively easy process to follow (for some but not all network measures) when the distribution of only one measure is involved, but gets progressively more challenging if multiple properties of the network are to be retained. This is especially true when these distributions are correlated. We illustrate a burbil example in the main text, but also examine it briefly here.

---

Our example uses the overall population association network of our main burbil study population

---

Please note that for this example, the version in the main text of the paper is not completely identical to the version presented here although all of the code is identical and the general patterns are the same.

```
#We can calculate the degree distribution for our burbil study population as follo
ws
deg<-igraph::degree(full_net2)
#Plot histogram of the degree distribution
hist(deg,las=1,xlab="Degree",cex.lab=1.5,col="grey",main="")
```

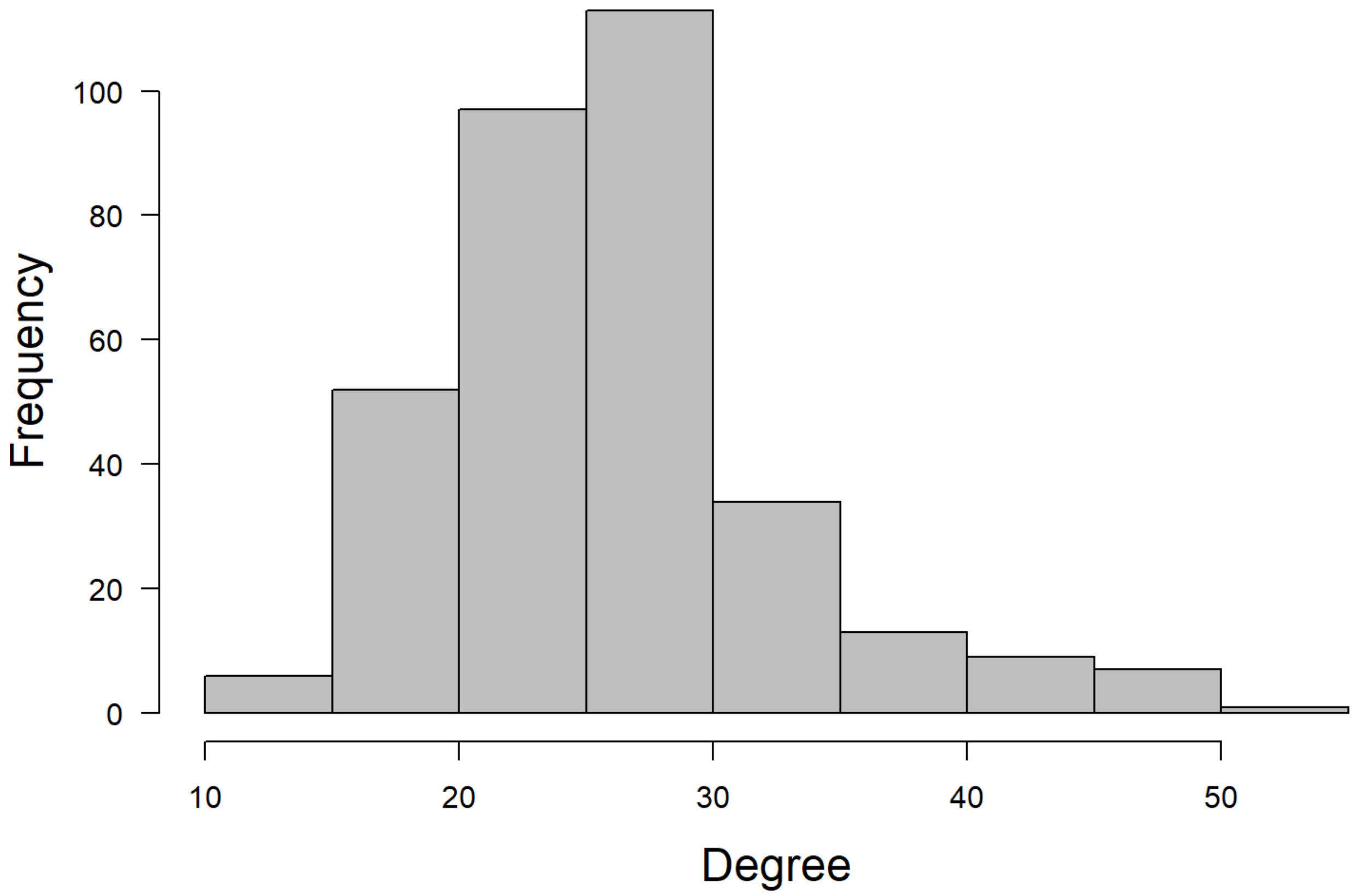

```
#In a similar way we can also calculate the distribution of clustering coefficient
s for the network
clu<-igraph::transitivity(full_net2,type="weighted")
#Plot histogram of the distribution of clustering coefficients
hist(clu,las=1,xlab="Clustering coefficient",cex.lab=1.5,col="grey",main="")
```

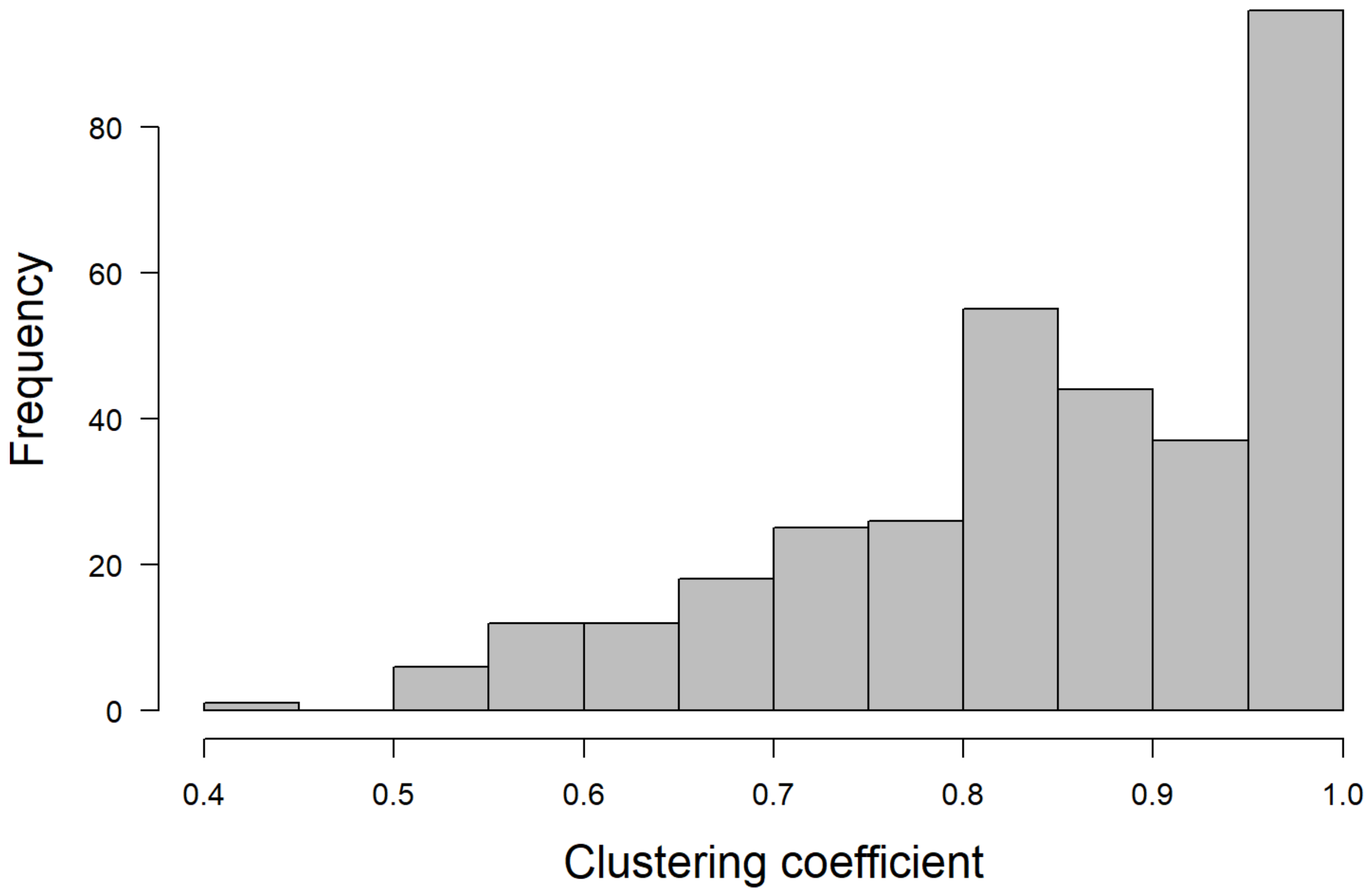

```
#Our degree distribution is approximately Poisson distributed
mean(deg)
```

```
## [1] 26.42771
```

```
var(deg)
```

```
## [1] 44.90412
```

```
#The mean and variance are similar and degree is a count of the edges an individua
l has

#Therefore we can fit a Poisson distribution to our data (we could do the same wit
h other distributions such as the Negative Binomial if preferred)
fit<-glm(deg~1)

#We can then store the parameter for the Poisson distribution
mdeg<-coef(fit)

#Then using the sample_degseq() function introduced previously we can generate a n
etwork with the degree distribution drawn from that Poisson distribution

#N.B. Some proposed degree distributions generated by the Poisson distribution are
not realisable and so we have to use a While loop to keep trying until we generate
a suitable degree sequence.
```

```
ndeg<-rpois(length(V(full_net2)),mdeg)
while(class(try(igraph::sample_degseq(ndeg,method="simple.no.multiple")))=="try-er
ror"){
  ndeg<-rpois(length(V(full_net2)),mdeg)
}
```

```
## Error in igraph::sample_degseq(ndeg, method = "simple.no.multiple") :
##   At games.c:937 : No simple undirected graph can realize the given degree sequ
ence, Invalid value
## Error in igraph::sample_degseq(ndeg, method = "simple.no.multiple") :
##   At games.c:937 : No simple undirected graph can realize the given degree sequ
ence, Invalid value
## Error in igraph::sample_degseq(ndeg, method = "simple.no.multiple") :
##   At games.c:937 : No simple undirected graph can realize the given degree sequ
ence, Invalid value
```

```
rdsn2<-igraph::sample_degseq(ndeg,method="simple.no.multiple")

#Plot the reference network generated
plot(rdsn2,vertex.label=NA,vertex.size=5)
```

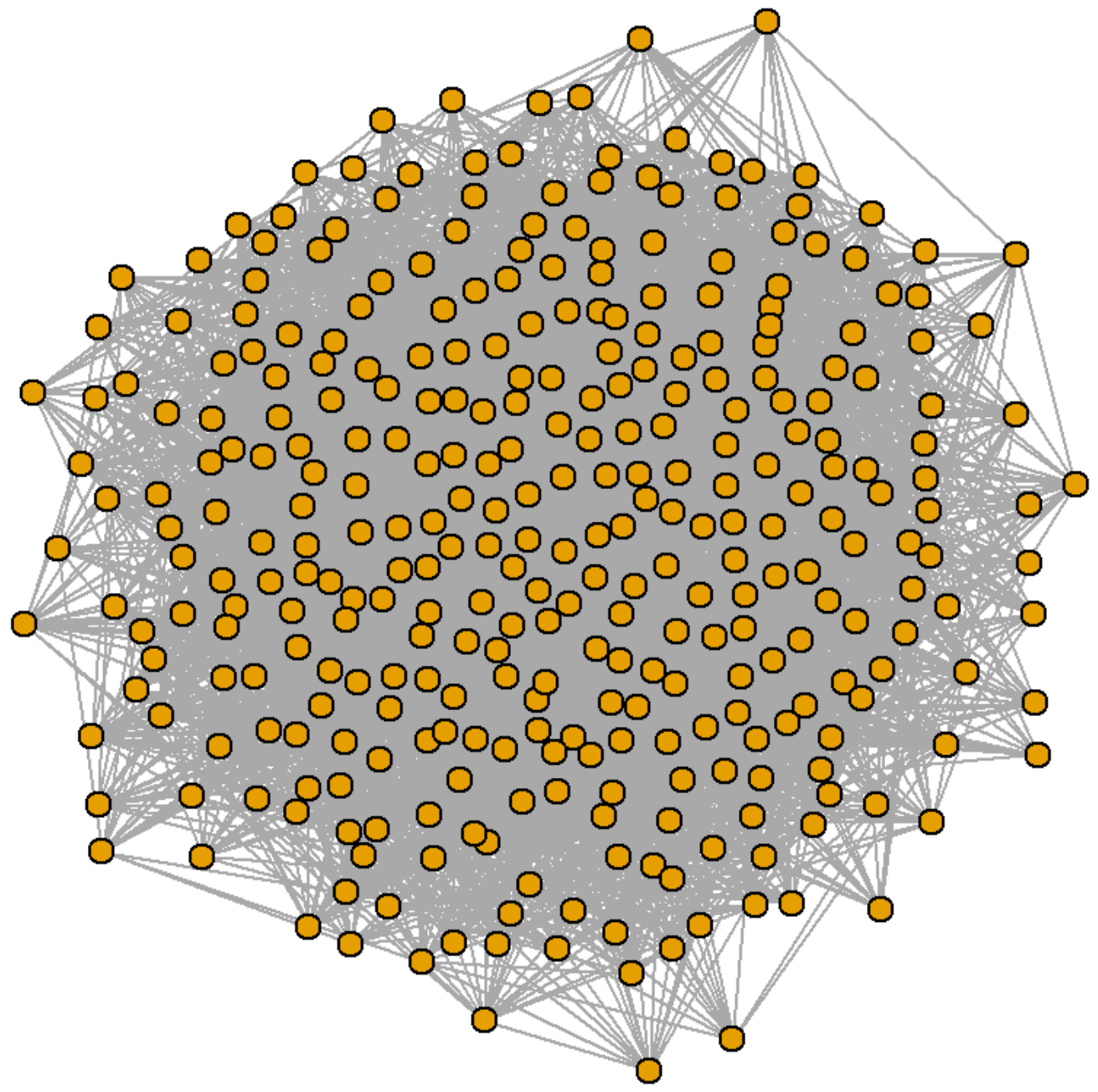

```
#Note that the network we generated here no longer has discernible community struc
ture
```

However, if we try to do the same with the distribution of clustering coefficients then we realise that it is not possible. Algorithms simply don't exist that would allow us to generate networks with a given distribution of many network measures. This is a key drawback of using distribution-based reference models.

Another potential pitfall when using distribution-based (or even resampling-based) reference models is that it can be important to consider covariance between the values of different measures.

```
#As an example of this pitfall, we calculate the distribution for the edge weights
for our burbil study population
hist(full_net[full_net>0],las=1,xlab="Degree",cex.lab=1.5,col="grey",main="")
```

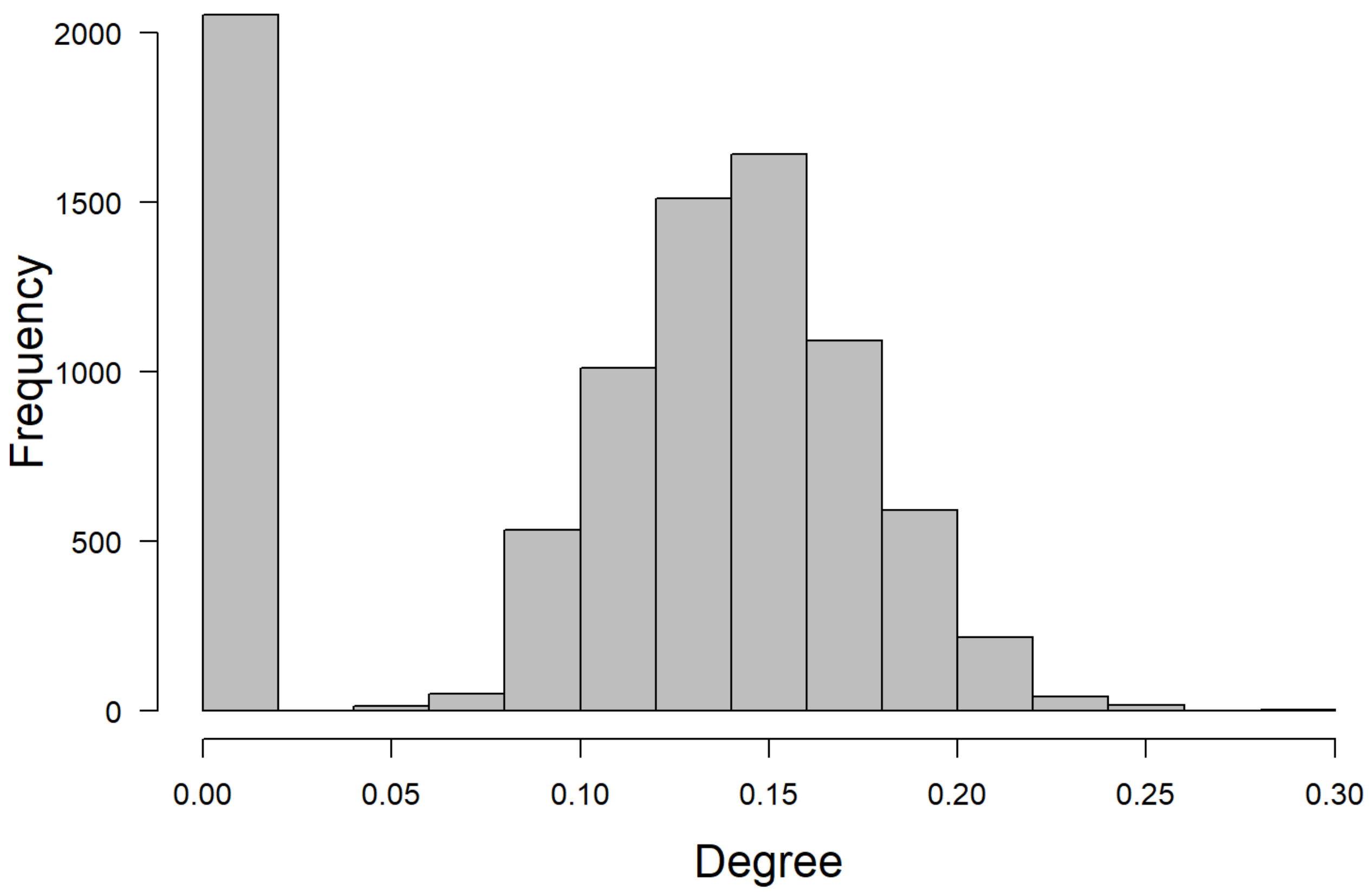

```
#This distribution is a little complex (it looks like there are two different stat
istical models generating it)

#We can examine these two processes by splitting the histogram. The two processes
 in this particular case (if you haven't worked it out) are caused in part by with
in-group versus between-group associations
par(mfrow=c(1,2))
hist(full_net[full_net>0.025],las=1,xlab="Degree",cex.lab=1.5,col="grey",main="")
hist(full_net[full_net>0&full_net<0.025],las=1,xlab="Degree",cex.lab=1.5,col="gre
y",main="")
```

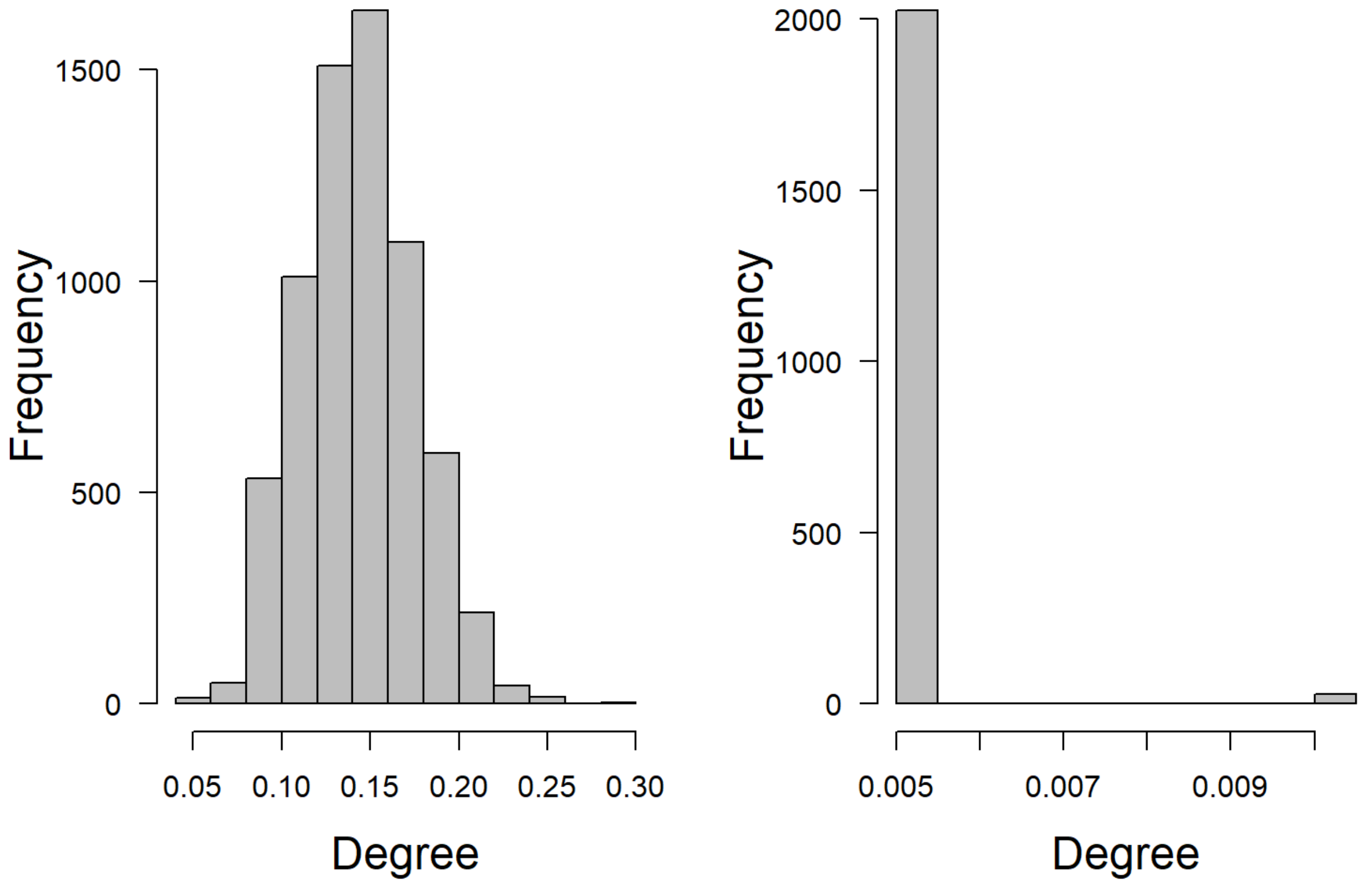

```
par(mfrow=c(1,1))

#We regenerate the distribution here in a slightly simplified form
#Calculate number of edges
ne<-gsize(full_net2)
#Calculate proportion of edges less than 0.025
pes<-sum(full_net>0&full_net<0.025)/ne

#mean and standard deviation of within-group edge weights
meb<-mean(full_net[full_net>0.025])
sdeb<-sd(full_net[full_net>0.025])

#set up vector to store edges for new graph
new_edgeweights<-rep(NA,gsize(rdsn2))

#fill in vector - we are effectively generating a normal distribution of within-gr
oup edge weights and making all between-group edge weights equal to 0.005
for(i in 1:gsize(rdsn2)){
  tb<-rbinom(1,1,pes)
  if(tb==1){
    new_edgeweights[i]<-0.005
  }
  if(tb==0){
    new_edgeweights[i]<-rnorm(1,meb,sdeb)
  }
}

#We can check our new edge weight degree distribution against the original one
```

```
par(mfrow=c(1,2))
hist(full_net[full_net>0&upper.tri(full_net)==TRUE],las=1,xlab="Degree",cex.lab=1.
5,col="grey",breaks=seq(0,0.3,0.01),main="",ylim=c(0,2000))
hist(new_edgeweights,las=1,xlab="Degree",cex.lab=1.5,col="grey",breaks=seq(0,0.3,
0.01),main="",ylim=c(0,2000))
```

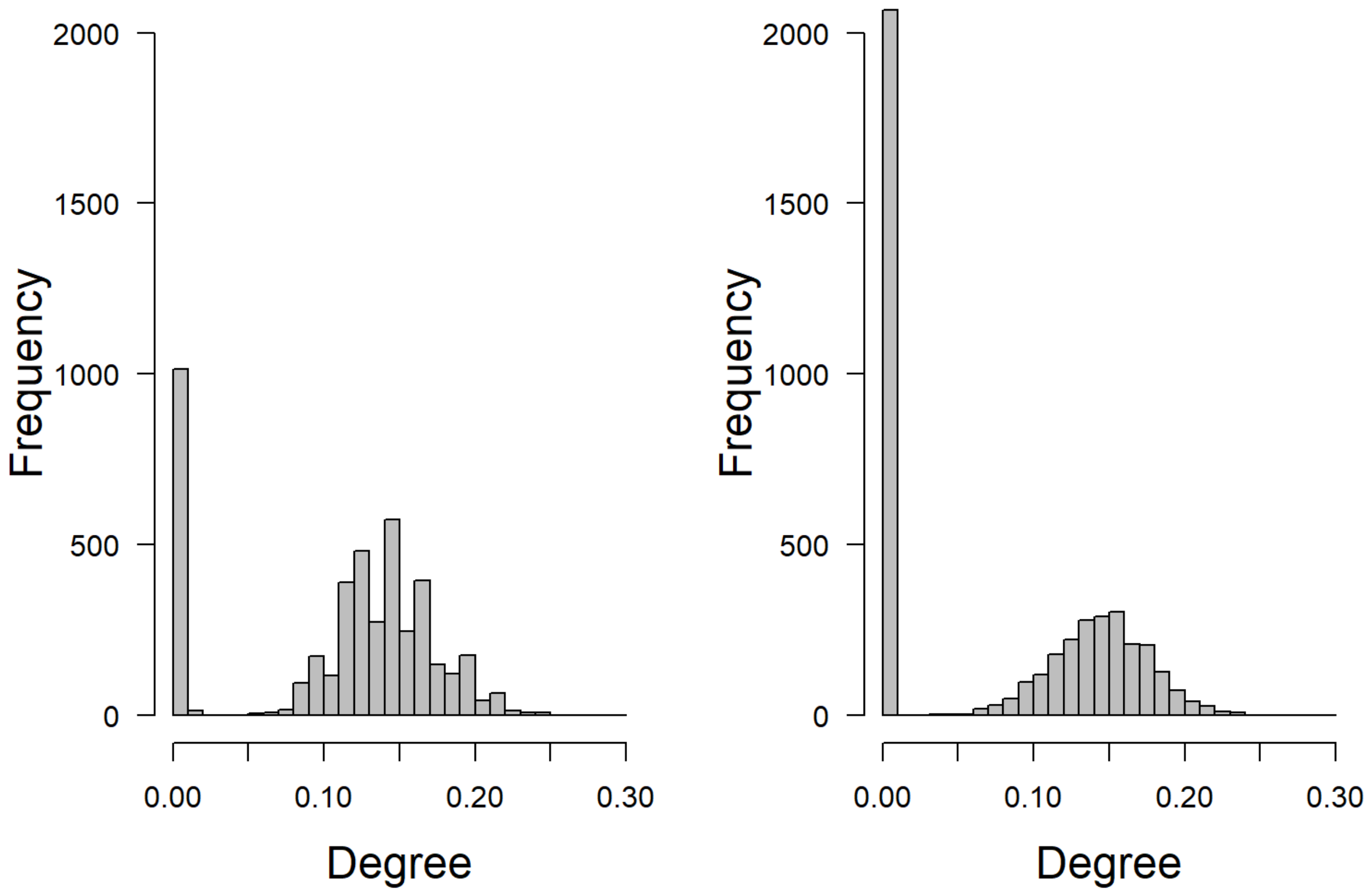

```
par(mfrow=c(1,1))

#We are feeling pretty pleased with ourselves, we have done a pretty good (albeit
 not perfect job) of fitting the edge weight distribution. The main weakness of ou
r current model is that it overstimates the number of very weak between-group conn
ections, which is unsurprising as we set all of these to the same very low value.
 We can use this for our new graph

#Plot newly generated, weighted reference network
edge_attr(rdsn2)$weight<-new_edgeweights
plot(rdsn2,vertex.label=NA,vertex.size=5,edge.width=(edge_attr(rdsn2)$weight*8)^3)
```

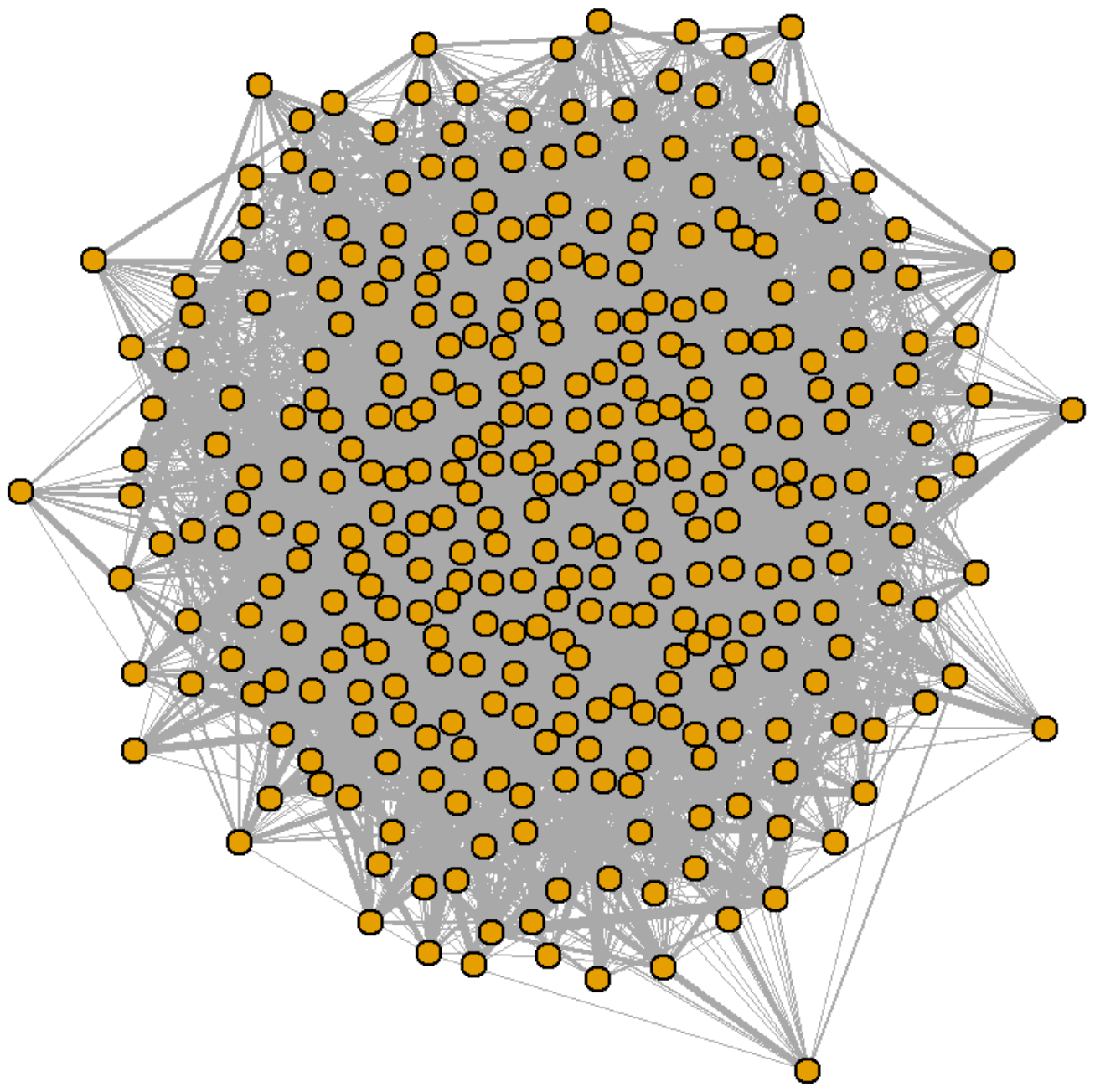

However, we then evaluate and realise there is are problems with what we've done. Not only have we underestimated the strength of some between-group associations (presumably between members of nearby groups) but between-group edges tend to be much weaker and individuals with high degree may have more between-group connections reducing their mean edge weight. Similarly edge weights may also be biased by other things that influence associations such as nose colour. We failed to consider this covariance between edge weights and degree.

```
#So we plot the relationship here
plot(deg,rowMeans(full_net),pch=16,xlab="Degree",ylab="Mean Association Strength",
cex.lab=1.5)
```

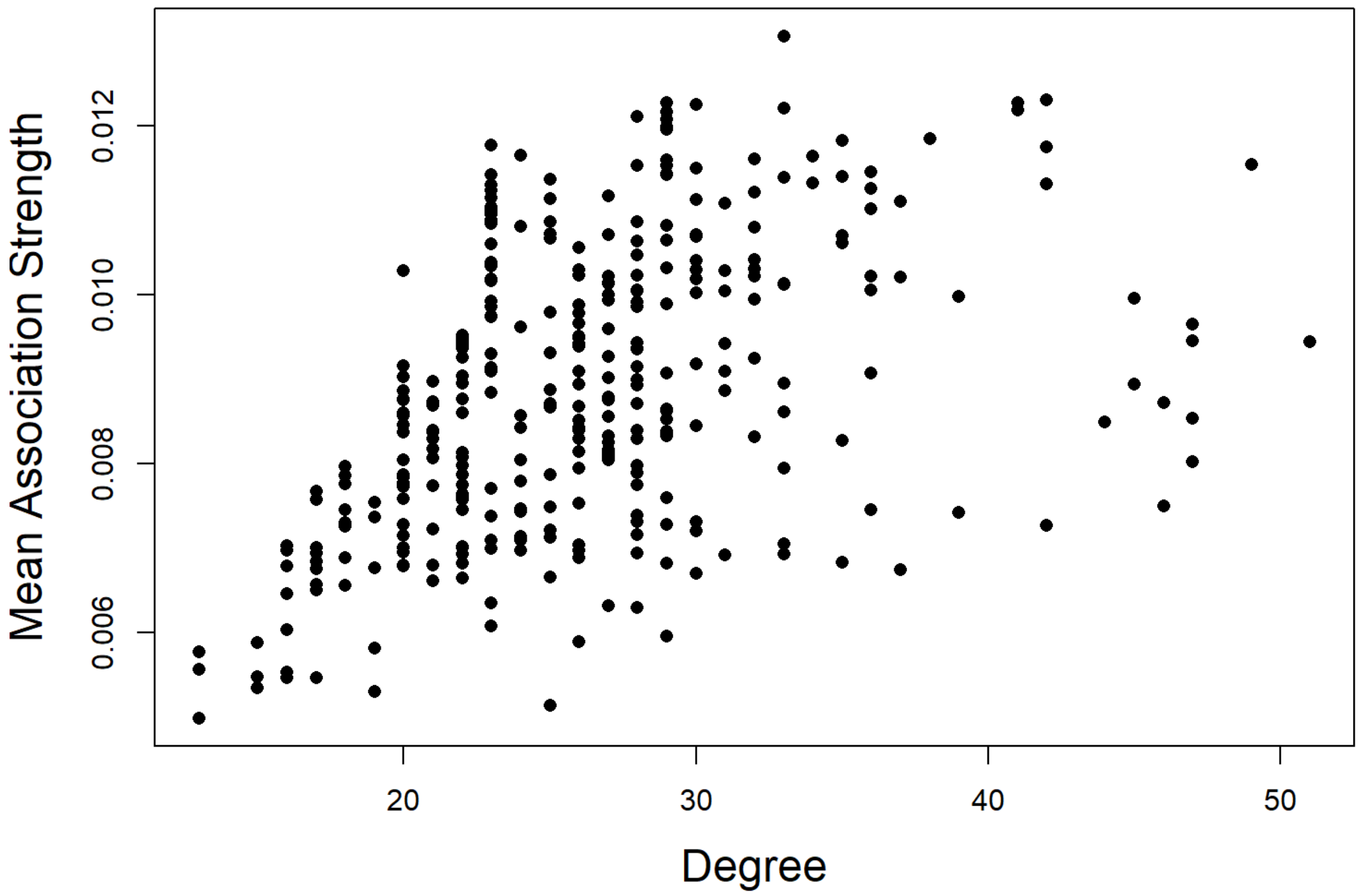

From the plot we can work out that there is a complicated relationship between degree and the mean association strength and we need a complex reference model to capture this relationship properly.

---

Distribution-based reference models are complicated! For some network measures they might not be possible at all. While for others it might be important to consider covariance between them in order to generate an appropriate reference model.

Distribution-based reference models could also be applied to raw data. You could fit a distribution to the relationship between individual traits or landscape features and the social or spatial data used to build you nets and rebuild your network from there. We don't cover that in this case study.

---

---

---

# Section 3.4 - GENERATIVE REFERENCE MODELS

Our final type of reference model is the generative reference model. We first briefly illustrate the use of some basic statistical models for the networks themselves, before showing how agent-based models can be used to generate reference distributions of networks

---

## Section 3.4.1 - Statistical network models

Statistical network models are well-covered elsewhere in the network structure. We touch on two commonly-used examples here: - a) Stochastic block models which can be used to generate a reference distribution related to the community structure of the graph; - b) Exponential random graph models which can be used to fit parameters to describe how the probability or weight of edges can be explained by structural properties of the network, nodal traits and dyadic traits.

```
#Note both these models are verbose during fitting and so we have hidden output an
d figures for this chunk of code

##Fit a stochastic block model to the association network
#We fit a block model for a weighted network, assuming edge weights have a Gaussia
n distribution as this is a reasonable assumption for our association network (see
previous sections)
sb<-blockmodels::BM_gaussian(membership_type="SBM_sym",adj=full_net,verbosity=0)
sb$estimate()
```

```
##Fit an ERGM to the dominance interaction data
#We first convert our dominance network to a network object for the ergm package i
n R
dom_el<-as.tnet(MAT_DOM)
```

```
## Warning in as.tnet(MAT_DOM): Data assumed to be weighted one-mode tnet (if
## this is not correct, specify type)
```

```
dom<-network(dom_el[,1:2])

#We then add edge weight as an attribute
network::set.edge.attribute(dom,"weight",as.vector(dom_el[,3]))

#We then add individual traits as node attributes
network::set.vertex.attribute(dom,"sex",as.vector(sexes[[1]]))
network::set.vertex.attribute(dom,"age",as.vector(ages[[1]]))
network::set.vertex.attribute(dom,"nose",as.vector(noses[[1]]))

#We can then fit a a count ERGM (with a Poisson reference distribution) to the net
work
#nonzero is a term to control for zero-inflation in edge counts (because many soci
al networks are sparse)
#Sum is an intercept-like term fo edge weights
#We can then fit an array of terms to test hypotheses about the network structure
 and associations between connection weights and individual traits
#See https://rdrr.io/cran/ergm/man/ergm-terms.html for full details on ERGM terms

dom_mod<-ergm(dom~nonzero+sum+mutual(form="nabsdiff")+cyclicalweights(twopath="mi
n",combine="max",affect="min")+transitiveweights(twopath="min",combine="max",affec
t="min")+nodematch("sex",diff=TRUE)+nodematch("age",diff=TRUE)+nodematch("nose",di
ff=TRUE)+nodeofactor("age")+nodeofactor("sex")+nodeofactor("nose"),reference=~Pois
son,response="weight",silent=TRUE)
```

```
## Warning: `set_attrs()` is deprecated as of rlang 0.3.0
## This warning is displayed once per session.
```

```
##To check that the model has converged we would run
#mcmc.diagnostics(dom_mod)
```

```
#(code not run here)
```

Now we can examine the fit of these statistical models and explore how to use them as reference distributions.

```
#For the stochastic block model, we can see how the fit of the model depends on th
e number of blocks or communities
plot(sb$ICL,pch=16)
```

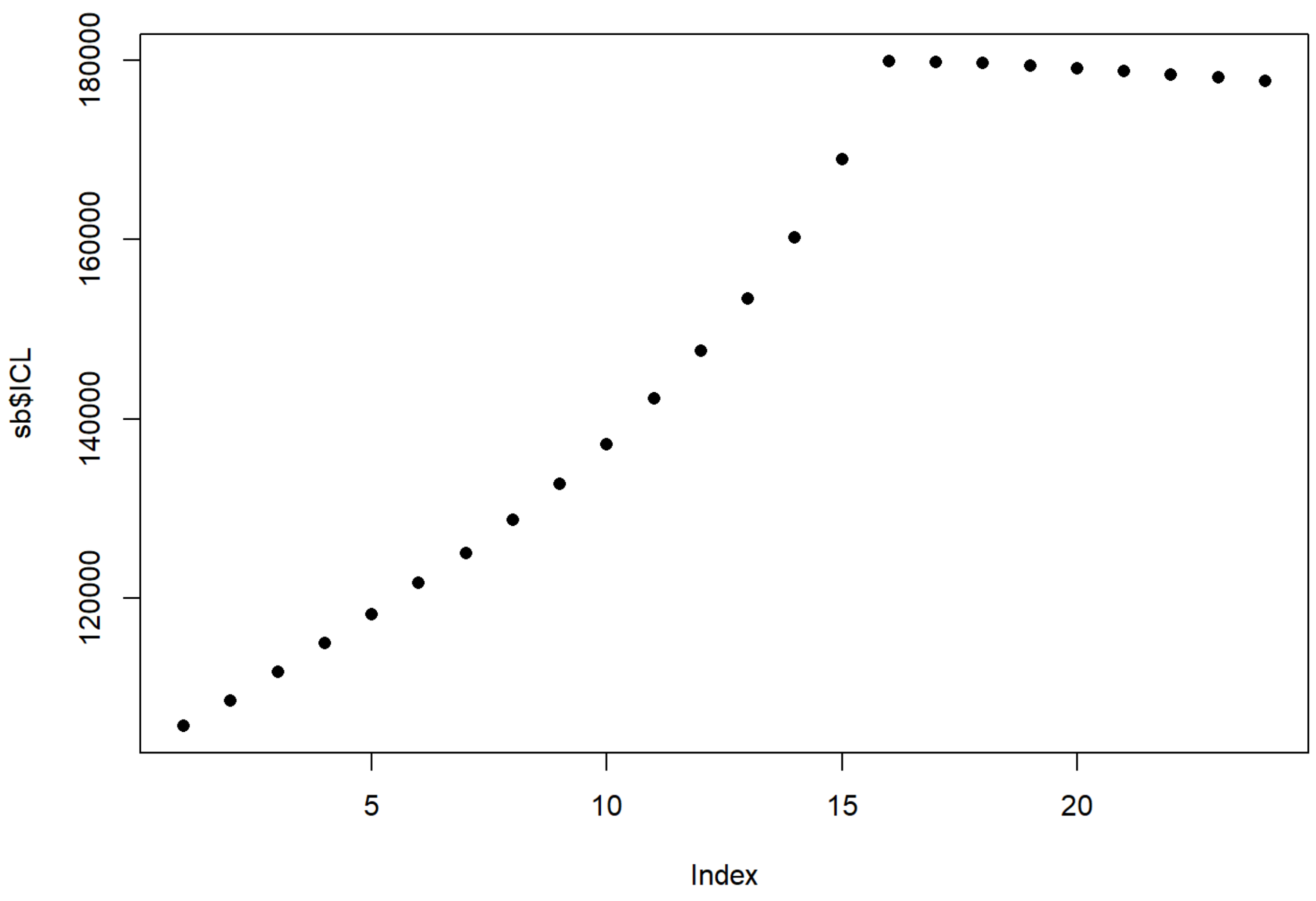

```
#We can see that the fit of the model doesn't really improve once 16 communities a
re included. This is unsurprising given we simulated 16 burbil groups and within-g
roup associations are so much more frequent than between-group associations.
#The best model fit is for 16 blocks/communities
which.max(sb$ICL)
```

```
## [1] 16
```

```
#We can examine the model predictions visually as follows
#The stochastic block model fits very closely to the observed network structure vi
sually
sb$plot_obs_pred(16)
```

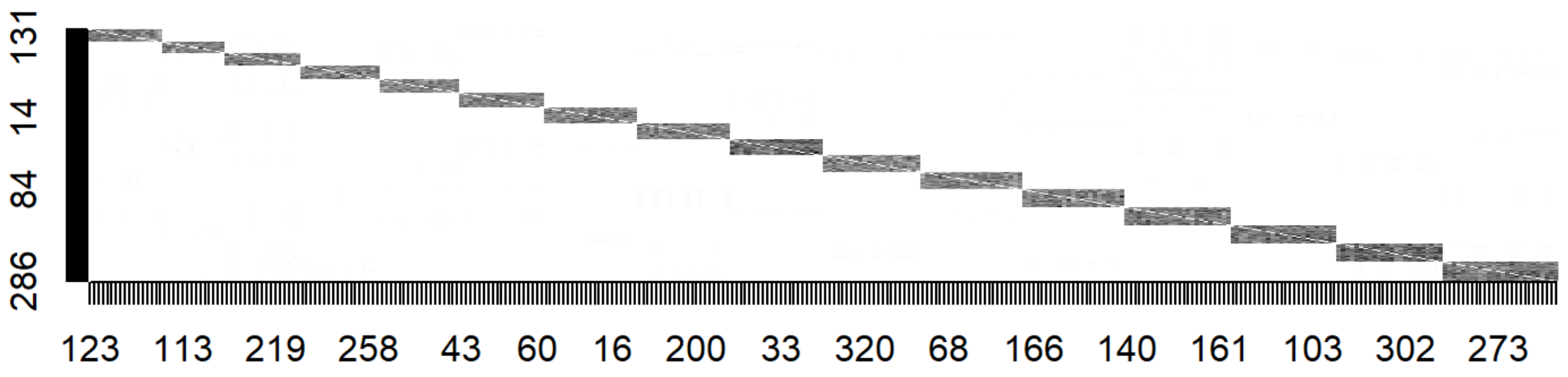

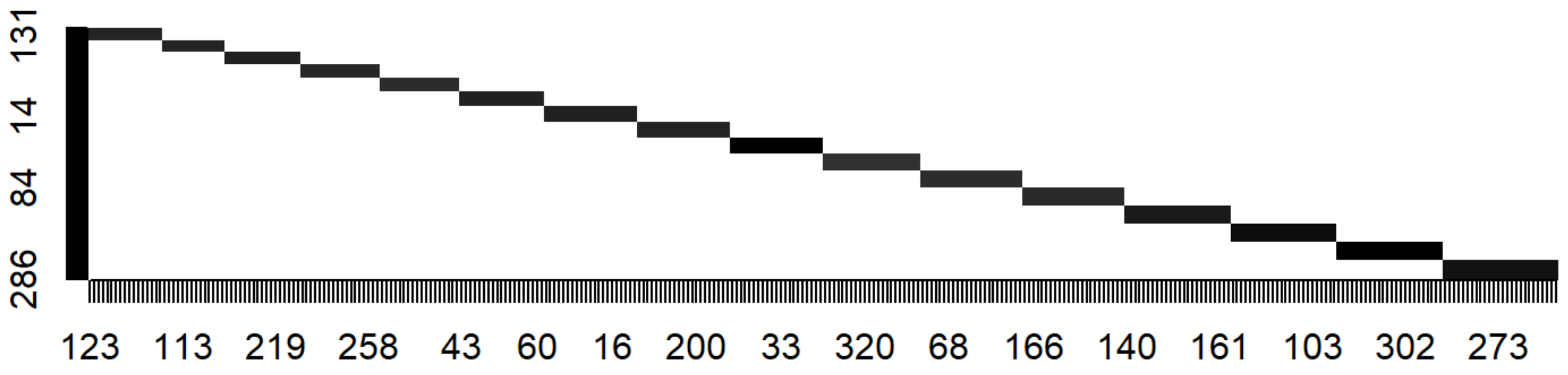

```
#We can check the fit of the block model further by working out the memberships it
applies and comparing the size of blocks to the size of the groups we initially ge
nerated
mems<-sign(round(sb$memberships[[16]]$Z,2))
table(unlist(gss))
```

```
##
## 14 17 18 19 21 22 23 24 26
##  1  2  2  1  3  1  2  3  1
```

```
table(colSums(mems))
```

```
##
## 14 17 18 19 21 22 23 24 26
##  1  2  2  1  3  1  2  3  1
```

```
#We can see full model parameters using the command below (not run here)
#sb$model_parameters[16]

############################

#For the ERGM we can print out a model summary much like we do for other statistic
al models
summary(dom_mod)
```

```
##
## ==========================
## Summary of model fit
## ==========================
##
## Formula:   dom ~ nonzero + sum + mutual(form = "nabsdiff") + cyclicalweights(tw
opath = "min",
##     combine = "max", affect = "min") + transitiveweights(twopath = "min",
##     combine = "max", affect = "min") + nodematch("sex", diff = TRUE) +
##     nodematch("age", diff = TRUE) + nodematch("nose", diff = TRUE) +
##     nodeofactor("age") + nodeofactor("sex") + nodeofactor("nose")
##
## Iterations:  5 out of 20
##
## Monte Carlo MLE Results:
##                                Estimate Std. Error MCMC % z value Pr(>|z|)
## nonzero                       -1.955482   0.409815      0  -4.772  < 1e-04
## sum                            2.616254   0.059477      0  43.988  < 1e-04
## mutual.nabsdiff               -0.213979   0.030494      0  -7.017  < 1e-04
## cyclicalweights.min.max.min   -0.078708   0.021052      0  -3.739 0.000185
## transitiveweights.min.max.min -0.014112   0.046224      0  -0.305 0.760138
## nodematch.sum.sex.F           -0.014817   0.049582      0  -0.299 0.765058
## nodematch.sum.sex.M            0.090035   0.043901      0   2.051 0.040280
## nodematch.sum.age.AD          -0.259056   0.055446      0  -4.672  < 1e-04
## nodematch.sum.age.JUV          0.670305   0.073325      0   9.142  < 1e-04
## nodematch.sum.age.SUB         -0.001308   0.077338      0  -0.017 0.986507
## nodematch.sum.nose.ORANGE      0.235533   0.063111      0   3.732 0.000190
## nodematch.sum.nose.RED         0.185945   0.042246      0   4.402  < 1e-04
## nodeofactor.sum.age.JUV       -0.851502   0.077000      0 -11.058  < 1e-04
## nodeofactor.sum.age.SUB       -0.260616   0.043400      0  -6.005  < 1e-04
## nodeofactor.sum.sex.M         -0.180688   0.042170      0  -4.285  < 1e-04
## nodeofactor.sum.nose.RED      -0.045799   0.041396      0  -1.106 0.268567
##
## nonzero                       ***
## sum                           ***
## mutual.nabsdiff               ***
## cyclicalweights.min.max.min   ***
## transitiveweights.min.max.min
## nodematch.sum.sex.F
## nodematch.sum.sex.M           *
## nodematch.sum.age.AD          ***
## nodematch.sum.age.JUV         ***
## nodematch.sum.age.SUB
## nodematch.sum.nose.ORANGE     ***
## nodematch.sum.nose.RED        ***
## nodeofactor.sum.age.JUV       ***
## nodeofactor.sum.age.SUB       ***
## nodeofactor.sum.sex.M         ***
## nodeofactor.sum.nose.RED
## ---
## Signif. codes:  0 '***' 0.001 '**' 0.01 '*' 0.05 '.' 0.1 ' ' 1
##
##      Null Deviance:      0  on 420  degrees of freedom
##  Residual Deviance: -13035  on 404  degrees of freedom
##
## Note that the null model likelihood and deviance are defined to be
```

```
## 0. This means that all likelihood-based inference (LRT, Analysis
## of Deviance, AIC, BIC, etc.) is only valid between models with the
## same reference distribution and constraints.
##
## AIC: -13003    BIC: -12938    (Smaller is better.)
```

```r
#We can also simulate networks based on the ERGM fit to provide a reference distri
bution for further hypothesis testing (for example, by seeing how goodness of fit
 changes for different regions of the network)

#Here we simulate 10 networks
ref_doms<-simulate(dom_mod,10)

#A quick plot to show the 10 reference networks
#n.b we are plotting using the network package here for speed. We could convert to
igraph if desired
par(mfrow=c(2,5),mar=c(0,0,0,0))
for(i in 1:length(ref_doms)){
  plot(ref_doms[[1]])
}
```

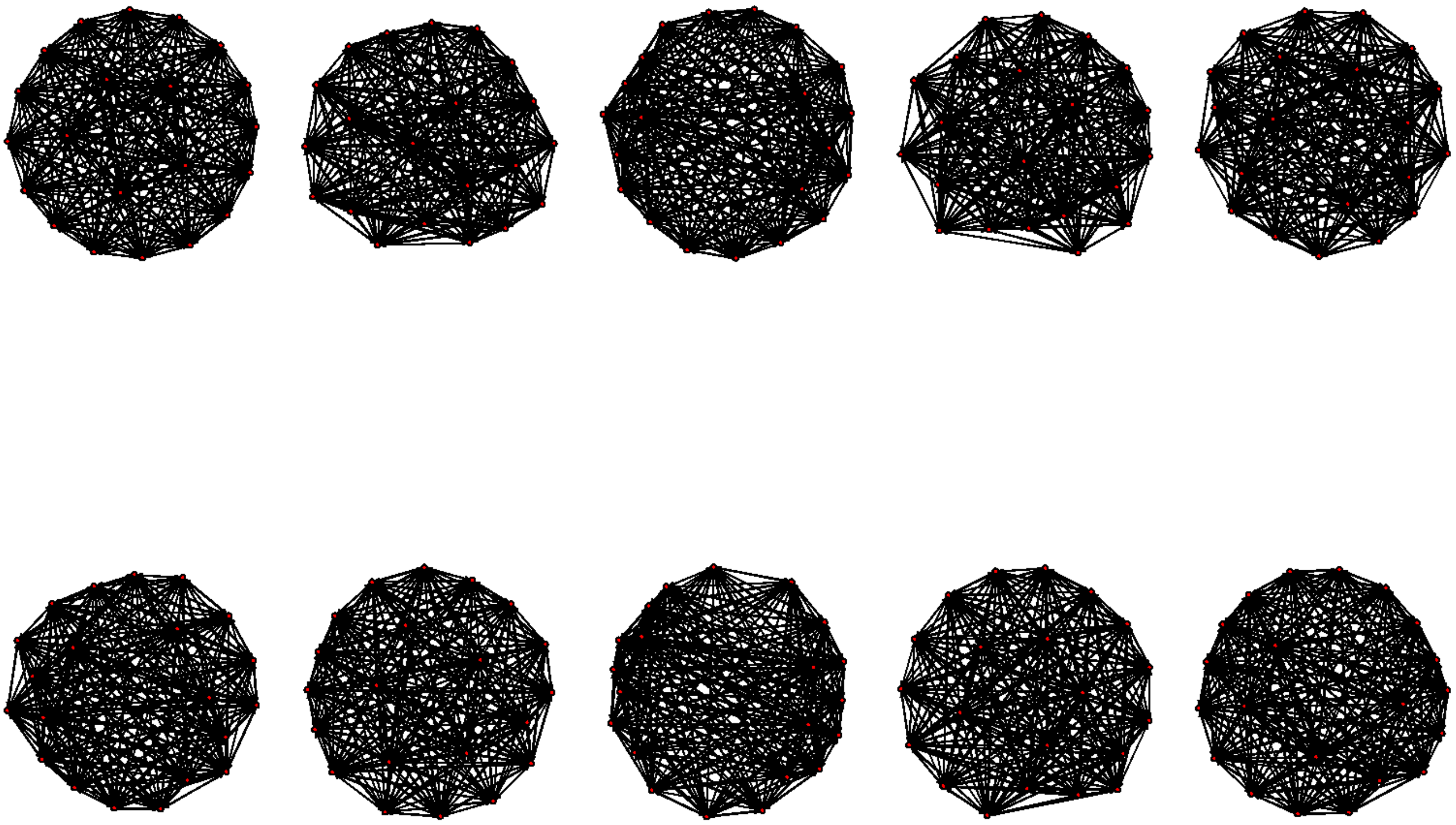

```r
par(mfrow=c(1,1),mar=c(5,6,2,2))

#Here is the conversion into adjacency matrices
ref_mats<-as.sociomatrix(ref_doms,attrname="weight",simplify=FALSE)
```

```
#Print an adjacency matrix to demonstrate
ref_mats[[1]]
```

```
##     1  2  3  4  5  6  7  8  9 10 11 12 13 14 15 16 17 18 19 20 21
## 1   0  3  6  2  1  5  5  3  6  7  2 20  2  6 11 12  7 25 14  4  2
## 2   8  0  5 17  9 11  9  8  5 10  4  9  9 14 10 21  7 10 12  4  9
## 3   8 18  0 12 15 14 11 11 14 24 17 26 13 17 27 15 19 15  9 17  6
## 4  13 11  5  0  4  7  9  6  6  9  7  5  9 22  9  8  6 13 10 11  6
## 5  14 16  4 15  0 14  7 12  3 19  6 16  5 13 14 24 18 18 13  5  2
## 6  10 10  5 21 12  0  9  6  9  8  5 11  7  6  6  9  6 11 10  5  7
## 7  11 16  6 19 22 18  0 10  4 16  9 17 14 16 12 14 15 26 19 19  8
## 8  11 20  6 21  9 14  7  0  8 15  6 23 10 20 14 13 24 13 12  8  7
## 9  20 14 12 18 11 23 12  6  0 21 18 22  8 15 26 18 10 14 15 20  0
## 10  6  6  6  4  4  2  8  3  3  0  6  8  0 17  6  7  6 10  6  0  0
## 11 20 18  2 26 15 12 15 15  3 19  0 13 17 17 18 22 26 26 11 23 20
## 12  7  3  2  4  3  6  4  0  3  5  3  0  2  9 10  4  6  7 13  4  4
## 13 11 22  8 12 13  7 21 13  7 19  6 11  0 34 10 14 15 17 15 10  7
## 14  7  4  3  6  4  3  6  4  2 11  2  8  6  0  7  5  5  5 10  3  2
## 15 19  6  5  9  1  4  5  1  4 10  5 26  3 13  0 10  3 13 12  0  2
## 16  7  6  4  4  5  8  5  7  2 13  1 15  0 14  2  0  2 21 13  3  5
## 17 13 15  9 11 15  7  5 10  8  7  7  6  8 13 15 11  0 14  8 11  5
## 18  5  4  4  3  4  3  4  5  0 22  6 16  6 13  4 11  4  0  3  4  0
## 19  7  3  2  5  5  0  3  7  5 16  7  8  4 19 10 10  1 11  0  5  3
## 20 12 11  7  3  8 10  9 17  7 15  9 16  7 13 10 15 21 11 16  0 11
## 21 18 25 10 13 11 20  8 12  8 19  3 13 22 26 12 11 22 21 13 21  0
```

## Section 3.4.2 - Agent-based reference models

Agent-based models offer a powerful way to develop reference distributions that depend on behavioural rules rather than the structure of the observed network. You can program individuals to behave in a particular way and record their interactions and associations to generate a simulated network.

This is of course how we generated our burbil society in the first place. Therefore, in order to demonstrate the use of agent-based models we are going to reuse some of our previous code and encourage you to examine the consequences of changing key parts of it.

First we fit a spatially-explicit agent-based model

Second we fit a spatially-explicit model applied at a subgroup level

Third we develop a socially-explicit agent-based model to see whether it is better able to explain burbil association patterns.

*Note that we only fit produce one simulation of each agent-based model here. However, stochastic agent-based models such as this can also be used to build reference distributions of test statistics if run multiple times. Give it a go if you fancy!*

In this example our question is: How Are between-group association networks structured by space-use?

Our test statistics will be the correlation between the network generated using the ABM and the observed between-group network (a Mantel test), the summed difference in values between the reference network and observed network, which can highlight any bias in the edge weights of the reference network and the summed absolute difference in values between the reference network and observed network, which shows how similar the reference network is to the observed network (smaller value is a better fit).

*Note that these are the same test statistics used in one of our resampling examples*

*Note also that we have learned our lesson and creating networks of summed associations between burbils from different groups rather than the entire network*

*Finally note that we add parameters as we go along, e.g. dist_eff defined in the first code chunk is used in all three*

```
#Here we set the standard deviation for how far burbil subgroups tend to travel fr
om their home range centre (we will assume we know these for now)
#Note that we have used the value we originally used to generate the data here. Fe
el free to change the value and see what effect it has
#This will be used for all three reference models
dist_eff<-2

#First we need our group locations (printed below)
print(group_locs)
```

```
##       x  y
## 18    4  4
## 22    8  4
## 26   12  4
## 30   16  4
## 82    4  8
## 86    8  8
## 90   12  8
## 94   16  8
## 146   4 12
## 150   8 12
## 154  12 12
## 158  16 12
## 210   4 16
## 214   8 16
## 218  12 16
## 222  16 16
```

```
#We now create the observed between-group network
group_net<-matrix(0,nr=dim(group_locs)[1],nc=dim(group_locs)[1])
for(i in 1:nrow(full_net)){
  for(j in 1:ncol(full_net)){
    if(g_tot[i]!=g_tot[j]){
      group_net[g_tot[i],g_tot[j]]<-group_net[g_tot[i],g_tot[j]]+full_net[i,j]
    }
  }
}

#And we can then plot the observed between-group network
gnet<-graph.adjacency(group_net,mode="undirected",weighted=TRUE)
plot(gnet,vertex.color="light blue",edge.width=(edge_attr(gnet)$weight)^2)
```

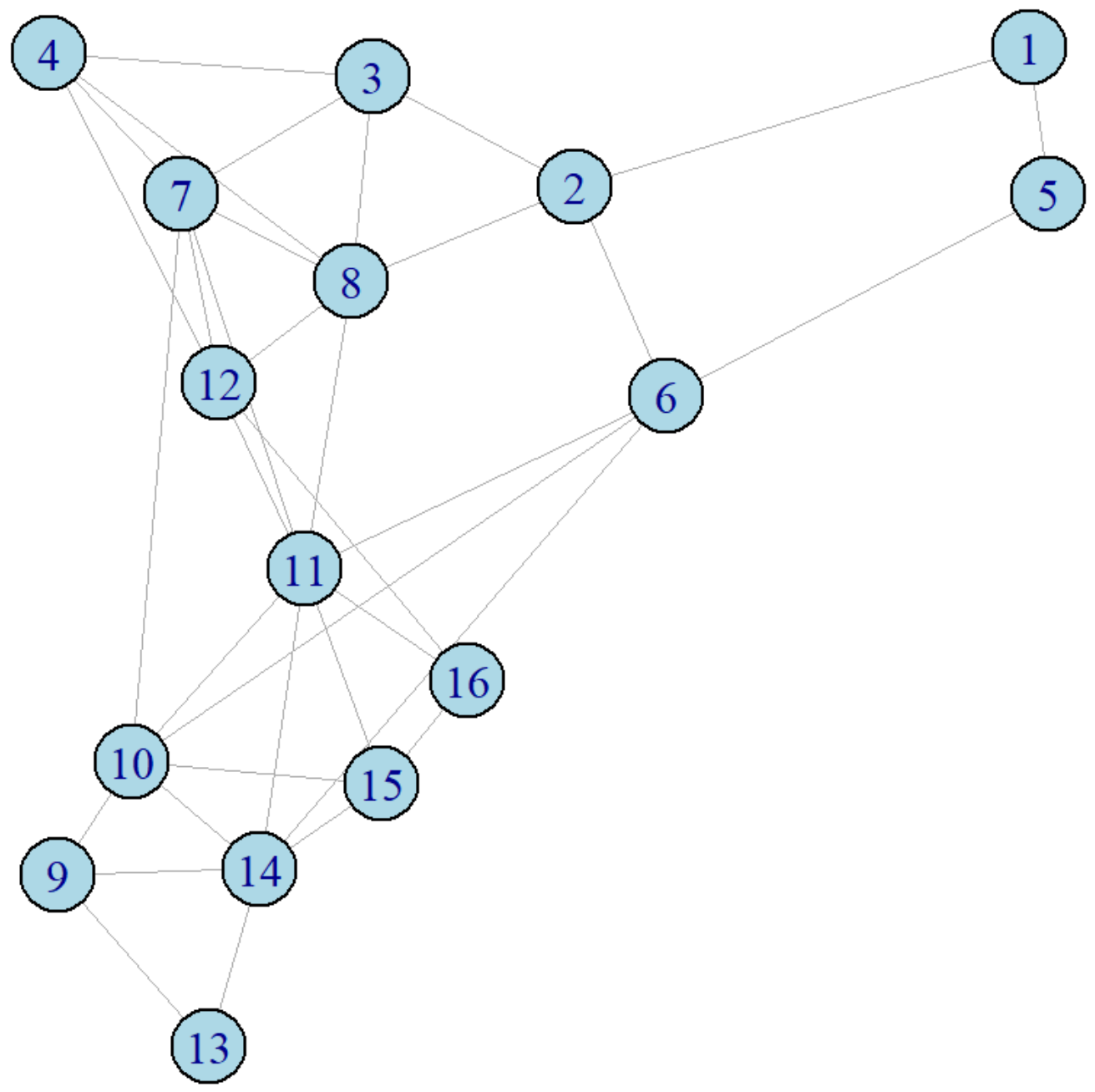

```
###########################################################


##We now generate our reference model with a truly spatially-explicit ABM (i.e. we
remove the clan effect and allow individuals to be observed independently and not
 necessarily as subgroups)

#We assume that each individual is observed 100 times but this assumption can be c
hanged if desired

#Empty list to store new locations
R_indiv_locs<-list()

#Assign individual locations
for(i in 1:nrow(full_net)){
  tx<-round(rnorm(100,group_locs[g_tot[i],1],dist_eff))
  ty<-round(rnorm(100,group_locs[g_tot[i],2],dist_eff))
  R_indiv_locs[[i]]<-cbind(tx,ty)
}

#Generate full network for associations between individuals
R_fn<-matrix(NA,nr=nrow(full_net),nc=ncol(full_net))
for(i in 1:nrow(R_fn)){
  for(j in 1:ncol(R_fn)){
    R_fn[i,j]<-sum(rowSums(R_indiv_locs[[i]]==R_indiv_locs[[j]])==2)/100
  }
}
diag(R_fn)<-0
```

```
#Generate network of summed between-group associations
R_gn<-matrix(0,nr=dim(group_locs)[1],nc=dim(group_locs)[1])
for(i in 1:nrow(R_fn)){
  for(j in 1:ncol(R_fn)){
    if(g_tot[i]!=g_tot[j]){
      R_gn[g_tot[i],g_tot[j]]<-R_gn[g_tot[i],g_tot[j]]+R_fn[i,j]
    }
  }
}

#Plot network generated
RGN<-graph.adjacency(R_gn,mode="undirected",weighted=TRUE)
plot(RGN,vertex.color="light blue",edge.width=(edge_attr(RGN)$weight)^2)
```

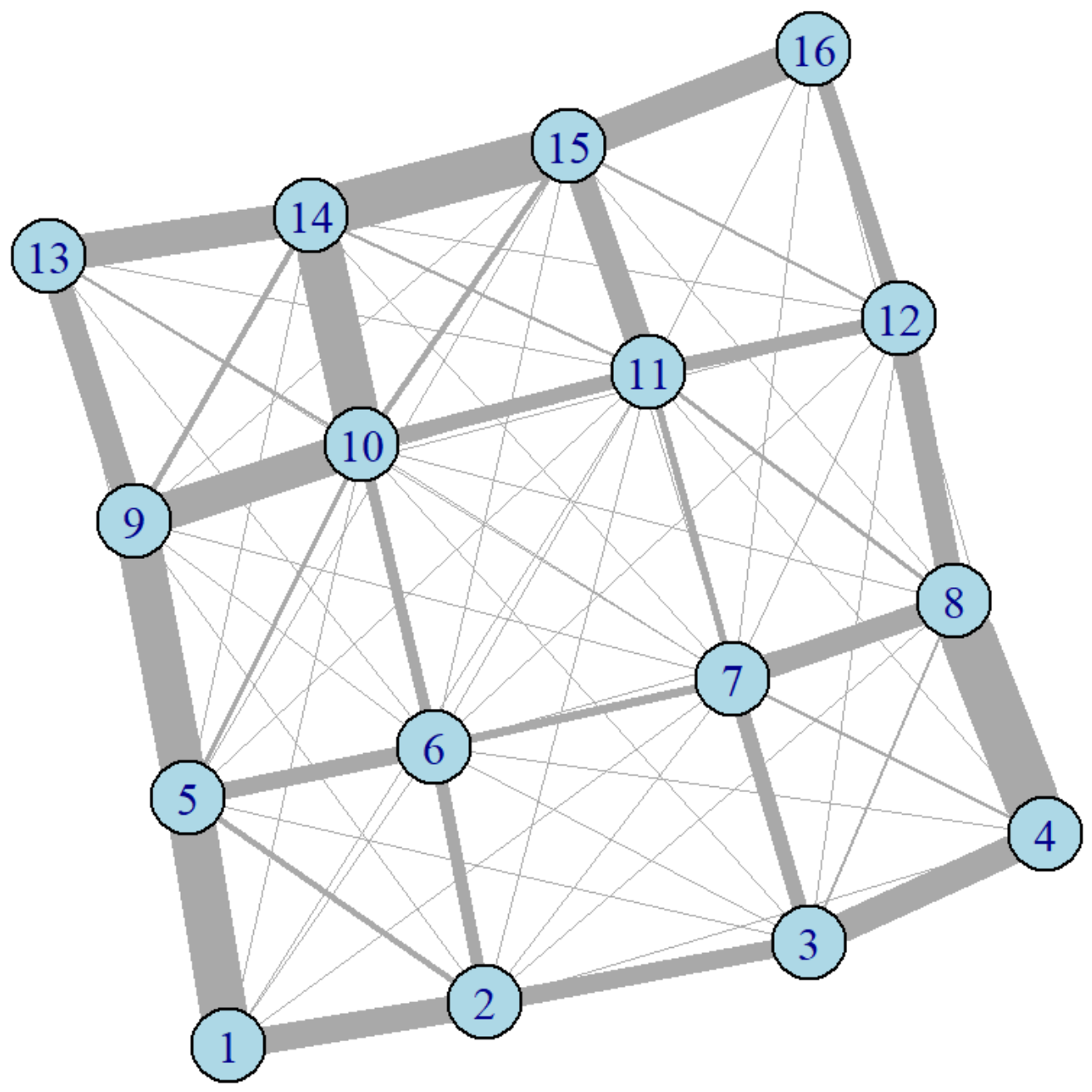

```
#Calculate values for the test statistics
vegan::mantel(R_gn,group_net)
```

```
##
## Mantel statistic based on Pearson's product-moment correlation
##
## Call:
## vegan::mantel(xdis = R_gn, ydis = group_net)
##
## Mantel statistic r: 0.6596
##       Significance: 0.001
##
```

```
## Upper quantiles of permutations (null model):
##   90%   95% 97.5%   99% 
## 0.128 0.171 0.216 0.246 
## Permutation: free
## Number of permutations: 999
```

```
sum(R_gn-group_net)
```

```
## [1] 179.6763
```

```
sum(abs(R_gn-group_net))
```

```
## [1] 180.4819
```

Note for the first reference model, that while the network is fairly well correlated, the values of edge weights recorded are very different and upward biased

---

The first reference model therefore does not explain our observed between-group network well at all. So we now go through and re-simulate subgroups (assuming we have knowledge about their typical properties) and assign a location to every subgroup instead of making the model purely individual-based.

We maintain group sizes from the original population

```
#We have copied/pasted code from where we first generated our GBIs and then change
d object names

#Create a list to store individual IDs
Rindss<-list()

#Create a list to store group sizes
Rgss<-list()

#Create a list to store the sex of each individual
Rsexes<-list()

#Create a list to store the age of each individual
Rages<-list()

#Create a list to store the nose colour of each individual
Rnoses<-list()

#Create a list to store information on which day a subgroup is observed on
Rdaysl<-list()

#Create a list to store a group-by-individual matrix for each burbil group
Rgbis<-list()

#Set the mean number of subgroups observed for each group each day
Rsg_mn<-5

#Set the strength of assortativity based on nose colour
#Set a number between 0 and 1
```

```
Rsg_ass<-0.2

#Genereate association data within each burbil group!
for(j in 1:n_groups){

#individual identities
Rinds<-seq(1,n_inds[j],1)
Rindss[[j]]<-Rinds

#group size
gs<-length(Rinds)
Rgss[[j]]<-gs

#sex
sex<-sample(c("M","F"),gs,replace=TRUE)
Rsexes[[j]]<-sex

#age
age<-sample(c("AD","SUB","JUV"),gs,replace=TRUE,prob=c(0.6,0.2,0.2))
Rages[[j]]<-age

#nose
nose<-sample(c("RED","ORANGE"),gs,replace=TRUE,prob=c(0.7,0.3))
Rnoses[[j]]<-nose

#################################

#Define number of subgroups on the first day
n_sg<-rpois(1,Rsg_mn-1)+1

#find halfway point
max_red<-floor(n_sg/2)

#Sample subgroups on the first day
subgroups1<-sample(n_sg,sum(nose=="RED"),replace=TRUE,prob=c(rep(0.5+Rsg_ass,max_r
ed),rep(0.5-Rsg_ass,n_sg-max_red)))
subgroups2<-sample(n_sg,sum(nose=="ORANGE"),replace=TRUE,prob=c(rep(0.5-Rsg_ass,ma
x_red),rep(0.5+Rsg_ass,n_sg-max_red)))

subgroups<-rep(NA,gs)
subgroups[nose=="RED"]<-subgroups1
subgroups[nose=="ORANGE"]<-subgroups2

#Store relevant information in the group-by-individual matrix and days vector
Rgbi<-matrix(0,nc=gs,nr=n_sg)
Rgbi[cbind(subgroups,seq(1,gs,1))]<-1
Rdays<-rep(1,nrow(Rgbi))

#Repeat process over 100 days of observations
for(i in 2:100){

  n_sg<-rpois(1,Rsg_mn-1)+1

  #find halfway point
  max_red<-floor(n_sg/2)

  subgroups1<-sample(n_sg,sum(nose=="RED"),replace=TRUE,prob=c(rep(0.5+Rsg_ass,max
```

```
_red),rep(0.5-Rsg_ass,n_sg-max_red)))
  subgroups2<-sample(n_sg,sum(nose=="ORANGE"),replace=TRUE,prob=c(rep(0.5-Rsg_ass,
max_red),rep(0.5+Rsg_ass,n_sg-max_red)))

  subgroups<-rep(NA,gss[[j]])
  subgroups[nose=="RED"]<-subgroups1
  subgroups[nose=="ORANGE"]<-subgroups2

  tgbi<-matrix(0,nc=gs,nr=n_sg)
  tgbi[cbind(subgroups,seq(1,gs,1))]<-1
  Rdays<-c(Rdays,rep(i,nrow(tgbi)))
  Rgbi<-rbind(Rgbi,tgbi)
}

#We edit the group-by-individual matrix and days vector to delete any "empty" grou
ps
Rgbi2<-Rgbi[rowSums(Rgbi)>0,]
Rdays<-Rdays[rowSums(Rgbi)>0]
Rgbi<-Rgbi2

Rdaysl[[j]]<-Rdays
Rgbis[[j]]<-Rgbi

}

Rsglocs<-list()
for(i in 1:n_groups){
  tx<-rep(NA,dim(Rgbis[[i]])[1])
  ty<-rep(NA,dim(Rgbis[[i]])[1])
  Rsglocs[[i]]<-data.frame(tx,ty)
  names(Rsglocs[[i]])<-c("x","y")
  Rsglocs[[i]]$x<-group_locs[i,1]+round(rnorm(dim(Rgbis[[i]])[1],0,dist_eff))
  Rsglocs[[i]]$y<-group_locs[i,2]+round(rnorm(dim(Rgbis[[i]])[1],0,dist_eff))
}

#We now calculate the full population association network
R_fn2<-matrix(0,nr=n_tot,nc=n_tot)

#Counts up between-group associations
for(i in 1:100){
  for(j in 1:(n_groups-1)){
    for(k in (j+1):n_groups){
      tA<-paste0(Rsglocs[[j]][,1],"-",Rsglocs[[j]][,2])
      tB<-paste0(Rsglocs[[k]][,1],"-",Rsglocs[[k]][,2])
      tA2<-tA[Rdaysl[[j]]==i]
      tB2<-tB[Rdaysl[[k]]==i]
      tt<-match(tA2,tB2)
      if(sum(is.na(tt))<length(tt)){
      #if(group_clans[j]==group_clans[k]){same<-rbinom(1,1,p_wc)} ###N.B.We have r
emoved clan effects
      #if(group_clans[j]!=group_clans[k]){same<-rbinom(1,1,p_bc)} ###N.B.We have r
emoved clan effects
      same<-1
      if(same==1){
        paste(i,j,k)
        for(m in length(tt)){
          if(is.na(tt[m])==FALSE){
```

```
            tsg1<-which(tA==tA2[m]&Rdaysl[[j]]==i)
            tsg2<-which(tB==tB2[tt[m]]&Rdaysl[[k]]==i)
            tid1<-which(Rgbis[[j]][tsg1,]==1)
            tid2<-which(Rgbis[[k]][tsg2,]==1)
            tid1a<-inds_tot[g_tot==j&gi_tot%in%tid1]
            tid2a<-inds_tot[g_tot==k&gi_tot%in%tid2]
            R_fn2[tid1a,tid2a]<-R_fn2[tid1a,tid2a]+1
            R_fn2[tid2a,tid1a]<-R_fn2[tid1a,tid2a]
          }
        }
      }
      }
    }
  }
}

#Create association network
for(i in 1:(nrow(R_fn2)-1)){
  for(j in (i+1):nrow(R_fn2)){
    R_fn2[i,j]<-R_fn2[i,j]/(200-R_fn2[i,j])
    R_fn2[j,i]<-R_fn2[i,j]
  }
}
for(i in 1:n_groups){
  R_fn2[inds_tot[g_tot==i],inds_tot[g_tot==i]]<-get_network2(Rgbis[[i]])
}

#Create between-group network
R_gn2<-matrix(0,nr=dim(group_locs)[1],nc=dim(group_locs)[1])
for(i in 1:nrow(R_fn2)){
  for(j in 1:ncol(R_fn2)){
    if(g_tot[i]!=g_tot[j]){
      R_gn2[g_tot[i],g_tot[j]]<-R_gn2[g_tot[i],g_tot[j]]+R_fn2[i,j]
    }
  }
}

#Plot between-group network from spatially-explicit reference model with subgroups
RGN2<-graph.adjacency(R_gn2,mode="undirected",weighted=TRUE)
plot(RGN2,vertex.color="light blue",edge.width=(edge_attr(RGN2)$weight)^2)
```

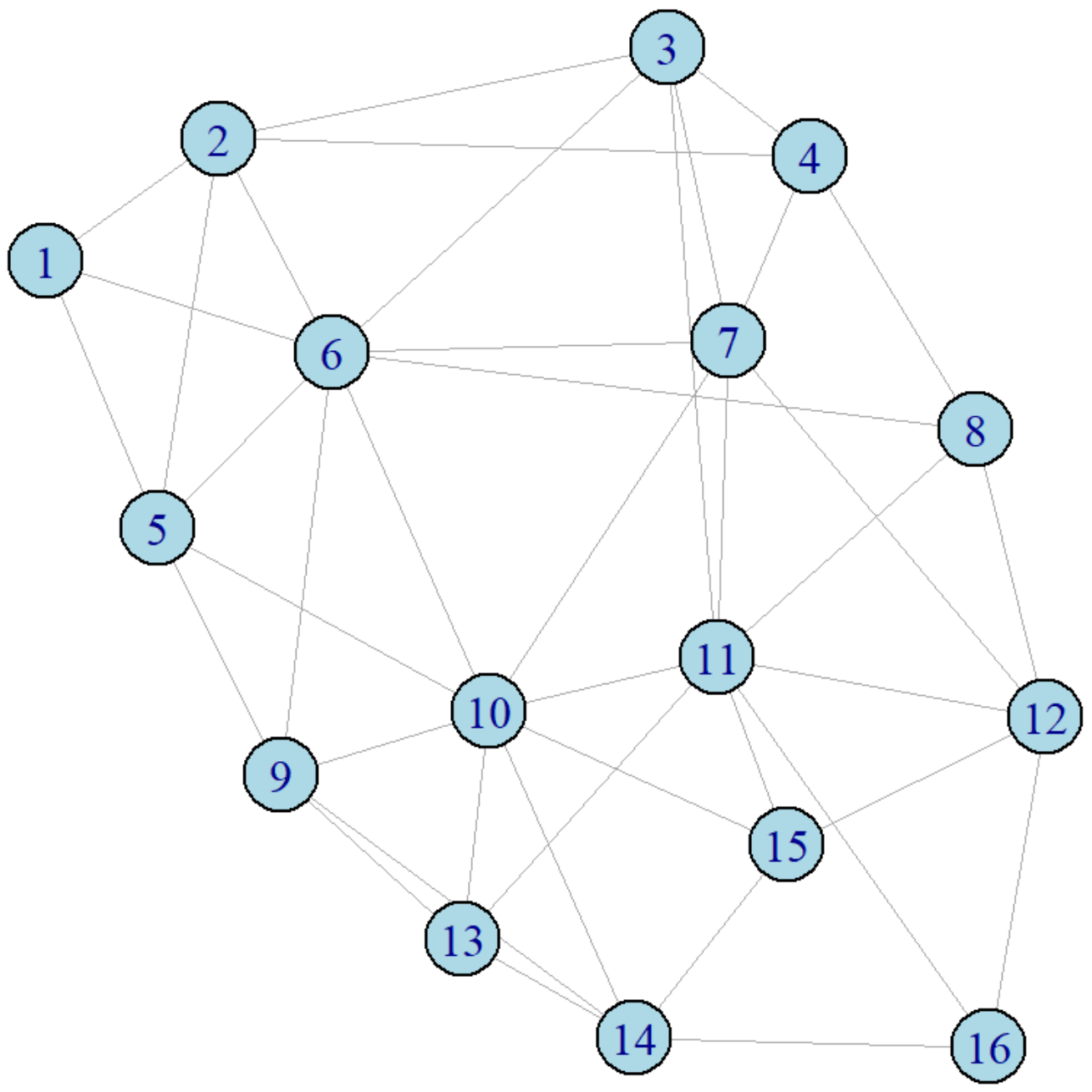

```
#Calculate test statistics
vegan::mantel(R_gn2,group_net)
```

```
##
## Mantel statistic based on Pearson's product-moment correlation
##
## Call:
## vegan::mantel(xdis = R_gn2, ydis = group_net)
##
## Mantel statistic r: 0.5581
##       Significance: 0.001
##
## Upper quantiles of permutations (null model):
##   90%   95% 97.5%   99%
## 0.113 0.161 0.197 0.237
## Permutation: free
## Number of permutations: 999
```

```
sum(R_gn2-group_net)
```

```
## [1] 5.552106
```

```
sum(abs(R_gn2-group_net))
```

```
## [1] 13.3913
```

We can see from our test statistics that the correlation with the observed network is much poorer, but the edge weights are much more similar, although still seemingly overestimated on average

---

We can now develop a third reference model that is socially-explicit, that is we included an effect of clan membership on whether between-group interactions occur between subgroups at the same location (you'll recall this is how we simulated our networks in the first place)

```
#We add our socially explicit parameters here. Note we have retained them as the o
riginal values used to create our burbil world. But please feel free to change the
m to see what effect it had on the network structure
Rp_wc<-p_wc
Rp_bc<-p_bc

#We now calculate the full population association network
R_fn3<-matrix(0,nr=n_tot,nc=n_tot)

#Counts up between-group associations
for(i in 1:100){
  for(j in 1:(n_groups-1)){
    for(k in (j+1):n_groups){
      tA<-paste0(Rsglocs[[j]][,1],"-",Rsglocs[[j]][,2])
      tB<-paste0(Rsglocs[[k]][,1],"-",Rsglocs[[k]][,2])
      tA2<-tA[daysl[[j]]==i]
      tB2<-tB[daysl[[k]]==i]
      tt<-match(tA2,tB2)
      if(sum(is.na(tt))<length(tt)){
      if(group_clans[j]==group_clans[k]){same<-rbinom(1,1,Rp_wc)}
      if(group_clans[j]!=group_clans[k]){same<-rbinom(1,1,Rp_bc)}
      #same<-1
      if(same==1){
        paste(i,j,k)
        for(m in length(tt)){
          if(is.na(tt[m])==FALSE){
            tsg1<-which(tA==tA2[m]&daysl[[j]]==i)
            tsg2<-which(tB==tB2[tt[m]]&daysl[[k]]==i)
            tid1<-which(Rgbis[[j]][tsg1,]==1)
            tid2<-which(Rgbis[[k]][tsg2,]==1)
            tid1a<-inds_tot[g_tot==j&gi_tot%in%tid1]
            tid2a<-inds_tot[g_tot==k&gi_tot%in%tid2]
            R_fn3[tid1a,tid2a]<-R_fn3[tid1a,tid2a]+1
            R_fn3[tid2a,tid1a]<-R_fn3[tid1a,tid2a]
          }
        }
      }
      }
    }
  }
}

#Create association network
for(i in 1:(nrow(R_fn3)-1)){
  for(j in (i+1):nrow(R_fn3)){
```

```
    R_fn3[i,j]<-R_fn3[i,j]/(200-R_fn3[i,j])
    R_fn3[j,i]<-R_fn3[i,j]
  }
}
for(i in 1:n_groups){
  R_fn3[inds_tot[g_tot==i],inds_tot[g_tot==i]]<-get_network2(gbis[[i]])
}

#Create between-group network
R_gn3<-matrix(0,nr=dim(group_locs)[1],nc=dim(group_locs)[1])
for(i in 1:nrow(R_fn3)){
  for(j in 1:ncol(R_fn3)){
    if(g_tot[i]!=g_tot[j]){
      R_gn3[g_tot[i],g_tot[j]]<-R_gn3[g_tot[i],g_tot[j]]+R_fn3[i,j]
    }
  }
}

#Plot between-group network from socially-explicit reference model
RGN3<-graph.adjacency(R_gn3,mode="undirected",weighted=TRUE)
plot(RGN3,vertex.color="light blue",edge.width=(edge_attr(RGN3)$weight)^2)
```

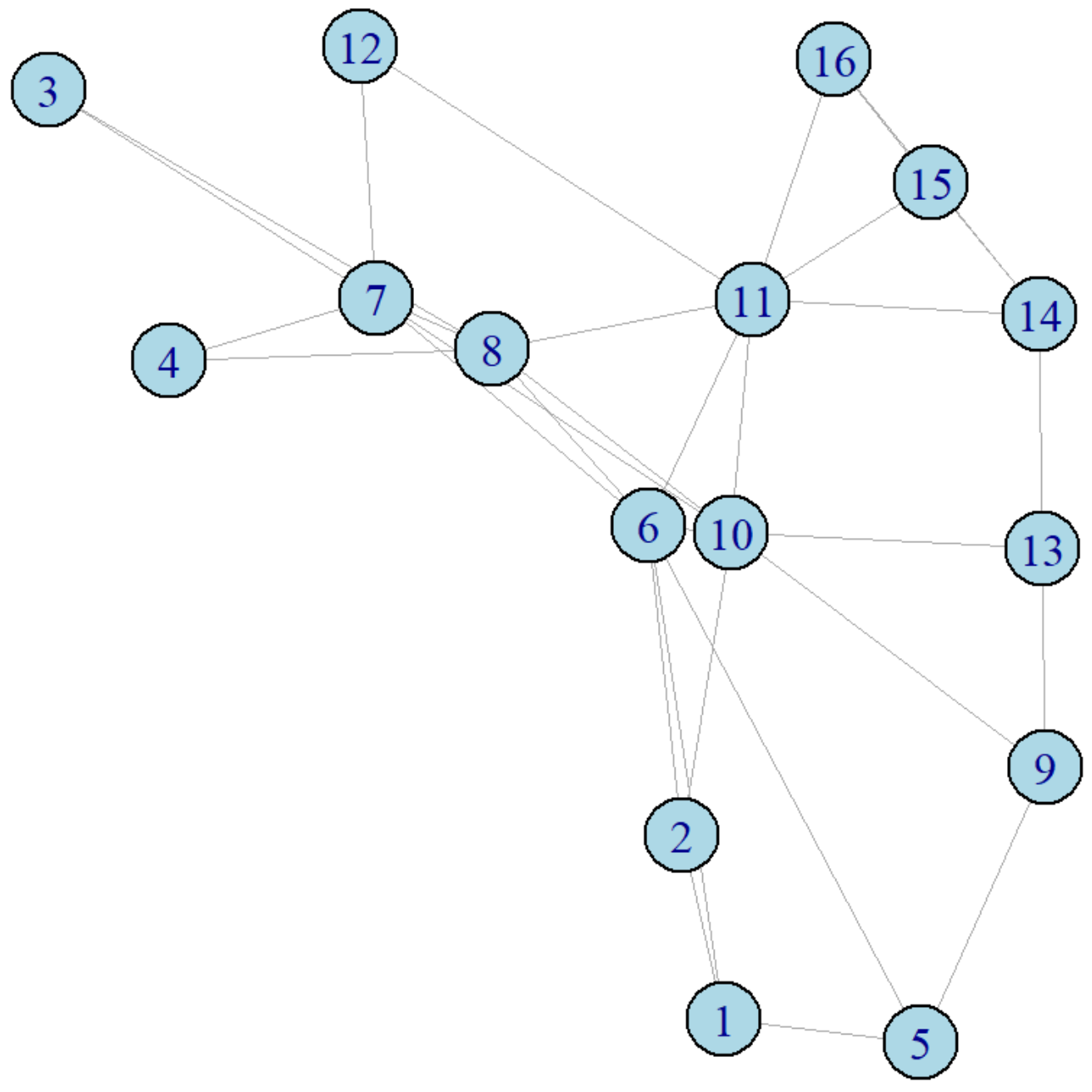

```
#Calculate test statistics
vegan::mantel(R_gn3,group_net)
```

```
##
## Mantel statistic based on Pearson's product-moment correlation
```

```
##
## Call:
## vegan::mantel(xdis = R_gn3, ydis = group_net)
##
## Mantel statistic r: 0.3646
##       Significance: 0.001
##
## Upper quantiles of permutations (null model):
##   90%   95% 97.5%   99%
## 0.114 0.156 0.203 0.247
## Permutation: free
## Number of permutations: 999
```

```
sum(R_gn3-group_net)
```

```
## [1] -1.628445
```

```
sum(abs(R_gn3-group_net))
```

```
## [1] 10.3543
```

The third reference model does a much better job of explaining the observed burbil association network, indicating that including the clan membership is an important factor driving between-group network structure

---

When you use an agent-based model then you may want to pick specific values of key parameters and generate distributions of test statistics (as we have done here). However, you can also use values of your test-statistic to fit agent-based models to your observed network using your chosen test statistic. For example, you could use a Markov Chain or even approximate Bayesian computation (ABC) to produce estimates for parameter values that generate networks most similar to the observed network according to the test statistic you have selected.

---

THE END

---